\documentclass[letterpaper,twocolumn,10pt]{article}

\newif\ifarxiv
\arxivtrue

\ifarxiv
\usepackage[T1]{fontenc}
\usepackage[utf8]{inputenc}
\usepackage{pslatex}
\usepackage{multicol}
\usepackage{mathptmx}  %
\usepackage[kerning,spacing]{microtype}

\usepackage{breakurl}           %
\usepackage{url}                %
\usepackage{xcolor}             %
\usepackage[]{hyperref}         %
\fi

\usepackage{amsmath,amsfonts}
\usepackage{algorithmic}
\usepackage{graphicx}
\usepackage{textcomp}
\usepackage[table]{xcolor}
\usepackage{booktabs}
\usepackage{float}
\usepackage[tableposition=top]{caption}
\usepackage{multirow}
\usepackage{soul}
\usepackage{bm}
\usepackage[numbers,sort&compress]{natbib}
\usepackage[noabbrev,nameinlink,capitalise]{cleveref}
\usepackage{tabularx}
\usepackage{./styles/dirtytalk/dirtytalk}
\usepackage{cryptocode}
\usepackage{balance}

\usepackage{appendix}
\usepackage[round-mode=places, round-pad=false, round-precision=2]{siunitx}

\makeatletter
\def\old@comma{,}
\def\old@semicolon{;}
\catcode`\,=13
\def,{%
  \ifmmode%
    \old@comma\discretionary{}{}{}%
  \else%
    \old@comma%
  \fi%
}
\def\;{%
  \ifmmode%
    \old@semicolon\discretionary{}{}{}%
  \else%
    \old@semicolon%
  \fi%
}
\makeatother

\makeatletter
\AddToHook{cmd/x@supertabular/after}
   {%
     \let\orig@startpbox@action\@startpbox@action
     \let\@startpbox@action\@startpbox
   }
\AddToHook{cmd/ST@restore/after}
   {%
     \let\@startpbox@action\orig@startpbox@action
   }
\makeatother

\ifundef{\ts}{\newcommand{\ts}{\textsuperscript}}{}

\ifundef{\definition}{\newtheorem{definition}{\iflanguage{ngerman}{Definition}{Definition}}}{}
\ifundef{\theorem}{\newtheorem{theorem}{Theorem}}{}

\newcounter{game}

\Crefname{game}{Game}{Games}
\crefname{game}{game}{games}

 \usepackage[shortcuts]{extdash}

 \def\authnotes{0}
\def\anj#1{\ifnum\authnotes=1 \textcolor{red}{AL: #1} \else \fi}
\def\ea#1{\ifnum\authnotes=1 \textcolor{brown}{EA: #1} \else \fi}

\usepackage{url}

\usepackage{flushend}
\usepackage{./styles/supertabular/supertabular}
\usepackage[shortcuts]{extdash}
\usepackage{hyphenat}
\usepackage{csquotes}
\usepackage{tipa}

\definecolor{nature_bluish_green}{RGB}{0, 158, 115}
\definecolor{nature_orange}{RGB}{230, 159, 0}
\usepackage{bytefield}

\DeclareRobustCommand{\hlgray}[1]{{\sethlcolor{gray!25}\hl{#1}}}

\begin{document}

\date{}

\ifarxiv
\title{\Large \bf Trust Nothing: RTOS Security without Run-Time Software TCB (Extended Version)}
\author{
{\rm Eric Ackermann, Sven Bugiel}\\
CISPA Helmholtz Center for Information Security\\
Saarbr\"ucken, Germany\\
\{eric.ackermann,bugiel\}@cispa.de
} %
\fi
\maketitle

\begin{abstract}
    Embedded devices face an ever-expanding threat landscape: 
    vulnerabilities in application software, operating system kernels, and peripherals threaten the embedded device integrity.
    Existing computer-architectural defenses fully consider at most two of these threat vectors in their security model.
    
    This paper aims at addressing this gap using a novel capability architecture.
    To this end, we combine a \emph{token} capability approach suitable for building an untrusted operating system with protection against malicious devices without requiring hardware changes to peripherals.
    
    First, we develop and evaluate a full FPGA implementation of our capability architecture around legacy hardware components.
    Further, we present a soft real-time operating system based on Zephyr that has \emph{no run-time software TCB}. 
    To this end, we \emph{disaggregate} Zephyr's subsystems into small, mutually isolated components.
    All subsystems that exist at run time, including scheduler, allocator and DMA drivers, and all peripherals are \emph{fully untrusted}.
    We believe that our work offers a foundation for more rigorous security-by-design in tomorrow's security-critical embedded devices.

    \end{abstract}

\section{Introduction}
\label{sec:introduction}
Owing to advances in computer architectures, inflexible and expensive hardware circuitry in embedded devices is increasingly being replaced with flexible software running on generic inexpensive systems-on-chip (SoCs).
While reducing engineering cost and time to market, this \emph{softwarization} has lead to an increase of \emph{security vulnerabilities} in embedded devices that compromise integrity and availability of critical services.
Vulnerabilities are commonly found in \emph{application software}, the \emph{operating system kernel}, or \emph{peripheral devices with Direct Memory Access (DMA) facilities}.

Application software vulnerabilities in embedded devices are widespread: An exemplary study of printers found known vulnerabilities in 80\% of the analyzed firmware images ~\cite{cuiWhenFirmwareModifications2013}.
On a broader scale, existing attack taxonomies list thousands of known CVEs for embedded devices~\cite{pappEmbeddedSystemsSecurity2015,themitrecorporationMITREEMB3D2025}.
Exploiting vulnerabilities in application software can be used as a stepping stone to compromise the entire device.
To this end, adversaries can exploit vulnerabilities in the operating system kernel:
In RTOS such as Zephyr, separation of kernel and user space is opt-in and not supported on all compatible SoCs~\cite{silvaOperatingSystemsInternet2019}. Hence, \emph{any software vulnerability causes the attacker to gain complete control over the device}.
Even in Linux, where kernel and application software are separated using a memory management unit (MMU), vulnerabilities that allow kernel exploitation continue to be found~\cite{linDirtyCredEscalatingPrivilege2022,shameli-sendiUnderstandingLinuxKernel2021,chenSLAKEFacilitatingSlab2019,xuCollisionExploitationUnleashing2015,chenLinuxKernelVulnerabilities2011}.
Finally, \emph{devices within the SoC} are also a possible threat vector:
\citeauthor{beniaminiProjectZeroAir2017}~\cite{beniaminiProjectZeroAir2017} demonstrates an over-the-air exploitation of a broadcom mobile WiFi chip, turning a benign DMA device into a hardware trojan.
\citeauthor{markettosThunderclapExploringVulnerabilities2019}~\cite{markettosThunderclapExploringVulnerabilities2019} showcase an implant that poses as a benign Thunderbolt-enabled peripheral, but performs DMA attacks via PCIe encapsulation. 

All existing defenses against the three threat vectors---application software, kernel, devices---can only provide security against two out of three threat vectors.
Generally, kernels utilize MMUs to provide security against compromised application software and devices.
However, MMU designs require trusting the kernel to configure the MMU correctly, making it impossible to achieve \emph{no run-time software TCB}.
The goal of designing operating systems without trusted kernels is far from a new idea---in fact, \citeauthor{dennisProgrammingSemanticsMultiprogrammed1966} \cite{dennisProgrammingSemanticsMultiprogrammed1966} already addressed this very problem with their concept of \emph{capability architectures} in 1966.
Several modern capability architectures capitalize on this idea to reduce or even eliminate the trusted part of the operating system~\cite{yuCapstoneCapabilitybasedFoundation2023,woodruffCHERICapabilityModel2014,kimRVCURERISCVCapability2025,esswoodCheriOSDesigningUntrusted2021}.
However, they share the limitation that they \emph{exclude devices from the threat model}.
Thereby, they trade a minimal \emph{software TCB} for a larger \emph{hardware TCB} or need to \emph{trust their drivers}~\cite{chengAdaptiveCHERICompartmentalization2025}.
As a result, no existing architecture achieves security in the presence of both untrusted OSes and devices simultaneously.

Recently, \citeauthor{ackermannWorkinProgressNorthcapeEmbedded2024}~\cite{ackermannWorkinProgressNorthcapeEmbedded2024} proposed the Northcape concept for a novel capability architecture that closes this gap.
Northcape considers all application software, the kernel and devices possible threats. The key idea is to implement capability memory access control at the \emph{system bus level}.
However, before our work, Northcape existed purely as a theoretical concept without implementation and proof of feasibility.

\paragraph*{Contribution}
Our work aims at answering the following research question:
\emph{How can an embedded system provide security against malicious application software, operating system kernels and peripheral devices without relying on a run-time software TCB while retaining soft real-time guarantees?}
We solve this question using a hardware-software-co-design that comprises both the \emph{Bredi SoC} and the \emph{Skadi RTOS}.

First, we present Bredi (Bre\textipa{D}i) in \Cref{sec:northcape-hardware}. Bredi is the first hardware implementation of Northcape, built around the cva6 RISC-V CPU~\cite{zarubaCostApplicationClassProcessing2019} and an unmodified ethernet subsystem~\cite{advancedmicrodevicesinc.AXI1G25G2024} with DMA~\cite{advancedmicrodevicesinc.AXIDMALogiCORE2022}, targeting an FPGA.
Bredi addresses two security shortcomings of Northcape:
First, akin to CHERITrEE~\cite{vanstrydonckCHERITrEEFlexibleEnclaves2023}, it relied on trusted software for interrupt handling.
Through modifications of cva6, we eliminate this trusted software, facilitating untrusted ISRs.
Real-time availability was also out of scope for Northcape.
Using non-maskable interrupts, we ensure the availability of critical services in the presence of attackers.
Bredi also increases the security of Northcape's \emph{exclusive access} and addresses \emph{capability tag guessing attacks} using our novel \emph{subsystem restrictions}.

Second, we present Skadi (Ska\textipa{D}i) OS in \Cref{sec:skadi-os}.
Skadi is a modified Zephyr RTOS~\cite{ZephyrprojectZephyr2025} that uses the operations provided by Northcape to enforce \emph{run-time isolation} of operating system components into \emph{subsystems} that are \emph{mutually distrusting}.
Using Bredi operations, Skadi facilitates \emph{full spatial and temporal isolation} for all data structures that are private to the subsystems or shared between the subsystems, including \emph{hardware-enforced exclusive access} to any data structure.
A \emph{subsystem call} primitive allows synchronous function invocation between subsystems by \emph{atomically} changing the set of accessible capabilities and control flow.
Skadi implements all operating system services as \emph{isolated subsystems}, leaving \emph{no kernel} and \emph{no run-time software TCB}. Peripheral devices, drivers and ISRs are also removed from the TCB.
\ifarxiv
We will release Skadi and Bredi under an OpenSource license in the near future.
\fi

Finally, we discuss the security and performance of our prototype in \Cref{sec:security} and \Cref{sec:evaluation}.
The results show that in an ideal scenario such as a compute benchmark, Skadi is as fast as an insecure reference.
For I/O-bound scenarios, while Skadi and Bredi currently impose an overhead on response times, the performance of the system remains practical for real-world use.
As \emph{security} was the focus of this work, Skadi and Bredi leave room for future performance optimization at the hardware and software levels.
Overall, we believe that Skadi and Bredi can contribute to a persistent improvement of the security posture of tomorrow's embedded devices such as network infrastructure devices by realizing security-by-design for both the hardware and software stack.

\section{Background: The Northcape Architecture}

We start with a brief introduction of object capability systems in \Cref{sec:object_cap_systems}.
We then introduce Northcape, covering capability representation and translation in \Cref{sec:nc_representation} and operations in \Cref{sec:nc_operations}.
In this section, we present Northcape as implemented in Bredi.
We discuss improvements made to Northcape~\cite{ackermannWorkinProgressNorthcapeEmbedded2024} in \Cref{sec:related_northcape}.
\ifarxiv
Finally, an in-depth discussion of Northcape can be found in \cref{sec:full_overview}.
\fi

\subsection{Object Capability Systems}
\label{sec:object_cap_systems}

First coined by \citeauthor{dennisProgrammingSemanticsMultiprogrammed1966}~\cite{dennisProgrammingSemanticsMultiprogrammed1966} in 1966, object capability architectures separate the physical memory of a computer into \emph{segments} with byte-granular starting and ending addresses.
Subjects (subsystems) need to hold explicit access rights (\emph{capabilities}) for each object (segment) that they access. \emph{Unforgeability} of capabilities ensures security.

Capability architectures do not require privileged modes of operation.
Instead, an explicit \emph{subsystem call} instruction transfers control between subsystems with mutual isolation.

Finally, \emph{capability delegation} from one subsystem to another allows controlled sharing of \emph{specific} data structures, eliminating \emph{ambient authority}~\cite{millerCapabilityMythsDemolished2003}.
Thereby, delegation and subsystem calls facilitate cooperation between mutually isolated subsystems, enabling \emph{untrusted kernels}~\cite{esswoodCheriOSDesigningUntrusted2021}.

\subsection{The Northcape Capability Architecture}
\label{sec:nc_representation}

\begin{figure}[t]
    \includegraphics[width=\linewidth]{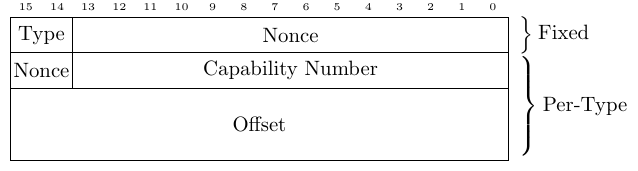}
    \caption{Northcape token format (32-bit offset type shown).}
    \label{fig:token-type}
\end{figure}
\medskip
\textbf{Representations.}
Two distinct representations of object capabilities have emerged in the literature: 1) \emph{Fat-pointer systems} like CHERI~\cite{woodruffCHERICapabilityModel2014} differentiate between capabilities and memory addresses.
Capabilities need to be loaded into explicit \emph{capability registers} and referenced as source or destination addresses in CPU instructions.
CHERI capabilities \emph{themselves} contain segment bounds and access permissions.
Unforgeability of capabilities relies on \emph{tagging} of memory locations that hold valid capabilities, requiring a separate \emph{tag bitmap}.
2) \emph{Token capability systems} like RV-CURE~\cite{kimRVCURERISCVCapability2025} separate capability metadata and representation.
RV-CURE embeds a capability identifier into the high order bits of a pointer. The identifier is resolved to segment bounds via a central metadata table (CMT).
Unforgeability of capabilities relies on a \emph{MAC tag}.

Similar to RV-CURE, Northcape is a token-based capability system.
As illustrated in \Cref{fig:token-type}, Northcape separates 64-bit pointers into an \emph{offset} into a capability in the low-order bits and a \emph{nonce} and \emph{capability number} in the high bits.
The number identifies a CMT entry, which holds the metadata.

Northcape makes the sizes of offset and number \emph{variable} by introducing \emph{offset types}, facilitating a tradeoff between the maximum size and number of supported capabilities.
We support offsets from 8 to 32 bits in 8-bit increments, resulting in $2^{14}$ capabilities with 32 bit offset down to $2^{38}$ capabilities with 8 bit offset \emph{at the same time}.
We accounted for changing these values with minimal effort in our prototype.

The separation of Northcape capability tokens into offset, number and nonce makes it possible to use them interchangeably with C pointers.
Crucially, one can \emph{interpret} a capability token as the \emph{starting position} of an object in memory.
The fields of the objects can then be addressed by incrementing the offset.
Therein, this adjustment of the offset happens \emph{transparently}, facilitating \emph{backwards compatibility} with legacy pointer arithmetic.
In contrast, RV-CURE and CHERI require the use of explicit \emph{instructions} for accessing capabilities~\cite{kimRVCURERISCVCapability2025,woodruffCHERICapabilityModel2014}.

\ifarxiv
\begin{figure}[t]
    \includegraphics[width=\linewidth]{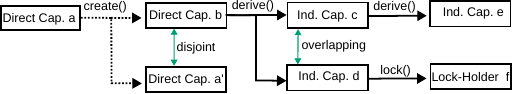}
    \caption{Types of capabilities in the Northcape system.}
    \label{fig:capability-types}
    \label{fig:cap_flow}
\end{figure}
\fi

\medskip
\textbf{Capability Types.}
\ifarxiv
In the following, we give a textual description of Northcape's capability types and their relationships as illustrated in~\cref{fig:cap_flow}.
\fi

Northcape can ensure \emph{sub-object protection}, a common feature in modern object capability systems~\cite{woodruffCHERICapabilityModel2014,yuCapstoneCapabilitybasedFoundation2023,kimRVCURERISCVCapability2025}.
Therefore, it distinguishes \emph{direct capabilities} and \emph{indirect capabilities}:
Direct capabilities have ownership of the segment that they resolve to.
Northcape operations ensure that all existing direct capabilities are \emph{disjoint}.

Indirect capabilities grant access to a subsegment of a direct capability.
Indirect capabilities for the same direct capability may overlap, and can be created \emph{recursively}:
One can create an indirect capability from an indirect capability, decreasing segment bounds (\emph{monotonicity}~\cite{woodruffCHERICapabilityModel2014}).
Northcape tracks the parent of each indirect capability for \emph{temporal security}.

Using the $lock$ operation, subsystems can gain \emph{exclusive access} to capabilities.
To achieve this, $lock$ creates a \emph{lock-holder capability}.
Like indirect capabilities, lock-holder capabilities can permit access to a subset of a parent direct or indirect capability.
However, while the lock-holder exists, \emph{access to all overlapping capabilities is impossible}.

\medskip
\textbf{Capability Unforgeability.}
A successful forgery requires \emph{guessing} a nonce:
This has a likelihood of $\frac{1}{2^{16}}$, as Northcape uses a 16-bit random nonce.
RV-CURE uses the same design and relies on the low likelihood of a successful guess~\cite{kimRVCURERISCVCapability2025}.

We provide a key improvement here:
Optionally, capabilities can carry a \emph{subsystem-ID bound restriction}, permitting access to a capability only for a certain subsystem.
This needs to be indicated when the capability is created, or can be added to an existing capability without such protection using $restrict$.

When a subsystem-ID bound capability is accessed, Northcape ensures that the 32-bit subsystem ID in the capability matches the active subsystem.
To this end, Northcape tracks the ID of the active subsystem across subsystem calls.
Thereby, when a \emph{different subsystem} guesses a nonce, it is unable to access the capability due to its mismatching subsystem ID.
Importantly, the ID does not form an ambient authority since all access is still only possible through capabilities~\cite{millerCapabilityMythsDemolished2003}.

\begin{figure}[t]
    \includegraphics[width=\linewidth]{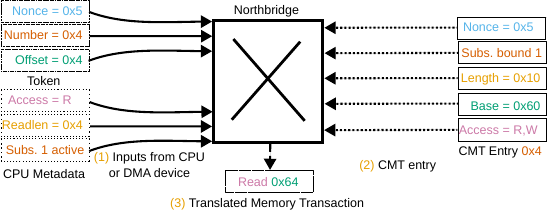}
    \caption{Translation of a Northcape capability token.}
    \label{fig:token-translation_front}
\end{figure}
\medskip
\textbf{Capability Translation.}
\Cref{fig:token-translation_front} illustrates the resolution of capabilities to physical addresses:
Software on a CPU or a DMA controller starts a memory transaction via the system bus \textcolor{nature_orange}{(1)}.
The Northcape-enabled northbridge will \emph{interpret} the provided address as a capability token and parse the number, offset and nonce components.
The number field uniquely identifies a CMT entry.
The northbridge will retrieve this entry \textcolor{nature_orange}{(2)}.
First, it will ensure \emph{unforgeability} by verifying the nonce and any subsystem ID restrictions.
It will then determine whether the requested \emph{offset} and \emph{length} remain within the bounds of the capability, ensuring \emph{spatial security}.
If the resolved capability is indirect or a lock-holder, the northbridge will \emph{recurse} until reaching the \emph{direct capability}, ensuring all parents are valid and unlocked.
This is crucial for \emph{temporal security}, specifically \emph{revocation} and \emph{exclusive access}.
Finally, the northbridge will compute a physical address as the \emph{base} stored in the CMT entry plus the token's \emph{offset} \textcolor{nature_orange}{(3)}.

\subsection{Northcape Capability Operations}
\label{sec:nc_operations}

\ifarxiv
\begin{table*}
    \centering
    \setlength{\aboverulesep}{0pt}
    \setlength{\belowrulesep}{0pt}
    \footnotesize
    \caption{Detailed Northcape operations.}
    \label{tab:northcape_operations_detailed}
    \begin{tabular}{p{3.5cm}|p{2.5cm}|p{10.5cm}}
    \textbf{Syntax}  & \textbf{Short Description} & \textbf{Full Description}\\
    \toprule
    $create(c_a,l,r,p)\longrightarrow c_a',c_b$ &Split a direct capability into two non-overlapping direct capabilities. & Splits a capability $c_a$ into two non-overlapping pieces. Creates one new capability $c_b$ and modifies the metadata of $c_a$ accordingly.
    Caller can specify a length $l$ as well as restrictions $r$ and permissions $p$ for the segment.
    $p$ and $r$ have to be as strict or stricter than the permissions and restrictions of $c_a$, and $l$ must be less than or equal the length of $c_a$.
    Invalidates $c_a$ if $l=l_a$.
    Can only be used for direct capabilities, requiring $c_a$ to not be locked and have a reference count of zero.\\
    \hline
    $merge(c_a,c_b,r,p)\longrightarrow c_m$ & Combine two adjacent direct capabilities into a direct capability. & This operation is the opposite of $create$.
    It creates a new capability $c_m$ from physically adjacent capabilities $c_a$, $c_b$ with new restrictions $r$ and new permissions $p$.
    Permissions $p$ and new restrictions $r$ can be more permissive than those of the input capabilities.
    The input capabilities are destroyed.
    Requires the capabilities to have a reference count of zero and not to be locked.\\
    \hline
    $derive(c_a,l,o,r,p)\longrightarrow c_i$ & Create an indirect capability from a direct or indirect capability. & The $derive$ operation is used to create indirect capabilities from an input capability $c_a$, which are exclusively used by all subsystems and devices except the memory allocator.
    An indirect capability has a (grand)parent that is a direct capability and possibly overlapping siblings.
    To this end, $c_a$ can be either a direct or another indirect capability.
    The output capability $c_i$ is always indirect and can have an additional offset $o$ and a reduced length $l$ in comparison to its parent.\\
    \hline
    $clone(c_a,r,p)\longrightarrow c_i$ & Variant of $derive$ that preserved capability bounds. & A synonym for $derive$ that internally calls $derive$ with an offset $o$ of 0 and a length $l$ that equals the length of the segment.
    Useful for removing permissions from or adding restrictions to a capability a subsystem wants to share, especially in cases where the segment length of the capability is not known to the caller of $clone$. \\
    \hline
    $lock(c_a,r,p)\longrightarrow c_l$ & Gain exclusive access to a capability and return lock-holder capability.&
    Attempts to gain \emph{exclusive access} to a capability.
    The input capability $c_a$ can be either direct or indirect.
    In case the \emph{lockable} permission is set for $c_a$'s grandparent direct capability, a lock-holder capability $c_l$ is created.
    The \emph{lockable} permission prevents abusing $lock$ for DoS.
    As long as it is alive, $c_l$ can be used to access the same (part of the) segment visible through $c_a$.
    At the same time, other capabilities that overlap with $c_a$ cannot be used.\\
    \hline
    $drop(c_a)\longrightarrow b$ & Destroy indirect or lock-holder capability if it has no living children. &
    Destroys capability $c_a$ if it does not have any references and returns whether this was successful.
    Can only be used on indirect and lock-holder capabilities.
    If $c_a$ is a lock-holder capability returned by $lock$, the parent direct capability is unlocked as a side effect.
    Thereby, the caller of $drop$ needs not know what type of capability a token resolves to.\\
    \hline
    $revoke(c_a,r,p)\longrightarrow c_a'$ & Destroy capability, create new capability and overwrite segment with zeros. & Destroys the CMT entry associated with $c_a$ and creates a new direct capability $c_a'$ for the segment identified by the capability with full permissions and removed restrictions.
    \emph{Overwrites} the segment with 0-bytes. Returns the new capability.
    Can only be used on direct capabilities.
    Permissions $p$ and new restrictions $r$ can be more permissive than those of the input capability.\\
    \hline
    $inspect(c_a)\longrightarrow b,l,r,p$ & Determine capability metadata &
    Reads the metadata for any capability $c_a$.
    Resolves $c_a$ recursively to determine base and length if necessary (for lock-holder capabilities, which do not directly encode these information) and the lockable permission.
    The operation acts slightly differently depending on any subsystem-id-bound restrictions on $c_a$: If there is no such restriction or it matches the caller of the operation, the operation returns all mentioned metadata.
    If $c_a$ has a set-subsystem-id restriction, the operation returns \emph{only the $R,W,X,I$ permissions} and the restrictions.
    In other cases, $c_a$ returns an error code.\\
    \hline
    $restrict(c_a,r,p,o^+,l^-)\longrightarrow b$ & Make capability bounds and permissions more restrictive. &
    Makes a capability more restrictive in-place.
    This can be applied to all types of capabilities with slightly different semantics:
    For indirect capabilities, an additional offset $o^+$ and a length minuend $l^-$ can be specified.
    The bounds of the indirect capabilities are reduced accordingly.
    For other capabilities, these parameters are ignored.
    Also, all permissions that are not specified in \emph{both} $p$ and $c_a$'s metadata are dropped.
    The lockable permission is only dropped if $c_a$ is a direct capability; otherwise, attempting to set or clear it has no effect.
    $r$ is only applied when $c_a$ does not currently have restrictions.\\
    \hline
    $calls(c_a)$ & Do a subsystem call. &This operation describes performing a subsystem call.
    On input a capability $c_a$ with set-subsystem-id restriction, set the \emph{non-architectural} subsystem identifier register to the indicated identity and transfer control to the code in the identified segment \emph{atomically}.\\
    \bottomrule
    \end{tabular}
\end{table*}

We will introduce Northcape's operations using three motivating examples: sharing of memory, dynamic memory allocation, and control transfer.
The full list of operations with detailed descriptions is given in~\cref{tab:northcape_operations_detailed}.
\fi

\medskip
\textbf{Secure Memory Sharing.}
Consider DMA transfers: A driver indicates a location $addr$ and length $l$ to a device.
The device is assumed to transfer up to $l$ bytes to or from $addr$ before signalling completion.
A different subsystem can perform a subsystem call into the DMA driver to initiate a transaction.
The call can pass a capability token as an argument. The calling subsystem can \emph{delegate} any subset of any capability that it has access to to the DMA driver.
This is done using the $derive$ or $clone$ operations, which create \emph{indirect capabilities} from a capability. Therein, $derive$ allows shrinking capability bounds while $clone$ preservers them.
Additionally, bounds can be shrunk and access rights can be removed using $restrict$.
The $inspect$ operation allows the DMA driver to determine metadata of a capability for sanity checks.

The DMA driver can delegate (a subset of) the received capability to the DMA device. Northcape will ensure \emph{spatial security}, i.e., confine the DMA device to the bounds of the capability with byte granularity.
After completion of the transfer, the driver can destroy the delegated capability using $drop$, which only succeeds for capabilities without living children and invalidates the capability, ensuring \emph{temporal security}.

In some cases, such as the DMA descriptors used by Xilinx' DMA~\cite{advancedmicrodevicesinc.AXIDMALogiCORE2022}, the DMA and its driver take turns reading and writing the same object.
Here, the DMA driver can use $lock$ to gain exclusive access to the descriptor, preventing time of check to time of use (TOCTOU) attacks by the device.

\medskip
\textbf{Memory Management.}
As noted earlier, memory is owned exclusively by \emph{direct capabilities}.
This is useful for slab allocators.
The allocator can be provided a large direct capability as an initial heap.
When subsystems request memory via a subsystem call, the allocator can use $create$ to slice off direct capabilities from the heap.
Thereby, allocated capabilities can be locked without the allocator losing access to the heap.

Freed memory can be returned to the heap using a $merge$ operation that combines two adjacent capabilities into one and restores access permissions.
Together, $create$ and $merge$ are sufficient for implementing a slab allocator~\cite{yurchenkoAlgorithmDynamicSegmented1981}.

Finally, different from $create$, the $revoke$ operation can be used on a direct capability \emph{that is locked or has children}.
It \emph{overwrites} the associated segment with zeros, ensuring that no secrets are leaked, and creates a new output capability with no children.
$revoke$ is useful, e.g., for reclaiming memory of subsystems after a crash by an \emph{untrusted allocator}.

\medskip
\textbf{Control Transfer with Mutual Isolation.}
Finally, Northcape provides a \emph{subsystem call} operation $calls$, which allows \emph{atomically} changing both the program counter of a CPU and the active subsystem ID, changing the sphere of protection.
$calls$ can only be invoked on the \emph{first byte} of capabilities with special \emph{set-subsystem-ID} restrictions:
Consequently, $calls$ allows transfers to explicit \emph{entry points} while ensuring a subsystem ID cannot be \emph{hijacked} by a different subsystem.

Set-subsystem-ID restrictions need to be assigned when a capability is created.
Crucially, subsystems can only create set-subsystem-ID restrictions for their own subsystem ID.
Otherwise, it would be trivial to hijack foreign subsystem IDs.
We make one exception to this rule:
a special subsystem ID 0 can create \emph{arbitrary} set-subsystem-ID restrictions.
Conceptually, subsystem ID 0 is exclusively used by the trusted Skadi Loader, which needs this privilege to bootstrap the remaining subsystems.
The loader makes subsystem ID 0 inaccessible at the end of load time, preventing abuse.
\section{Security Model}
\label{sec:security-model}

\citeauthor{lefeuvreSokSoftwareCompartmentalization2025}~\cite{lefeuvreSokSoftwareCompartmentalization2025} propose the following security model for a compartmentalization mechanism:
    \textbf{\textsf{C}} (Confidentiality): A subject cannot \emph{read} out of its protection domain,
    \textbf{\textsf{I}} (Integrity): A subject cannot \emph{write} out of its protection domain,
    \textbf{\textsf{A}} (Availability): A subject cannot \emph{prevent} other protection domains from executing normally.
We adapt this security model to the combined system comprising Skadi and Bredi.
Therein, our subjects refer to \emph{subsystems}.
The \emph{protection domain} (in capability terminology: \emph{sphere of protection}~\cite{dennisProgrammingSemanticsMultiprogrammed1966}) of a subsystem entails all capabilities that were \emph{explicitly delegated} to it.
This comprises per-subsystem code and data segments as well as any capabilities received as arguments to subsystem calls.
For \textbf{\textsf{C}} and \textbf{\textsf{I}}, we guarantee \emph{spatial} and \emph{temporal} security without requiring explicit \emph{memory sweeps}~\cite{xiaCHERIvokeCharacterisingPointer2019}.
Finally, we extend the definition of \textbf{\textsf{A}} to include \emph{soft real-time execution}.

We further require \emph{exclusive access} to capabilities for which properties \textbf{\textsf{C}} and \textbf{\textsf{I}} are to be protected:
Subsystems can assume that they have exclusive access to private code and data segments by default, excluding subsets that they delegated to other subsystems explicitly.
Subsystems can gain exclusive access to a shared capability using the \emph{lock} operation, after which the mechanism must guarantee \textbf{\textsf{C}} and \textbf{\textsf{I}}.
Furthermore, subsystems can use the \emph{drop} operation to \emph{un-share} a capability, regaining \textbf{\textsf{C}} and \textbf{\textsf{I}} for that capability.

Let the hardware trusted computing base (TCB) refer to the CPU, capability-enforcing northbridge and the system memory including buses.
The software TCB comprises solely the \emph{Skadi loader}. The loader is destroyed at the end of the boot sequence, leaving \emph{no run-time software TCB}.
All \emph{software and hardware} components outside of the TCB are assumed openly malicious.
Side-channel and physical attacks are an orthogonal problem and thus excluded.

We further define the \emph{availability TCB (aTCB)} as the TCB in conjunction with all software procedures involved in \emph{servicing interrupts}.
Note that the aTCB components that are not also in the TCB are only trusted for \textbf{\textsf{A}}---they are allowed to attack security goals \textbf{\textsf{C}} and \textbf{\textsf{I}}, e.g, by manipulating registers.

The \textbf{\textsf{A}} property is in conflict with our \textbf{\textsf{I}} property:
A software subsystem which disables interrupts and proceeds into an infinite loop can prevent other subsystems from executing ever again.
Hence, we need to be able to \emph{preempt} subsystems and let other critical subsystems run.
However, being able to \emph{preempt} such an adversary requires being able to prevent it from disabling interrupts.
In order to resolve this conflict without having to fall back to using CPU-enforced privilege rings, we relax our \textbf{\textsf{A}} goal:
The property is only required if either the adversary \emph{does not disable interrupts} or the availability-critical subsystem is an \emph{interrupt handler}.
We explicitly allow the availability-critical subsystem to \emph{not be the first-level interrupt handler}, however, relying on the aTCB we defined above to guarantee availability on the interrupt handler chain.

\section{Bredi SoC Implementation}
\label{sec:northcape-hardware}

This section covers our hardware implementation of Northcape, Bredi.
We will briefly introduce Bredi's core components.
We will then discuss three core features of Bredi that are crucial for achieving our security goals: 
First, \Cref{sec:Hierarchie_Skip} discusses how our caching hierarchy can hide the overhead of recursive capability resolution, which is critical for our availability goal (\textbf{\textsf{A}}).
Second, \Cref{sec:irq_sec} discusses changes to interrupt handling that were necessary to exclude ISRs from the TCB, a necessity for confidentiality (\textbf{\textsf{C}}), integrity (\textbf{\textsf{I}}) and availability (\textbf{\textsf{A}}).
Finally, \Cref{sec:isa_add} discusses custom CPU instructions needed for \textbf{\textsf{A}}.
\ifarxiv
A more detailed description of all components is given in \cref{sec:northcape-hardware-appendix}.
\fi

\begin{figure}[t]
    \includegraphics[width=\linewidth]{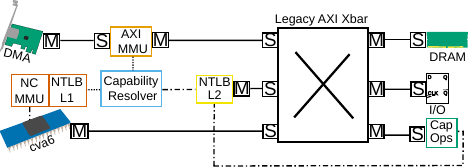}
    \caption{Bredi architecture. Solid lines denote AXI, dotted lines AXI-Stream, dash-dotted lines custom buses.}
    \label{fig:northcape-soc}
\end{figure}

Bredi is based on cva6~\cite{zarubaCostApplicationClassProcessing2019}, a Linux-capable in-order-commit application processor.
We configure the RV64G instruction set (64-bit words, full register set, full FPU support) with a PMP and the sv39 MMU~\cite{risc-vinternationalRISCVInstructionSet2025}.
We incorporate the proprietary DMA-based Xilinx AXI Ethernet subsystem~\cite{advancedmicrodevicesinc.AXIDMALogiCORE2022,advancedmicrodevicesinc.AXI1G25G2024} for evaluating DMA compatibility and performance (\textbf{\textsf{A}}).
We also use Xilinx' proprietary AXI crossbar~\cite{advancedmicrodevicesinc.AXIInterconnectV212022}.
Furthermore, we have implemented novel components for Northcape support as detailed in \cref{fig:northcape-soc}: a Northcape Memory Management Unit (NC MMU) with TLB (NTLB L1) for cva6, an AXI MMU, a Capability Resolver, Capability Operations (CapOps), and a level-2 Northcape TLB (NTLB L2).

\medskip
\textbf{AXI MMU.}
The AXI MMU enforces memory access restrictions on DMA devices, akin to an IOMMU or CHERI CapChecker~\cite{chengAdaptiveCHERICompartmentalization2025}.
It interfaces with  a DMA device via its front-side bus (AXI~4~\cite{armltd.AMBAAXIACE2011}), performs access control and translation from capability tokens to physical addresses as depicted in \cref{fig:token-translation_front} and forwards bus transactions to the AXI interconnect.
Remember that the translation is \emph{opaque} to the device.

The AXI MMU has no cache, relying on the capability resolver and NTLB L2 instead. This leads to a marginal performance degradation, but reduces chip area significantly.

\medskip
\textbf{cva6 MMU.}
Bredi's cva6 contains an instruction and a data Northcape MMU, performing the same tasks as the AXI MMU, but in cva6' pipeline.
The cva6 MMU contains a small, fully-associative TLB (NTLB L1) that maps capability tokens to bounds, permissions and restrictions and handles misses via the capability resolver.

\medskip
\textbf{Capability Resolver.}
All cva6 and AXI MMUs interface with a single Northcape Capability Resolver via a round-robin arbiter.
In doing so, the capability resolution and parsing logic only needs to exist once, irrespective of how many MMUs exist in the system, reducing chip area.
The resolver is pipelined, allowing it to accept one resolution request per cycle and send one resolution response per cycle.

\medskip
\textbf{Capability Operations Module.}
The penultimate component in the Northcape system is the operations module, which implements the Northcape operations that we introduced in \cref{sec:nc_operations} in a hardware finite state machine.
To this end, it is the only device in the system with write access to the CMT via the NTLB L2.
Devices and software can access it via MMIO.
The MMIO interface is \emph{locked} to a subsystem ID after a subsystem starts setting up an operation, ensuring that concurrent devices cannot disturb the interaction.

\medskip
\textbf{NTLB L2.}
\label{sec:ntlb-l2}
Finally, the operations module and resolver share the set-associative write-through dual-port NTLB L2.
In contrast to the NTLB L1, the NTLB L2 maps capability tokens to the \emph{full CMT entry}.
Thereby, both the capability resolver and operations unit can operate on NTLB L2 entries and need not access the CMT in memory, reducing memory accesses significantly.
The dual-port access alleviates the need for cache arbitration, benefiting availability (\textbf{\textsf{A}}).

\subsection{Contribution 1: Cache Hierarchy (\textbf{\textsf{A}})}
\label{sec:Hierarchie_Skip}

As detailed in \Cref{sec:nc_representation}, Northcape capabilities can be direct or indirect.
CMT entries provide all necessary information for translating a capability token and enforcing segment bounds, permission and restriction enforcement.
In effect, CMT entries can be \emph{cached} like page table entries in a TLB.

Hence, the cva6 MMU has a small, fully-associative first-level NTLB.
The NTLB L1 maps capability IDs to bounds, restrictions, and permissions and thereby facilitates completing the address translation process within one cycle.
Thus, cache hits \emph{do not block cva6' pipeline}, akin to the TLB used in the sv39 MMU ~\cite{risc-vinternationalRISCVInstructionSet2025}.
Our evaluation in \Cref{sec:compute-perf} shows that high NTLB L1 hit rates provide performance indistinguishable from a baseline scenario \emph{without memory protection}.

However, NTLB L1 entries can become \emph{stale} after completion of a Northcape operation such as \emph{lock} or \emph{drop}.
Stale entries can be used to read or write data outside of the \emph{sphere of protection} of a subsystem, jeopardizing \textbf{\textsf{C}} and \textbf{\textsf{I}}.
Our implementation destroys stale NTLB L1 entries at the end of each operation to ensure \textbf{\textsf{C}} and \textbf{\textsf{I}}.
To this end, the operations module broadcasts the IDs of all modified capabilities to the L1 NTLBs. If this capability is part of the cache, it is removed, prompting a new recursive resolution via the resolver.
In case of $revoke$ or $lock$, the entire NTLB L1 is invalidated, making sure no overlapping capabilities of the target are contained.

The NTLB L2 can also contain \emph{stale entries}.
However, the NTLB L2 cache is \emph{write-through}. All operations that only affect the input capability and its direct parents ($derive$, $drop$, $clone$, $create$, $merge$, $inspect$, $restrict$) \emph{cannot create stale entries}, which is why the cache does not need to be invalidated after these operations.
On the other hand, $lock$ and $revoke$ can affect \emph{any overlapping capability} of the input capability.
Hence, these operations can leave \emph{stale entries}, which need to be removed to ensure \textbf{\textsf{C}} and \textbf{\textsf{I}}.

A simple solution for the stale entries problem in the NTLB L2 is \emph{flushing the cache} on $lock$ and $revoke$.
We provide the following solution that benefits \textbf{\textsf{A}}:
After $lock$ or $revoke$, we taint \emph{all cache entries} as potentially stale.
After every NTLB L2 cache hit on a stale entry, we force the capability resolver to perform a full hierarchical resolution of the capability.
If the entry turns out to be valid, we remove the \emph{stale} taint.
Otherwise, we remove the stale entry from the cache.

\subsection{Contribution 2: Untrusted ISRs (\textsf{C},\textsf{I},\textsf{A})}
\label{sec:irq_sec}
Handling interrupts is a particular challenge for Bredi: On one hand, interrupt service routines are outside of the TCB.
As a result, we need to prevent ISRs from gaining access to register contents of interrupted subsystems or breaking subsystem-ID protection (crucial for \textbf{\textsf{C}}, \textbf{\textsf{I}}).
However, ISRs are critical for \textbf{\textsf{A}}: subsystems should not be able to prevent ISRs from executing.
Finally, for real-time operation (\textbf{\textsf{A}}), \emph{preemptive scheduling} is desirable.
We implement the following extensions to cva6 to fulfill these requirements: register stacking, interrupt vectoring, voluntary preemption, non-maskable interrupts.

\begin{figure}
    \centering
    \includegraphics[width=\linewidth]{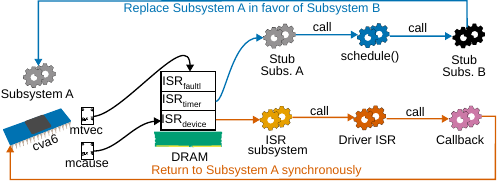}
    \caption{Vectored interrupts in Bredi cva6.}
    \label{fig:northcape-interrupts}
\end{figure}

\medskip
\textbf{Register Stacking (\textsf{C},\textsf{I}).}
We add a second copy of the register set to cva6.
For each issued instruction, we track whether it is part of the \emph{interrupt regime}: The interrupt regime begins with the first instruction of the ISR and ends with the return-from-interrupt \emph{mret} instruction.
Instructions in the interrupt regime use one copy of the register set, while instructions in the non-interrupt regime use the other.
In doing so, registers are never leaked from an interrupted subsystem to an ISR.

In addition to general-purpose and floating-point registers, CSRs are also duplicated for both regimes.
As a result, ISRs cannot modify CPU state for the running subsystem.
This also prevents ISRs from updating \emph{mepc}, the interrupted address, ensuring that the ISR always returns to the instruction that it interrupted.
Hence, ISRs cannot \emph{hijack} execution under the interrupted subsystem's ID.
At the same time, we made sure that within one regime, CSRs behave exactly as in the RISC-V specification~\cite{pattersonRISCVReaderOpen07}.
In effect, this ISA change only affects few lines of assembly in the interrupt handler.

The MMIO registers in the operations module are also duplicated for both regimes.
Accordingly, ISRs can perform capability operations as normal without interfering with the interaction of the interrupted subsystem with the module.

\medskip
\textbf{Interrupt vectoring (\textsf{C},\textsf{I}).}
We implement an interrupt vectoring scheme that matches the proposed FastIRQ extension~\cite{risc-vinternationalCoreLocalInterruptController2025}, see \cref{fig:northcape-interrupts}.
The CPU maps each interrupt cause to a Northcape capability token, depending on interrupt cause, using a \emph{vector table}.
Skadi populates the table with set-subsystem-ID tokens for the interrupt and exception subsystems.
Consequently, cva6 performs a \emph{subsystem call} into the handler when taking an interrupt, ensuring subsystem-ID isolation between interrupted subsystem and ISR.

Crucially, the subsystem ID in the cva6 MMU is also tracked separately for the interrupt and non-interrupt regime, ensuring immutability of the non-IRQ ID during interrupts.

\medskip
\textbf{Voluntary Preemption (\textsf{A}).}
The timer interrupt is \emph{exempt from register stacking}: taking timer interrupts does not transition cva6 into the IRQ regime.
Thus, opposite to device ISRs, we remain on the non-ISR register set, allowing us to \emph{reschedule}.
This mechanism is used for preemption in Skadi:
Each subsystem receives a write-only capability token that resolves to the vector table entry for the timer interrupt.
Subsystems register the address of a \emph{timer stub} in their code segment for the timer interrupt on each subsystem call.
Thereby, when a timer interrupt is taken, the timer stub of the current subsystem executes.
This timer stub can save and clear registers and perform a subsystem call into the scheduler, which can reschedule and return into any subsystem.
Subsystems can disable timer interrupts, making this scheme \emph{opt-in}.

\medskip
\textbf{Non-maskable Interrupts (\textsf{A}).}
Finally, interrupt causes can be set as \emph{non-maskable} using a write-once register.
Non-maskable interrupts (NMIs) are always taken, except if the CPU is already executing an NMI.
As a result, the timely execution of critical ISRs can be enforced.

\subsection{Contribution 3: Custom Instructions (\textsf{A})}
\label{sec:isa_add}

Capability access and subsystem calls are \emph{implicit} operations in Northcape.
In other words, capabilities are accessed using ordinary load, store and control flow instructions, akin to virtual addresses.
Thus, in contrast to capability systems with explicit capability instructions like CHERI~\cite{woodruffCHERICapabilityModel2014} and RV-CURE~\cite{kimRVCURERISCVCapability2025}, Northcape does not require custom instructions for \textbf{\textsf{C}} and \textbf{\textsf{I}}.
However, we provide three customizations for improving the \emph{performance} of the system, critical for \textbf{\textsf{A}}.

Bredi provides a custom instruction $calls$ for subsystem calls, which is a variant of \emph{jalr} that checks the validity of the destination address to be a subsystem call target and triggers an exception otherwise, eliminating one $inspect$ operation.

A second custom instruction zeros registers based on a bitmask, saving 18 cycles on a two-way subsystem call.

Finally, we provide custom CSRs for accessing the operations module, eliminating \emph{fences} that would be needed for the MMIO interface and saving tens of cycles per operation.

\section{Skadi RTOS Implementation}
\label{sec:skadi-os}

Our Skadi operating system takes advantage of Bredi in order to maximize the compartmentalization of the operating system.
We start with a design overview before detailing three core contributions: \Cref{sec:no-runtime-tcb} discusses how no run-time software TCB is achieved (critical for confidentiality \textbf{\textsf{C}}, integrity \textbf{\textsf{I}}), \Cref{sec:no-os-kernel} discusses the lack of a privileged kernel (\textbf{\textsf{C}}, \textbf{\textsf{I}}), and \Cref{sec:subsys-cc} discusses performance optimizations for Skadi's subsystem calls (needed for availability \textbf{\textsf{A}}).
\ifarxiv
A more detailed discussion of Skadi's implementation is provided in \cref{sec:skadi-os-appendix}.
\fi

Skadi is based on the Zephyr~\cite{ZephyrprojectZephyr2025} real-time operating system.
Zephyr is highly modular and configurable: it provides a minimal operating system kernel that implements scheduling and initialization.
In addition to that, a large selection of components including a network stack and device drivers can optionally be built.
Zephyr is optimized for embedded devices with little memory and low processing power.
As a consequence, Zephyr follows the unikernel paradigm, i.e., kernel, optional components and applications are linked into a single binary and execute in a single address space.
On boards with hardware support for memory isolation, e.g., an MPU, Zephyr can separate application logic from kernel and subsystems.
However, compromise of kernel or subsystems will inevitably cause compromise of the whole system~\cite{silvaOperatingSystemsInternet2019}.

Skadi takes advantage of Zephyr's highly modular design by introducing mutual run-time isolation of the existing components as \emph{subsystems} using subsystem calls.
We further break up the kernel into initialization logic, which mostly becomes part of the Skadi loader, the scheduler, which becomes an unprivileged subsystem, and interrupt handling, which also becomes a subsystem.
As a result, Skadi has \emph{no privileged kernel}.
\ifarxiv
\begin{figure}
    \centering
    \includegraphics[width=\linewidth]{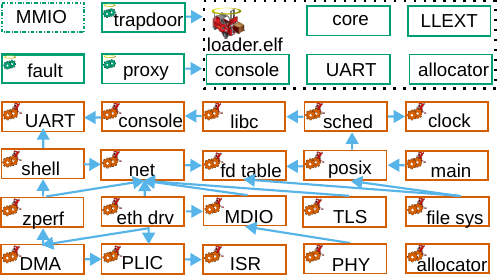}
        \caption{Compartmentalized subsystems in Skadi as well as loader architecture (dashed box). Arrows represent subsystem calls. First two rows of subsystems and loader are trusted: the fault subsystem is only trusted to halt the system after an exception, whereas \emph{trapdoor} and \emph{proxy} become inaccessible at the end of load time. MMIO capabilities are created by the loader and delegated to specific subsystems.}
        \label{fig:subsystems-skadi}
\end{figure}
We depict exemplary Skadi subsystems and their dependencies in~\cref{fig:subsystems-skadi}.
\fi

\subsection{Contribution 1: No Run-Time TCB (C,I)}
\label{sec:no-runtime-tcb}

Skadi subsystems are compiled individually into relocatable ELF binaries.
Thus, during the boot process, the subsystems need to be \emph{relocated} into their own protected address space.
Hence, we provide a minimal privileged firmware: the \emph{Skadi Loader}.
The loader is conceptually part of the \emph{Bredi SoC} and is assumed to be \emph{physically protected} in an immutable ROM.

After system reset, the Skadi loader starts execution at subsystem ID 0.
At this point, only Northcape's \emph{root capability} exists, which comprises the entire physical address space and is encoded such that type, ID and nonce are zero.
In effect, the loader can use it indistinguishably with ordinary 32-bit physical addresses initially.
The loader adds a subsystem-ID restriction of 0 to the root capability to make it inaccessible to subsystems and DMA devices.

The loader then proceeds to \emph{relocate} subsystems:
It first allocates capabilities for code, data and bss segments using $create$ operations as outlined in \Cref{sec:nc_operations}.
The loader proceeds to decompress the LZ4~\cite{colletLz4Lz4Extremely2025}-compressed ELF segments into said capabilities.
This is followed by a standard ELF run-time linking and relocation process: We compile subsystems using RISC-V's \emph{large code model}~\cite{RiscvelfpsabidocRiscvelfadocMaster}, where code and data symbols are referenced via relocation tables co-located with each function. 
Relocation table entries are 64 bits wide and natively fit Northcape tokens.
Hence, subsystems can import and export data and subsystem calls as simple ELF symbols, which the loader can resolve via standard linker symbol tables.
This allows us to use a standard compiler and linker, requiring no modifications to the toolchain. In contrast, RV-CURE and CHERI require custom compiler support~\cite{kimRVCURERISCVCapability2025,woodruffCHERICapabilityModel2014}.

Subsystems can also import \emph{MMIO mappings} by importing an ELF symbol that encodes the physical address and length of the MMIO region, for which the loader creates a capability accordingly.
We maintain compatibility with Zephyr's build-time device tree parsing by declaring the imported MMIO regions as \emph{functions} and using their \emph{address}.

Subsystems can register \emph{initialization functions} to perform setup work.
To this end, the loader will assign each subsystem a monotonic subsystem ID from Northcape's 32-bit subsystem ID space.
It will then create a set-subsystem-ID capability for that subsystem ID, resolving to a simple \emph{initialization trampoline}.
It will also apply \emph{subsystem-ID bound restrictions} to the capabilities holding the subsystem's code and data, ensuring they cannot be used by other subsystems.
The loader can perform a \emph{subsystem call} into the initialization trampoline, ensuring mutual isolation from the subsystem.
The trampoline then jumps into the code segment of the subsystem, which begins execution at its assigned subsystem ID, but is unable to compromise the loader or any other subsystem.

As soon as all subsystems are loaded and initialized, the loader will destroy itself by transitioning into the \emph{trapdoor} subsystem.
The \emph{trapdoor} subsystem performs $revoke$ on what remains of the root capability.
This invalidates all set-subsystem-ID capabilities with subsystem ID 0, making it impossible to re-gain execution at that subsystem ID.
For that reason, all secrets accessible to the loader---such as direct capabilities for subsystem code and data---become inaccessible.
The \emph{trapdoor} then performs a subsystem call into the \emph{scheduler} without valid return address.
From that point on, only \emph{unprivileged} subsystems with non-zero ID remain.
As a result, Skadi manages to achieve \emph{no run-time software TCB}.

\subsection{Contribution 2: No OS Kernel (C,I)}
\label{sec:no-os-kernel}

Skadi has no kernel.
Instead, all kernel functionality is realized via compartmentalized \emph{subsystems} that cooperate using \emph{subsystem calls}.
Crucially, subsystem calls require no trusted intermediary like the \emph{Switcher} in CHERIoT RTOS~\cite{amarCHERIoTRTOSOS2025}.

A \emph{scheduler} subsystem maintains thread control blocks and implements scheduling policies.
This is also where \emph{cancellation points} such as \emph{k\_fifo\_get} are implemented.
Waiting at a cancellation point is done via a \emph{subsystem call}.

The \emph{voluntary preemption} mechanism explained in \Cref{sec:irq_sec} allows us to implement \emph{preemptive scheduling}.
Remember that each subsystem call updates the handler for the \emph{timer interrupt}.
Thus, each subsystem can provide a \emph{timer stub} in its own code segment and ensure it is registered whenever the timer interrupt is enabled.
This timer stub can then perform a subsystem call into the yield endpoint, which can decide whether the system should reschedule.
In case the active thread changes, the scheduler can perform a normal \emph{context switch} on its own stack and then simply return to a different subsystem.
Otherwise, it returns to the timer stub.

Device interrupts and exceptions are handled via separate subsystems.
The loader registers the correct subsystems for each interrupt cause.
The \emph{fault} subsystem is responsible for safely halting the system when encountering an exception. It has no other function and is registered only for exceptions.
The \emph{ISR} subsystem contains the CPU-specific first-level interrupt service routine.
On Bredi, it will defer interrupts directly to the second-level interrupt controller (PLIC) via a subsystem call, which will call the interrupting device driver's ISR.

Typical library functionality is provided either \emph{inline} or as a subsystem:
The \emph{libc} subsystem implements standard C library functions like \emph{printf}.
Basic string functions like \emph{memset} can optionally be inlined into each subsystem for increasing performance.
We also implement an untrusted allocator subsystem following the strategy introduced in \Cref{sec:nc_operations}.

Finally, all drivers are subsystems.
Drivers can export device structs akin to Zephyr's device model~\cite{zephyrprojectmembersandindividualcontributorsDeviceDriverModel2025}.
The loader will derive corresponding capabilities and make them available to other subsystems, which can use them to call the device API via subsystem calls.
Drivers can register with the PLIC subsystem for servicing IRQs.
Finally, DMA drivers can \emph{delegate} capabilities to devices akin to the strategy in \Cref{sec:nc_operations}.
Crucially, DMA devices can \emph{only} use capabilities explicitly delegated to them.
Thereby, Skadi provides byte-granular spatial and temporal security for DMA devices.

\subsection{Contribution 3: Calling Convention (A)}
\label{sec:subsys-cc}

\begin{figure}[t]
    \centering
    \includegraphics[width=\linewidth]{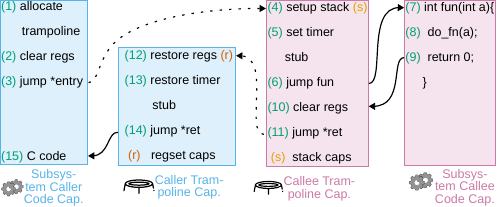}
    \caption{Full subsystem call. Solid lines denote control flow, whereas dashed lines denote subsystem calls.}
    \label{fig:subsys-call}
\end{figure}
Our subsystem calling convention is detailed in \cref{fig:subsys-call}:
The caller allocates a return trampoline \textcolor{nature_bluish_green}{(1)}, zeros all registers to prevent information leakage \textcolor{nature_bluish_green}{(2)} and jumps into the \emph{callee trampoline} \textcolor{nature_bluish_green}{(3)}.
The callee trampoline allocates a stack \textcolor{nature_bluish_green}{(4)} and sets the timer stub \textcolor{nature_bluish_green}{(5)}.
After that, the C function is invoked \textcolor{nature_bluish_green}{(6-9)}.
Finally, the callee trampoline zeros its registers \textcolor{nature_bluish_green}{(10)} and returns \textcolor{nature_bluish_green}{(11)}.
The caller trampoline then restores the register set and timer stub \textcolor{nature_bluish_green}{(12-13)} before returning to the C code \textcolor{nature_bluish_green}{(14-15)}.
Crucially, code and data of the trampolines are part of each subsystems' code and data segment, which is why they require \emph{no trusted intermediary}.
The used set-subsystem ID capabilities are created by the Skadi loader during relocation.

A core detail of our subsystem call convention is the \emph{allocation} of per-subsystem register sets and stacks.
Each Skadi subsystem allocates a build-time-configurable number of stack frames and register sets into a pool.
Allocation uses either an atomic bitmap or a uniprocessor per-IRQ-regime bitmap, offering a size-performance tradeoff while ensuring SMP and IRQ safety.
Allocation uses the $\mathcal{O}(1)$ \emph{ctz} (count trailing zeros) instruction.
If stack frames are available, we can allocate one in a single cycle.
\Cref{sec:microbench} shows that this facilitates fast subsystem calls.
If all stack frames are occupied, we can spin or trigger an exception.

Previously proposed subsystem calling conventions use a per-thread stack instead, and share it between caller and callee.
While reducing memory usage compared to Skadi, this necessitates clearing the stack on return~\cite{amarCHERIoTRTOSOS2025,skorstengaardReasoningMachineLocal2018} or relying on \emph{uninitialized capabilities}~\cite{gulmezMonCHERIMitigating2025,yuCapstoneCapabilitybasedFoundation2023}, incurring run-time cost.

Trampolines are generated using function-like macros that accept the call name and signature.
Therefore, modification of existing Zephyr components merely requires adding wrapper macros for subsystem calls.
Using \texttt{\_Generic}, we distinguish void trampolines and clear return registers accordingly.

We \emph{deduplicate} and \emph{prune} the caller trampolines, reducing \emph{code size}: caller trampoline macros generate an invocation of an external function whose name encodes the call name and signature.
At build time, for each subsystem, we generate a source file that implements the caller trampolines according to the names of unresolved functions that match the schema.

\section{Security Discussion}
\label{sec:security}

\begin{table*}
    \centering
    \setlength{\aboverulesep}{0pt}
    \setlength{\belowrulesep}{0pt}
    \small
    \caption{Summary of specific security guarantees provided and their key mechanism in hardware and software.}
    \label{tab:sec_goals}
    \begin{tabular}{p{2.6cm}|p{7cm}|p{6.9cm}}
    \textbf{Goal}  & \textbf{Bredi} & \textbf{Skadi}\\
    \toprule
    Spatial Memory Isolation (\textbf{\textsf{C}},\textbf{\textsf{I}})  & subsystem ID protection and restrictions, nonce, monotonicity of operations (\cref{sec:nc_representation}) & Private code and data, no run-time software TCB, no kernel (\cref{sec:no-runtime-tcb}, \cref{sec:no-os-kernel})\\
    \hline
    Temporal Memory Isolation (\textbf{\textsf{C}},\textbf{\textsf{I}}) & $lock$ exclusive access, tracking of parent-child relationships and recursive resolution (\cref{sec:nc_representation}) & no run-time software TCB, no kernel, untrusted scheduler and allocator (\cref{sec:no-runtime-tcb}, \cref{sec:no-os-kernel})\\
    \hline
    DMA isolation (\textbf{\textsf{C}},\textbf{\textsf{I}}) & AXI MMU (\cref{sec:northcape-hardware}) & untrusted DMA driver (\cref{sec:nc_operations}, \cref{sec:no-os-kernel})\\
    \hline
    ISR isolation (\textbf{\textsf{C}},\textbf{\textsf{I}})  & register stacking, interrupt vectoring (\cref{sec:irq_sec}) & ISR table managed by loader (\cref{sec:no-os-kernel})\\
    \hline
    Preemption (\textbf{\textsf{A}})  & voluntary preemption hardware support (\cref{sec:irq_sec}) & timer stub in trampolines (\cref{sec:subsys-cc})\\
    \hline
    NMIs (\textbf{\textsf{A}}) & NMI hardware support (\cref{sec:irq_sec}) & ISR subsystem (\cref{sec:no-os-kernel})\\
    \hline
    Fast operations (\textbf{\textsf{A}}) & cache hierarchy (\cref{sec:Hierarchie_Skip}), instructions (\cref{sec:isa_add}) & subsystem calling convention (\cref{sec:subsys-cc})\\
    \bottomrule
    \end{tabular}
\end{table*}

Skadi and Bredi can address the attack vectors that we motivated at the beginning of this work---\emph{application software}, \emph{operating system kernels}, and \emph{peripheral devices}.
The key idea to solve this challenge is to combine \emph{untrusted operating system} approaches from capability literature~\cite{amarCHERIoTCompleteMemory2023,amarCHERIoTRTOSOS2025,esswoodCheriOSDesigningUntrusted2021} with backwards-compatible enforcement of capabilities at the system bus level.
Thereby, we incorporate security against untrusted DMA devices~\cite{ackermannWorkinProgressNorthcapeEmbedded2024,markettosPositionPaperDefending2021}.
\Cref{tab:sec_goals} lists the detailed security guarantees provided by Skadi and Bredi and their key implementation mechanisms.
In a nutshell, the strong isolation of Skadi components facilitates confidentiality (\textbf{\textsf{C}}) and integrity (\textbf{\textsf{I}}) for the entire \emph{sphere of protection} of the subsystem.
As the loader, the only trusted software component, \emph{is destroyed after load time}, Skadi accomplishes this \emph{without run-time software TCB}.
A similar trusted loader is used, e.g., by CheriTrEE~\cite{vanstrydonckCHERITrEEFlexibleEnclaves2023} for the same purpose.
As Skadi's \emph{compartmentalization} is crucial for achieving \textbf{\textsf{C}} and \textbf{\textsf{I}} without compromising availability (\textbf{\textsf{A}}), in the remainder of this section, we argue why compartmentalization cannot be broken.

In order to protect \textbf{\textsf{C}} and \textbf{\textsf{I}} of subsystems' private code and data, we relocate subsystems' ELF segments into subsystem-id-bound capabilities.
Guessing attacks on these capabilities are prevented by \emph{subsystem-ID bound restrictions}: if a subsystem with different subsystem ID guesses a nonce, it still encounters an exception on access if it does not manage to hijack the subsystem ID as well.
Our subsystem call trampolines are designed to prevent being able to change the subsystem ID without transferring control on both call and return.
Without using subsystem calls, and without being able to use subsystem ID 0, there is no way to change the subsystem ID, preventing hijacking.
Due to \emph{register stacking} and \emph{IRQ regimes}, ISRs cannot interfere with interrupted subsystems.

Dynamically allocated memory is protected by the unforgeability of the capability tokens and optionally exclusive access via $lock$ or subsystem ID bound restrictions.
Not adding subsystem restrictions to allocated memory by default allows sharing the memory with other subsystems or devices.
The allocator cannot use $revoke$ to break \textbf{\textsf{C}} or \textbf{\textsf{I}}, as $revoke$ overwrites the data and invalidates overlapping capabilities.

Finally, \textbf{\textsf{A}} relies on NMIs.
We assume that the interrupt subsystem in conjunction with software ISRs do not stall interrupt execution, as they are in the aTCB.
Subsystem calls between ISR subsystems were designed to be real-time capable.
Adversaries outside of the ISR subsystems cannot prevent them from executing, as interrupts can be made unmaskable.
To this end, the IRQ mask is \emph{write-once}.

\section{Performance Evaluation}
\label{sec:evaluation}

Our evaluation captures the performance overhead imposed by Bredi and Skadi.
We implement our prototype on a Digilent Genesys 2 FPGA board.
The L1 and L2 NTLBs have sizes 16 (instruction) / 32 (data) and 512, respectively.
The L1 NTLB is fully associative, while the L2 NTLB uses 8-times associativity.
Our benchmarks cover chip area, code size, compute and networking performance, and real-time suitability.
We compare Zephyr and Skadi on Bredi.

Furthermore, where possible, we try to evaluate equivalent scenarios on Linux v5.10.7.
\emph{This is an apples-to-oranges comparison, as many APIs and system services differ significantly between Zephyr and Linux.}
We chose to provide this comparison anyway: The comparison with Linux shows that Skadi's performance is within the same order of magnitude as a commonly used general-purpose operating system, confirming our claims about Skadi's \emph{real-world practicality}.

\subsection{Hardware: Chip Area and Frequency}
\label{sec:hardware-area}

We synthesize Bredi  using Xilinx Vivado 2024.2~\cite{advancedmicrodevicesinc.Vivado2024} at 50 MHz akin to the original cva6 design~\cite{zarubaCostApplicationClassProcessing2019}.
In \cref{tab:resource_cost}, we report the number of FPGA primitives required for the components of Bredi as well as Bredi-cva6 and the original cva6.
Our design meets all timing requirements.

\begin{table*}
    \centering
    \small
    \sisetup{
      group-separator = {\,},
  group-minimum-digits = 4,
  round-mode = places,
  round-precision = 2,
  detect-all,
  table-align-text-pre = false,
  table-align-text-post = false,
  zero-decimal-to-integer = false,
    }
    \setlength{\aboverulesep}{0pt}
    \setlength{\belowrulesep}{0pt}
    \caption{Area in lookup tables (LUTs), flip flops (FFs), multiplexers (MUXs), and Block RAM (BRAMs) for Bredi and \hlgray{cva6}.}
    \label{tab:resource_cost}
    \begin{tabular}{l|S[table-format=6.0] >{\text{(}\hspace{1em}\itshape}S[table-format=3.2]<{\text{)}\hspace{1em}}|S[table-format=6.0] >{\text{(}\hspace{1em}\itshape}S[table-format=2.2]<{\text{)}\hspace{1em}}|S[table-format=6.0] >{\text{(}\hspace{1em}\itshape}S[table-format=2.2, table-text-alignment=center]<{\text{)}\hspace{1em}}|S[table-format=6.0] >{\text{(}\hspace{1em}\itshape}S[table-format=2.2, table-text-alignment=center]<{\text{)}\hspace{1em}}|S[table-format=6.0] >{\text{(}\hspace{1em}\itshape}S[table-format=3.2, table-text-alignment=center]<{\text{)}\hspace{1em}}}
    \multicolumn{1}{c|}{\textbf{Component}} & \multicolumn{2}{c|}{\textbf{LUTs \emph{(\%)}}} & \multicolumn{2}{c|}{\textbf{FFs \emph{(\%)}}} & \multicolumn{2}{c|}{\textbf{F7 MUXs \emph{(\%)}}} & \multicolumn{2}{c|}{\textbf{F8 MUXs \emph{(\%)}}} & \multicolumn{2}{c}{\textbf{BRAMs \emph{(\%)}}}\\
    \toprule
    AXI MMU         & 4085 &  2.00  &  832  & 0.20 &    2   & 0.01  & 0 & 0.00 & 0 & 0.00\\
    \hline
    Resolver        & 1640 & 0.80   & 506   & 0.12 & 58     & 0.06 &  0 & 0.00 & 0 & 0.00\\
    \hline
    Operations      & 9280 &  4.55  & 4675 & 1.15 &  251    & 0.25 &  0 & 0.00  &2& 0.45\\
    \hline
    L2 NTLB         & 20830 &  10.22  & 2230  & 0.55 & 357    & 0.35 & 57 & 0.11& 70 & 15.73\\
    \hline
    Bredi cva6 & 63245 & 31.03 & 39971 & 9.81 & 4367   & 4.29 &  1580 & 3.10 & 28 & 6.29\\
    \hline
    \rowcolor{gray!25}\cellcolor{white}Original cva6 & 46764 & 22.95 & 25817 & 6.33 & 2060   & 2.02 &  426 & 0.84 & 28 & 6.29\\
    \bottomrule
    \end{tabular}
\end{table*}

Our chip area and frequency results indicate that inclusion of Northcape into an FPGA or ASIC design is practical.
The L2 NTLB has the highest chip area, which is caused by the complex hazard avoidance logic and high associativity.

\subsection{Software: Code Size}

\begin{table}
    \centering
    \setlength{\aboverulesep}{0pt}
    \setlength{\belowrulesep}{0pt}
    \small
    \caption{Code size in byte for Skadi and \hlgray{Zephyr}.}
    \label{tab:code_size}
\ifarxiv
    \begin{tabular}{p{2.5cm}r|r|r}
\else
    \begin{tabular}{lr|r|r}
\fi
    \multicolumn{1}{l|}{\textbf{Category}} & \multicolumn{1}{c|}{\textbf{Hello World}} & \multicolumn{1}{c|}{\textbf{HTTP}} & \multicolumn{1}{c}{\textbf{zperf}}\\
    \toprule
     & \num{738890} & \num{2435932} & \num{3109444}\\\cline{2-4}
    \rowcolor{gray!25}\cellcolor{white}\multirow{-2}{*}{Disk size} & \num{49036} & \num{556952} & \num{842216}\\
    \midrule
    & \num{405550} & \num{1465822} & \num{3792040}\\\cline{2-4}
    \rowcolor{gray!25}\cellcolor{white}\multirow{-2}{*}{Text} & \num{26016} & \num{174838} & \num{389116}\\
    \midrule
     & \num{14796} & \num{56852} & \num{79852}\\\cline{2-4}
    \rowcolor{gray!25}\cellcolor{white}\multirow{-2}{*}{Data} &  \num{9524} & \num{54806} & \num{94526}\\
    \midrule
    &  \num{71376} & \num{149124} &\num{237008}\\\cline{2-4}
    \rowcolor{gray!25}\cellcolor{white}\multirow{-2}{*}{R/O data} &  - & - &-\\
    \midrule
    & \num{103192} & \num{1935060} &\num{1918092}\\\cline{2-4}
    \rowcolor{gray!25}\cellcolor{white}\multirow{-2}{*}{BSS} & \num{13493} & \num{327266} &\num{358519}\\
    \specialrule{0.7pt}{1pt}{1pt}
    \textbf{Sub\-systems} & 12 & 21 & 23\\
    \midrule
    \textbf{Callee Trampolines} & 219 & 450 & 490\\
    \midrule
    \textbf{Caller Trampolines} & 50 & 222 & 235\\
    \midrule
    \textbf{Stacks / Register Sets} & 4/4 & 16/16 &12/12\\
    \midrule
    \textbf{ELF Overhead} & \num{1872687} & \num{4608420} & \num{6137930}\\
    \midrule
    \textbf{Total memory} & \num{1774432} & \num{7258816} & \num{10289304}\\
    \bottomrule
    \end{tabular}
\end{table}

We detail the code size for three configurations of Skadi in~\cref{tab:code_size}:
a simple hello-world example, a basic socket-based HTTP server, and zperf (Zephyr's implementation of iperf).

We report total binary size (\emph{disk size}) and the size of the text, data, read-only data (rodata) and bss segments.
A large part of the binary also comprises metadata of subsystems---relocation sections, symbol tables etc., that are reclaimed after the load stage.
We provide the sum of the sizes of these metadata \emph{before LZ4 compression} in the \emph{ELF Overhead} row.
Finally, for Skadi, we provide the total allocated memory after initialization.
Zephyr solely uses the .bss segment.

Our code size evaluation shows that the overhead of Skadi's compartmentalization is significant.
As we explained in \cref{sec:skadi-os}, Zephyr is a \emph{unikernel} operating system:
In the default configuration, all operating system components live in the same address space.
On the other hand, Skadi's components are \emph{relocated at run time}:
Thus, all components need to carry additional metadata such as relocation entries, explaining the increase in disk size.
The increase in run-time memory use can be attributed to subsystem call trampolines, per-subsystem stacks and inlining of some library functions.

While Skadi has a high size overhead, it is still small enough for most modern application scenarios.
In particular, nowadays, the amount of memory on embedded systems is mostly limited by \emph{power budget} rather than \emph{cost}, with advancements in energy efficiency being made continuously~\cite{beniniEnergyawareDesignEmbedded2003}.
Thus, future embedded devices will likely comprise larger memories, making this drawback irrelevant in practice.

\subsection{System: Microbenchmarks}
\label{sec:microbench}
\begin{table}[t]
    \centering
    \setlength{\aboverulesep}{0pt}
    \setlength{\belowrulesep}{0pt}
    \setlength{\tabcolsep}{0pt}      %
    \small
    \caption{Microbenchmarks of Northcape operations.}
    \label{tab:microbenchmarks_short_genesys2}

    \begin{tabular}{l<{\quad}|@{\quad}r<{\,$\pm$\,}r<{\quad}|@{\quad}r<{\,$\pm$\,}r<{\quad}}
    \textbf{Benchmark} & \multicolumn{2}{c}{\textbf{CPU Cycles}} & \multicolumn{2}{c}{\textbf{Ops Cycles}}\\
    \toprule
    $create$ & \num{629.273684} & \num{4.387643} & \num{11.00} & \num{0.00}\\
    \hline
    $derive$ & \num{621.968421} & \num{6.613308} & \num{11.00} & \num{0.00}\\
    \hline
    $drop$ & \num{255.884211} & \num{4.502311} & \num{12.000000} & \num{0.00}\\
    \hline
    $unlock$ & \num{124.810526} & \num{1.292038} & \num{17.000000} & \num{0.00}\\
    \hline
    $merge$ &\num{407.126316} & \num{3.202769} & \num{16.000000} & \num{0.00}\\
    \hline
    $clone$ & \num{322.347368} & \num{5.295019} & \num{14.000000} & \num{0.00}\\
    \hline
    $revoke$ (1 MiB) & \num{39894.210526} & \num{59.620359} & \num{34934.536842} & \num{5.900599}\\
    \hline
    $lock$ & \num{402.684211} & \num{5.758770} &\num{16.000000} & \num{0.00}\\
    \hline
    $inspect$ & \num{284.284211} & \num{2.864164}& \num{8.00} & \num{0.00}\\
    \hline
    $restrict$ & \num{382.452632} &\num{3.959714} &\num{10.00} &\num{0.646833}\\
    \bottomrule
    \end{tabular}
\end{table}
\begin{table}[t]
    \centering
    \setlength{\aboverulesep}{0pt}
    \setlength{\belowrulesep}{0pt}
    \setlength{\tabcolsep}{1pt}      %
    \small
    \caption{Microbenchmarks of Northcape subsystem calls compared against legacy \hlgray{Zephyr system calls} via PMP.}
    \label{tab:microbenchmarks_subsys_short_genesys2}

    \begin{tabular}{l<{\hspace{1em}}|>{\hspace{1em}}r<{\,$\pm$\,}r<{\hspace{1em}}|>{\hspace{1em}}r<{\,$\pm$\,}r<{\hspace{1em}}}
    \textbf{Benchmark} & \multicolumn{2}{c}{\textbf{CPU Cycles}} & \multicolumn{2}{c}{\textbf{Instructions}}\\
    \toprule
    \rowcolor{gray!25}\cellcolor{white}syscall (one-way) & \num{364.157895} & \num{2.758015} & 140 & 0.00\\
    \hline
    \rowcolor{gray!25}\cellcolor{white}syscall (two-way) & \num{485.263158} & \num{2.749969} & 196 & 0.00\\
    \specialrule{0.7pt}{1pt}{1pt}
    subs. call (one-way) & \num{250.768421} & \num{12.262440} & 113 & 0.00\\
    \hline
    subs. call (two-way) &  \num{402.557895} & \num{13.427165} & 180 & 0.00\\
    \bottomrule
    \end{tabular}

\end{table}

We provide a microbenchmark of Skadi operations in \cref{tab:microbenchmarks_short_genesys2}.
We record the total duration of the corresponding driver function and the number of cycles used by the operations module.

For the subsystem call microbenchmark in \cref{tab:microbenchmarks_subsys_short_genesys2}, we measure one-way and two-way duration of subsystem calls in comparison with Zephyr system calls.
To this end, we enable the optional PMP-protected user space for the baseline.

All in all, all operations except $revoke$ complete in under 1,000 cycles including driver overhead (mainly packing arguments into registers).
For $revoke$, the bottleneck is zeroing out the segment in DRAM, a security necessity.
However, $revoke$ is only used on rare occasions, such as at a few boot steps, making the overhead irrelevant in practice.

One-way and two-way subsystem calls take appx. 251 and 403 cycles and outperform legacy systemcalls at 364 and 485 cycles, respectively.
This is caused by the lower number of retired instructions for Skadi, courtesy of our custom instructions (\cref{sec:isa_add}).
Remember that subsystem calls are essential for \emph{mutual isolation}, while system calls only isolate \emph{the privileged kernel from the userspace, not vice versa}.

Still, subsystem calls are not free: The caller trampoline needs to save and restore the \emph{entire} callee-saved set and perform register set and stack allocation.
Function calls can rely on the existing stack pointer and only need to save and restore callee-saved registers that they modify, allowing them to be much faster.
In fact, RISC-V functions that solely operate on registers can have an overhead as low as two instructions (a call and a return)~\cite{pattersonRISCVReaderOpen07}.
Thereby, subsystem calls will inevitably cause a performance overhead, resulting in a security-performance tradeoff between Skadi and Zephyr.

\subsection{System: Compute Performance}
\label{sec:compute-perf}

\begin{table}
    \centering
    \setlength{\aboverulesep}{0pt}
    \setlength{\belowrulesep}{0pt}
    \small
    \caption{Compute benchmarks for Skadi, \hlgray{Zephyr} and \hlgray{Linux}.}
    \label{tab:compute_mem_benchmarks}
    \begin{tabular}{p{2.2cm}|>{\columncolor{gray!25}}r|>{\columncolor{gray!25}}r|r|r}
    \textbf{Benchmark}  & \textbf{Linux} & \textbf{Zephyr} & \textbf{Skadi} & \bm{$\Delta$}\\
    \toprule
    Coremark/sec  & \num{92.1003055} & \num{110.017053} & \num{110.144289} & \num{0.127236}\\
    \hline
    Stream Copy sec & \num{11.226838} & \num{6.044444} & \num{5.999914} & \num{-0.04453}\\
    \hline
    Stream Scale sec  & \num{12.753133} & \num{6.655556} & \num{6.646107} & \num{-0.009449}\\
    \hline
    Stream Add sec  & \num{17.534844} &\num{9.033333} &  \num{9.015753} & \num{-0.01758}\\
    \hline
    Stream Triad sec & \num{18.619079} & \num{9.500000}  &\num{9.500033} & \num{0.000033}\\
    \bottomrule
    \end{tabular}
\end{table}

We utilize eembc coremark~\cite{eembcCoremark2025} and Stream~\cite{mccalpinMemoryBandwidthMachine1995} for compute and memory performance.
Results can be found in \cref{tab:compute_mem_benchmarks}.
Extended results are available in \cref{tab:compute_mem_benchmarks_full} in the appendix.
Overall, the differences between Skadi and Zephyr on Bredi are negligible.
This can be explained by a 100\% hit rate on the L1 NTLB, which does not impose any latency on cva6' pipeline, and a lack of subsystem calls during the benchmark.

Linux' comparatively lower performance can be explained by TLB misses. We use cva6' default setting of 16 instruction and data TLBs, respectively, which are shared between kernel and user space.
\emph{We do not claim that this result is generalizable.} We solely provide the value to point out that Zephyr and Skadi perform in the same order of magnitude as Linux.

\subsection{System: Network Stack Performance}
\label{sec:network-overhead}

\begin{table}
    \centering
    \setlength{\aboverulesep}{0pt}
    \setlength{\belowrulesep}{0pt}
    \setlength{\tabcolsep}{0pt}      %
    \small
    \caption{Networking benchmarks for Skadi, \hlgray{Zephyr} and \hlgray{Linux} (omitted for MQTT). Number of subsystem calls (total for ping and iperf, otherwise per-iteration) provided for context.}
    \label{tab:system_network_performance}

    \begin{tabular}{l@{\quad}r<{\,$\pm$\,}r<{\,}|@{\quad}r}
    \textbf{Benchmark} & \multicolumn{2}{c}{\textbf{Avg$\pm$std.dev.}} & \bm{$\Delta$}\\
    \toprule
     & \num{2.555} & \num{0.075}\\
    \rowcolor{gray!25}\cellcolor{white} & (Zephyr) \num{0.876} & \num{0.021} & \cellcolor{white} \cellcolor{white}x~\num{2.916666667}\\
    \rowcolor{gray!25}\cellcolor{white}\multirow{-3}{*}{Ping latency~(ms)} &(Linux) \num{1.916}  & \num{0.152} & \cellcolor{white}x~\num{1.333507307}\\
    \hline
    Subsystem calls & 59788 & -\\
     \specialrule{0.8pt}{0pt}{0pt}
      & \num{1499.3333} & \num{002.5154887} & \\
      \rowcolor{gray!25}\cellcolor{white}&  (Zephyr) \num{4396.6667} & \num{0006.8064443} & \cellcolor{white}x~\num{0.341015911}\\
     \rowcolor{gray!25}\cellcolor{white}\multirow{-3}{*}{iperf TCP~(kbps)}& (Linux) \num{9031.1667} & \num{1733.1954} & \cellcolor{white}x~\num{0.166017675} \\
     \hline
    Subsystem calls & 1051756 & -\\
     \specialrule{0.8pt}{0pt}{0pt}
     &  \num{2138493.631579} &  \num{13707.449448} &\\
     \rowcolor{gray!25}\cellcolor{white}& (Zephyr) \num{534142.810526} &\num{8120.032767} & \cellcolor{white}x~\num{4.003599018}\\
    \rowcolor{gray!25}\cellcolor{white}\multirow{-3}{*}{RX duration~(ns)}& (Linux) \num{1346524.210526}  & \num{136558.719135} & \cellcolor{white}x~\num{1.588158323}\\
     \hline
    Subsystem calls & \num{290.052632} & \num{11.426072}\\
    \specialrule{0.8pt}{0pt}{0pt}
     & \num{1516145.021053} & \num{11670.623603} &\\
     \rowcolor{gray!25}\cellcolor{white}& (Zephyr) \num{530697.96} & \num{13202.113} & \cellcolor{white}x~\num{2.85688873}\\
    \rowcolor{gray!25}\cellcolor{white}\multirow{-3}{*}{TX duration~(ns)} & (Linux) \num{1050532.631579} & \num{203162.692163} & \cellcolor{white}x~\num{1.443215542}\\
     \hline
    Subsystem calls & \num{235.084211} & \num{2.430781}\\
    \specialrule{0.8pt}{0pt}{0pt}
     & \num{2057174000.842105} & \num{5772516.868643} &\\
    \rowcolor{gray!25}\cellcolor{white}\multirow{-3}{*}{MQTT-TLS~(ns)} & \num{1993484845.473684} & \num{30590.976095} & \cellcolor{white}x~\num{1.031948653}\\
    \hline
    Subsystem calls & \num{2058.800000} & \num{147.169834}\\
    \bottomrule
    \end{tabular}

\end{table}

We measure ICMP echo round-trip time using \emph{ping}, TCP throughput using \emph{iperf} and \emph{zperf}, network stack latency using hardware timestamps, and an MQTT scenario similar to CHERIoT~\cite{amarCHERIoTCompleteMemory2023} to determine networking performance.
For the network stack latency, we use \emph{hardware timestamps} collected by a ha1588 hardware time stamp unit (TSU) to determine the exact time when a network packet is sent or received via the MAC-PHY-interface.
We then determine the delay between TX or RX time and the time the packet is returned from the socket or written to the socket, respectively, relative to ha1588's time.
On Linux, RX and TX delays are measured via \emph{standard software timestamps}, causing a slight undercount.
Our MQTT scenario is based on Zephyr's MQTT publisher sample with TLS enabled, measuring the time to publish three messages over 100 iterations.
\emph{MQTT} is omitted on Linux, as the MQTT implementation is Zephyr-specific.

We report the most important results in \cref{tab:system_network_performance}. Full results are given in \cref{tab:system_network_performance_full} in \cref{sec:extra-eval}.
The performance overhead that Skadi and Northcape impose on our ICMP and zperf scenarios are significant.
In particular, we measure an appx. 3x-4x overhead between the baseline and Skadi scenarios.
This is a consequence of the strong compartmentalization provided by Skadi as evidenced by the high number of subsystem calls.
However, we believe that the performance in absolute numbers still shows that Skadi is sufficiently fast for most embedded applications.
In particular, all measurements except TCP throughput have an overhead less than factor two in comparison to Linux.
As Zephyr's TCP throughput is lower than Linux' as well, we assume that Zephyr's network stack just provides lower performance on our platform in general.

We see two avenues for future optimization.
First, our isolated compartments conceptually match Zephyr's subsystems as closely as possible.
This is the most fine-grained compartmentalization of the system, whereas in practical scenarios, some compartments could potentially be merged into one, reducing subsystem calls. This is done in Linux~\cite{linuxcontributorsLinuxDriversNet2025} for the Ethernet driver and DMA driver, for example.
Second, it might be possible to eliminate subsystem calls between the Ethernet and DMA drivers, networking subsystem and scheduler at the expense of breaking existing APIs and driver models.

\subsection{System: Real-Time Capability}
\label{sec:rt-capability}

\begin{table}
    \centering
    \setlength{\aboverulesep}{0pt}
    \setlength{\belowrulesep}{0pt}
    \setlength{\tabcolsep}{0pt}
    \small
    \caption{IRQ benchmarks results for Skadi and \hlgray{Zephyr}.}
    \label{tab:real_time_performance}
    \begin{tabular}{l@{\quad}r<{\,}c<{\,}r<{\,}|@{\quad}r}
    \textbf{Benchmark} & \multicolumn{3}{c}{\textbf{Avg$\pm$std.dev.}~(ns)} & \bm{$\Delta$}\\
    \toprule
    &\num{68423.157895} & $\pm$ & \num{1529.023950}\\
    \rowcolor{gray!25}\cellcolor{white}\multirow{-3}{*}{independent} & \num{24127.157895} & $\pm$ & \num{32149.031897} & \cellcolor{white}x~\num{2.835939409}\\
    \hline
    Subsystem calls & \num{1144.684211} & $\pm$ & \num{18.550224} &\\
    \specialrule{0.8pt}{0pt}{0pt}
    & \num{136000} & &\\
    \rowcolor{gray!25}\cellcolor{white}\multirow{-3}{*}{monotonic period} & \num{70000} & & & \cellcolor{white}x~\num{1.942857143}\\
    \specialrule{0.8pt}{0pt}{0pt}
    & \num{58826.315789} & $\pm$ & \num{18714.180441} \\
    \rowcolor{gray!25}\cellcolor{white}\multirow{-3}{*}{monotonic} & \num{8392.421053} & $\pm$ & \num{12120.038276} & \cellcolor{white}x~\num{7.009457154}\\
    \hline
    Subsystem calls & \num{7.021053} & $\pm$ & \num{0.143560} &\\
    \bottomrule
    \end{tabular}
\end{table}

For evaluating the real-time capability of Skadi, we cover a middle ground of suitable evaluation approaches in the literature:
We cover Zephyr's scheduler microbenchmark~\cite{garcia-martinezComprehensiveApproachPerformance1996} as well as periodic interrupt response time and maximum interrupt rate without missed interrupts~\cite{arocaRealTimeOperating2009}.

In order to test our availability guarantees, on Skadi, we make our external interrupt \emph{non-maskable} and \emph{disable} interrupts while busy-waiting.
The Zephyr scenarios run with interrupts enabled, as they would otherwise not complete.

Results of our IRQ experiments can be found in~\cref{tab:real_time_performance}.
Extended results and our scheduler microbenchmark can be found in \cref{tab:real_time_performance_full} and \cref{tab:macrobenchmark_full} in the appendix.

All in all, Skadi incurs an overhead both in terms of constant latency and in latency range and standard deviation.
For the IRQ scenarios, Skadi's overhead is between 3 and 7 times.
Similar to the networking overhead determined in \cref{sec:network-overhead}, this is caused by compartmentalization (see the number of subsystem calls needed).
However, in absolute numbers, Skadi is still able to respond to interrupts in roughly $\frac{1}{10}$ of a millisecond, with a comparable standard deviation.

The results for the scheduler benchmark indicate that Skadi's strong compartmentalization imposes an overhead mostly between 3 and 10 times.
One notable outlier is thread cancellation---if this occurs, Skadi needs to invoke a callback in each subsystem  to release any stacks that might still be associated with this thread.
Again, considering especially the absolute numbers, we believe that the results indicate that Skadi is suitable as a soft real-time system.

\section{Related Work}
\label{sec:related-work}

We cover three areas of related work:
First, we discuss the differences to the previously published Northcape concept~\cite{ackermannWorkinProgressNorthcapeEmbedded2024}.
Then, we compare Bredi and Skadi with related capability architectures and finally non-capability approaches.

\subsection{Improvements to Northcape}
\label{sec:related_northcape}

\citeauthor{ackermannWorkinProgressNorthcapeEmbedded2024}~\cite{ackermannWorkinProgressNorthcapeEmbedded2024} proposed Northcape as a purely theoretical capability system without hardware implementation and operating system.
Additionally, Bredi's realization of Northcape improves the concept in three key areas.

\emph{Preventing guessing of capability tokens and subsystem identifiers.}
Northcape originally used a \emph{secret} subsystem identifier in an \emph{architectural register tid} (task ID).
Each \emph{locked} capability would then only be accessible for the subsystem that locked it, using \emph{tid} to distinguish subsystems.
Thereby, exclusive access was always \emph{per-subsystem}.
This mechanism has a crucial drawback: \emph{the tid can be leaked or guessed}, allowing circumvention of exclusive access.
We removed the \emph{tid} register and re-designed both locking and restrictions:
They are now independent mechanisms.
Hence, \emph{lock-holder tokens} and \emph{subsystem-id restrictions} are our contribution.

\emph{Extending the set of access permissions.}
In addition to the capability access permissions \emph{read}, \emph{write}, \emph{execute}, \emph{lockable}, Bredi also adds \emph{irq\_accessible}, \emph{cacheable\_data} and \emph{cacheable\_tlb}.
\emph{irq\_accessible} controls whether capabilities can be accessed in ISRs, which is used, e.g., for preventing overwriting the timer stub from ISRs, which would break ISR isolation.
The cache permissions can be used for hard real-time processing in specific subsystems.

\emph{Implicit subsystem calls.}
Finally, Bredi makes two changes to the $calls$ operation.
First, subsystem calls in Bredi can be accomplished using regular jump instructions.
To this end, the cva6 MMU ensures mutual isolation \emph{irregardless} of the instruction that commences a subsystem call:
Every attempt to subvert control flow by jumping over the first instruction or reading secret data from the set-subsystem ID capability is met with an exception.
$calls$ solely exists to optimize performance.
Second, Bredi's $calls$ does not return a context pointer---instead, the context is saved in the set-subsystem-ID capability in Skadi.
This increases software flexibility.

\subsection{Capability Systems}
Most capability systems do not consider attacks from devices (\emph{DMA attacks}) in their core concept (some simply because they precede DMA~\cite{winklerComputerCommunicationImpacts1972,levyCapabilityBasedComputerSystems2014,carterHardwareSupportFast1994}).
Capability systems such as CAPSTONE~\cite{yuCapstoneCapabilitybasedFoundation2023} and CHERI~\cite{woodruffCHERICapabilityModel2014} are focused on protecting \emph{software} from \emph{system-level software adversaries}.
Several related capability systems also focus solely on \emph{in-process compartmentalization}, i.e., isolating mutually distrusting components or libraries within a process address space in a trusted kernel setting~\cite{kimRVCURERISCVCapability2025,dinhduyCapacityCryptographicallyEnforcedInProcess2023,watsonCapsicumPracticalCapabilities2010,lemayCryptographicCapabilityComputing2021}.

CHERI was recently extended to support IOMMU-like functionality~\cite{chengAdaptiveCHERICompartmentalization2025}.
The proposed CapChecker is initialized with a set of CHERI capabilities, which are identified by a DMA device by a number.
DMA transactions are then interpreted relative to the capability, enforcing spatial memory safety.
Crucially, this work relies on trusted drivers, imposes a hard limit on the number of capabilities a device can use and requires hardware changes to the DMA device for full security.
Bredi and the underlying Northcape do not have these limitations, provide spatial and temporal memory safety, and allow both CPUs and devices to create capabilities.

Userspace drivers have also been proposed for CHERI platforms, using capabilities to access a safe subset of device MMIO without system call~\cite{metzgerDeprivilegingLowLevelGPU2025,dokuCAPIOSafeKernelBypass2025}.
However, this design requires a \emph{trusted} kernel:
A kernel-level driver needs to setup an IOMMU or Graphics Translation Table (i.e., the MMU of a GPU) to prevent the driver from abusing the device as a \emph{confused deputy}.
These mechanisms operate at \emph{page} granularity, failing to prevent attacks like Thunderclap~\cite{markettosThunderclapExploringVulnerabilities2019} in principle.

Furthermore, CHERIoT RTOS~\cite{amarCHERIoTRTOSOS2025} resembles Skadi, comprising a fully privileged boot-time-only loader that bootstraps unprivileged compartments.
However, in CHERIoT RTOS, switcher and allocator remain part of the run-time TCB~\cite{amarCHERIoTRTOSOS2025}.
The dIPC system~\cite{vilanovaDirectInterProcessCommunication2017} similarly proposes direct inter-process communication for the CODOMs architecture~\cite{vilanovaCODOMsProtectingSoftware2014} using trampolines and a trusted intermediary, akin to CHERIoT RTOS.
Furthermore, other untrusted operating systems in the CHERI domain rely on a privileged run-time nanokernel~\cite{esswoodCheriOSDesigningUntrusted2021} or do not consider DMA devices in the threat model~\cite{amarCHERIoTRTOSOS2025,almataryCompartOSCHERICompartmentalization2022}.

Related work on the $M^3$ operating system~\cite{paulsTrustminimizedPlatformSecure2025,haasArchitecturalApproachSecure2025,asmussenDistrustingCoresSeparating2025,hilleSemperOSDistributedCapability2019} takes a different approach: A computer system can be disaggregated into isolated \emph{tiles} that hold CPUs or accelerators.
The $M^3$ kernel can then facilitate interaction between tiles or between tiles and the system memory using a \emph{distributed} capability system. This is enforced by a per-tile access control component similar to the Bredi AXI MMU.
Crucially, like in other microkernel operating system, the $M^3$ microkernel forms a run-time software TCB. Also, there is no support for local \emph{subsystem calls} with mutual isolation, as processing cores use legacy memory protection like MMUs or PMPs.

Bastion~\cite{restucciaBASTIONFrameworkSecure2025} implements a similar architecture to $M^3$, but relies on custom firmware running on an OpenTitan instead of $M^3$ and uses access control lists (ACLs) for memory access control. The authors further propose a token capability solution as future work.

\subsection{Non-Capability Approaches}
In contrast to \emph{securely containing} vulnerabilities in kernel software, one can also \emph{strive to prevent} exploitable vulnerabilities in kernel code.
Microkernel operating systems do just that: they move all parts of the kernel that do not strictly require operating with elevated hardware privileges into userspace, reducing the attack surface of the overall system~\cite{elphinstoneL3SeL4What2013,carlvanschaikHighperformanceMicrokernelsVirtualisation2007,karlssonPartitioningKernelCapability2025}.
High-performance IPC is then used for communication between the compartments.
SeL4~\cite{kleinSeL4FormalVerification2009} combines this approach with formal verification of the microkernel, formally proving the absence of vulnerabilities in the TCB.
We consider SeL4 to be an orthogonal approach to the reduction of a run-time TCB: SeL4 aims at preventing exploitation of the software TCB, at the expense that changes to the TCB require costly re-verification~\cite{elphinstoneL3SeL4What2013}.
Skadi aims at eliminating the TCB, at the expense of performance, chip area and code size.
Both approaches could be combined in future work, e.g., by formally verifying the Skadi loader and Bredi hardware TCB.

\emph{Confidential computing} also strives to reduce the software TCB.
Commodity implementations of this technology are ARM CCA~\cite{armltdArmCCASecurity2021}, Intel TDX~\cite{chengIntelTDXDemystified2024} and AMD SEV~\cite{advancedmicrodevicesinc.AMDSecureEncrypted2025}. 
CHERI-TrEE, a system for confidential computing on the CHERI architecture, has also been presented~\cite{vanstrydonckCHERITrEEFlexibleEnclaves2023}.
For CCA, various approaches that provide security in the presence of compromised devices, operating systems and application software have been proposed~\cite{wangCAGEComplementingArm2024,zhangSHELTERExtendingArm2023,bertschiDevloreExtendingArm2024,huangHiveTEEScalableFinegrained2026}.
In particular, HiveTEE~\cite{huangHiveTEEScalableFinegrained2026} promises to isolate an arbitrary number of secure application compartments from untrusted devices, operating systems and hypervisors, similar to Skadi's security model.
However, as confidential computing technologies rely on a run-time trusted software \emph{Monitor} for configuration of hardware security primitives and transition between compartments, vulnerabilities in trusted software components continue to jeopardize the security of the overall system~\cite{schluterHeraclesChosenPlaintext2025}.
Furthermore, the underlying hardware protection mechanisms impose a limit on the achievable granularity (e.g., page or 16-byte MTE memory region for HiveTEE).
CHERI-TrEE provides byte-granular memory protection and subsystem calls courtesy of CHERI, but relies on a trusted software interrupt handler as its run-time TCB and does not consider untrusted devices~\cite{vanstrydonckCHERITrEEFlexibleEnclaves2023}.

The ongoing advances in confidential computing show that even in MMU-based systems, the necessary TCB can also be minimized.
However, using capability-based memory access control has two advantages:
Capabilities support \emph{delegation} and \emph{unprivileged modification using operations}, allowing components to share memory efficiently and securely without a trusted intermediary.
In turn, this facilitates \emph{subsystem calls}, the key building block for mutual separation of distrusting components in Skadi.
Also, capabilities provide \emph{byte-granular} security while MMUs work with larger \emph{pages}.
Overall, Skadi and Bredi can protect any software at \emph{function granularity} with \emph{mutual distrust} and \emph{byte granularity} from any other run-time software or hardware.

However, physical attacks like bus snooping or cold boot are out of scope for Bredi, while SGX, CCA, SEV and TDX can provide security against such attacks.
Future extensions to Bredi that address this shortcoming are possible.

Finally, Skadi provides solely a \emph{mechanism} for compartmentalization.
Hence, of the properties defined by \citeauthor{lefeuvreSokSoftwareCompartmentalization2025}~\cite{lefeuvreSokSoftwareCompartmentalization2025}, we are able to provide \emph{security and safety} and limited source code \emph{compatibility} with legacy Zephyr modules.
Crucially, the focus of our work was providing \emph{security without run-time TCB}, which none of the works surveyed by \citeauthor{lefeuvreSokSoftwareCompartmentalization2025} achieves.
Skadi could utilize automatic \emph{compartmentalization policies} to improve \emph{usability}~\cite{lefeuvreSokSoftwareCompartmentalization2025}.

\section{Conclusion and Future Work}
\label{sec:summary}
We have presented the Skadi operating system and Bredi SoC, the first implementation of the Northcape capability architecture with significant conceptual improvements.
In combination, Skadi and Bredi achieve security in the presence of \emph{malicious application software, operating system kernels and peripheral devices} for embedded devices \emph{without run-time software TCB} and \emph{with soft real-time suitability}.

Within the scope of this work, we focused primarily on the \emph{security} of our prototype.
\emph{Performance} could be improved further by adding custom compiler optimizations or using a different programming model that avoids domain switches~\cite{sartakovEActorsFastFlexible2018}.

Bredi and Skadi are a promising foundation for future research.
The high security provided by Skadi and Bredi in conjunction with their real-time operation make them suitable as enabling technology for, e.g., high-security network infrastructure devices like 5G base stations.
The \emph{hardware-backed subsystem ID} in conjunction with \emph{restrictions} are also a very powerful primitive for providing additional security guarantees to system peripherals on a \emph{per-capability} basis.

\appendix
\ifarxiv
\fi

\bibliographystyle{plainnat}
\bibliography{references_cleaned.bib}

\crefalias{section}{appendix}

\ifarxiv
\section{Northcape: Overview}
\label{sec:full_overview}
Northcape is a hardware capability architecture that was originally proposed by \citeauthor{ackermannWorkinProgressNorthcapeEmbedded2024}~\cite{ackermannWorkinProgressNorthcapeEmbedded2024}.
The key idea in Northcape is (conceptually) moving capability enforcement out of the CPU and into the system bus, making it applicable to both software and hardware devices.
Our proposed Skadi operating system relies on our implementation of Northcape (Bredi) to achieve its security guarantees.

In the original publication, Northcape was presented as a pure concept without implementation.
One of the contributions of this work is an implementation of Northcape in hardware.
To this end, minor changes to the original Northcape system were necessary.

This section introduces key concepts of the Northcape architecture.
The implementation of Northcape in Bredi is discussed in \cref{sec:northcape-hardware}.
We start with the terminology, followed with the implementation of capabilities, the operations and finally a discussion of subsystem calls.
For readability, we introduce Northcape in the form that we implemented.
The differences to the original Northcape architecture are discussed in~\cref{sec:related_northcape}.

\subsection{Terminology and Design Goals}
Northcape generalizes bus masters such as CPUs, NICs, accelerators etc. as \emph{data users}, and bus slaves including memory controllers and MMIO peripherals to \emph{data stores}.
Data users run one or more \emph{subsystems} which can be implicit (e.g. sending and receiving packets on a NIC) or explicit (conceptually distinct parts of software on a CPU).
One special software subsystem, the \emph{loader subsystem}, is conceptually part of the Northcape hardware and hence \emph{trusted}.
All other subsystems are \emph{untrusted}.
Data users can also run \emph{subsystems} that may provide \emph{subsystem calls} to other subsystems on the same device.
Subsystem calls are a generalization of system calls; they facilitate calls between unprivileged subsystems with \emph{mutual isolation}.
Data stores contain \emph{segments} identified by a \emph{physical address} interpreted by northbridge and slave and a byte-granular \emph{length}.
To this end, the northbridge maintains a \emph{capability metadata table} (CMT) in DRAM, whose entries contain these bounds and other metadata.
Finally, subsystems hold \emph{capability tokens} which authorize them to access segments within the scope of the \emph{permissions} specified at \emph{capability creation time}.
To this end, the capability system maintains a \emph{bijective mapping} between capability tokens and CMT entries.

\subsection{Implementation of Capabilities}

Northcape capabilities come in two different forms: a \emph{CMT entry} and a \emph{Capability Token}.
Conceptually, the capability token is the \emph{front end} of the capability system: it is used directly by software and devices.
To this end, it is encoded with backwards compatibility to 64-bit \emph{bus addresses}.
In contrast to fat-pointer architectures like CHERI~\cite{woodruffCHERICapabilityModel2014}, the token does not contain any metadata like segment bounds information or access permissions.
Instead, the capability hardware needs to \emph{resolve} a token to a CMT entry in the \emph{back end}.
The CMT entry encodes the bounds information and other metadata.
Given the provided capability token and CMT entry, the capability hardware can decide whether a memory access is permissible.

Note that to our knowledge, the idea of separating capabilities into capability tokens and CMT entries was pioneered by RV-CURE~\cite{kimRVCURERISCVCapability2025}.
While Northcape uses the same basic idea, the main difference is that Northcape capability tokens are used instead of \emph{bus addresses} and the resolution of tokens to CMT entries is implemented in the \emph{northbridge}.
In contrast, RV-CURE uses capability tokens in the place of \emph{virtual addresses} and implements the resolution \emph{in the CPU}.
In the remainder of this section, we will discuss both the representation of capability tokens and CMT entries and how they enable the capability operations discussed in \cref{sec:operations}.

\ifarxiv
\else
\paragraph{Capability Tokens}
\begin{figure*}[t]
    \includegraphics[width=\linewidth]{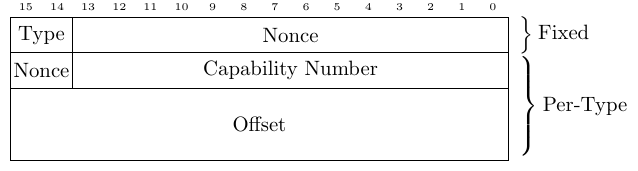}
    \caption{Parsing exemplary capability token $0x1518404200000004$ into \emph{offset length}, \emph{nonce}, \emph{ID} and \emph{offset} components.}
    \label{fig:northcape-cap-token}
\end{figure*}
\fi
\ifarxiv
In Northcape, a capability token $c$ conceptually consists of an \emph{identifier} $n$, an \emph{unpredictable nonce} $\sigma$ and an offset $o$, as illustrated in~\cref{fig:token-type}.
\else
In Northcape, a capability token $c$ conceptually consists of an \emph{identifier} $n$, an \emph{unpredictable nonce} $\sigma$ and an offset $o$, as illustrated in~\cref{fig:northcape-cap-token}.
\fi
$n$ identifies an entry in the CMT, referencing a \emph{segment} in a data store.
$o$ identifies a relative starting position in the segment for each access.
$\sigma$ serves the \emph{unforgeability} of the token, preventing \emph{use-after-reallocation} and \emph{replay} attacks.
To this end, it should be infeasible to \emph{guess} $\sigma$ even if $n$ and all user-controlled metadata of the CMT entry are known to the attacker.
In our implementation, $\sigma$ is selected from a sequence of random numbers.
Such a sequence can be computed efficiently by encrypting a monotonic nonce with a block cipher under an unpredictable key, akin to counter-mode encryption~\cite{salmonParallelRandomNumbers2011}.
We use the Qarma-64 block cipher, which was designed specifically to allow truncation of nonces to 16 bits and below ~\cite{avanziQARMABlockCipher2017}. 
The key used for Qarma as well as the initial nonce are derived from the output of a lightweight hardware-based random number generator~\cite{meiHighlyFlexibleLightweight2018}.
As nonces are never revealed to untrusted devices and software, nonces are increased sequentially after initialization.
Note that we chose not to rely on the random number generator directly for computing $\sigma$: $\sigma$ is very short, consisting of only 16 bits.
Thus, any bias in the random number generator can cause a value of $\sigma$ to be predictable.
By seeding the Qarma key and nonce individually from the random number generator, we can achieve a significantly higher bit security.

An additional similarity to RV-CURE is our encoding, where $n$ and $\sigma$ are stored in the high bits of a 64-bit pointer, leaving the lower bits for $o$. 
This ensures compatibility with legacy 64-bit pointers, including pointer arithmetic \emph{within one segment}~\cite{kimRVCURERISCVCapability2025,chisnallPDP11ArchitecturalSupport2015}.
A key difference to RV-CURE is that we provide no specific instructions for converting pointers into capability tokens. All pointers in the system are \emph{implicitly} capabilities and all accesses protected.

A challenge in designing the token representation was balancing the \emph{number of capabilities} that can be active at any given time and the \emph{maximum capability size}.
We use the following compromise: two additional header bits in the token identify a \emph{capability token type} with a different allocation of bits for the capability identifier and offset.
We offer 8, 16, 24 and 32 bit offsets.
Accounting for the token size and MAC, this leaves 38, 30, 22 and 14 bits for the ID field, respectively.
We allocate capability identifiers $[0,2^{14}-1]$ for tokens with 32 bits of offset, identifiers $[2^{14},2^{22}]$ for tokens with 24 bits of offset and so on.
Thereby, we can support up to $2^{38}$ capabilities and segments of up to $2^{32}$ bytes.

\paragraph{CMT entries}
\ifarxiv
\else
\begin{figure*}
    \includegraphics{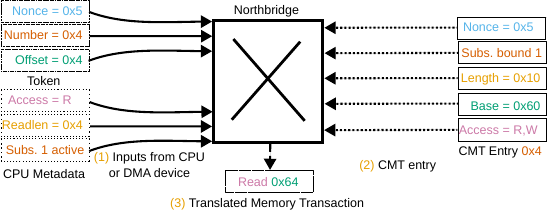}
    \caption{Translation of capability token components to address.}
    \label{fig:token-translation_front}
\end{figure*}
\fi
After resolving a capability token, the capability hardware needs to parse the associated CMT entry in order to decide whether the access is permissible.
Different metadata attributes are associated with capability tokens, as illustrated in~\cref{fig:token-translation_front}.
First, the \emph{capability type} encoded in a CMT entry is crucial for determining which operations are permissible for a segment.

We differentiate between two cardinal types of capabilities, \emph{direct} and \emph{indirect}, shown in \Cref{fig:capability-types}.

\paragraph{Direct Capabilities}
A \emph{direct} capability \emph{owns} the segment it is pointing to, i.e., the direct capability can be used to \emph{revoke} all indirect capabilities referencing the same physical segment.
Direct capabilities can also be \emph{sliced} for the creation of new, non-overlapping direct capabilities.

For the purposes of bootstrapping the system after a power cycle, the northbridge creates a well-known \emph{root capability} on reset.
This is a direct capability that owns one segment comprising the \emph{entire physical address space} and is encoded in a way such that na\"ive use of 32-bit physical addresses is \emph{interpreted} as accessing the root capability at the corresponding offset. 
Direct capabilities have the following access permission bits: read, write, execute, lockable and interrupt accessible.
Read, write and execute mirror the permissions found in most paging-based memory protection mechanisms.
The lockable permission indicates whether it is possible to gain exclusive access to the capability using the corresponding capability operation; this can be disabled to prevent denial of service using locking.
The interrupt accessible permission bit determines whether a capability can be accessed in an interrupt service routine.
The primary use of this bit is to protect the system from certain impersonation attacks explained in~\cref{sec:interrupt-architecture}.
As an added benefit, it can be used to enforce a designation of functions and data structures as \emph{IRQ safe} in hardware.
The interrupt accessible permission is only interpreted on CPU devices and depends on the read, write and execute permissions - if none of these permissions are set, the capability cannot be accessed at all.

\paragraph{Indirect Capabilities}
Indirect capabilities are \emph{derived} either from direct or other indirect capabilities, restricting access in both cases.
This approach is used to provide restricted sub-object capabilities, akin to CHERI and RV-CURE~\cite{woodruffCHERICapabilityModel2014,kimRVCURERISCVCapability2025}.

\paragraph{Northcape Memory Management}
Crucially, in the Northcape system, \emph{direct} capabilities are \emph{exclusively used} by the \emph{allocator}.
Thereby, in the face of memory pressure or crashes, the allocator can revoke capabilities, thus preventing subsystems from \emph{stealing} memory.
This invalidates the entry in the CMT and creates a new entry, while also \emph{zero-initializing the segment}.
The design of the CMT in conjunction with the implementation of the lookup ensures that all further uses of indirect capabilities referencing the revoked direct capability cause a \emph{bus error}.
Revocation further overwrites the physical segment with 0-bytes to ensure no secrets are leaked to the allocator.
Overwriting segments on revocation along with the \emph{locking} mechanism we will introduce shortly allows us to \emph{define a TCB that does not contain the allocator}.
A second benefit of this strategy is that it allows the allocator to allocate \emph{direct} capabilities for segments \emph{larger} than what was requested.
Efficient allocation algorithms for segmented memory commonly define a \emph{minimal} segment size to limit external fragmentation, possibly allocating a larger segment than requested~\cite{yurchenkoAlgorithmDynamicSegmented1981}.
In Northcape, the allocator can afterwards derive an \emph{indirect} capability with exactly the requested size, ensuring no over-reading or over-writing of the segment is possible in the application.
The allocator can also safely store metadata in a segment inaccessible to the application.

\paragraph{Indirect Capabilities Facilitate Subsystem Calls}
Indirect capabilities have a specialized form called \emph{execute-only} capability\footnote{Execute-only capabilities are not identified by a different type; instead, they are implemented using indirect capabilities bearing additional restrictions as outlined in \cref{sec:subsystem-calls}}.
Execute-only capabilities are at the core of Northcape's \emph{subsystem call} operation, which enables \emph{transitions between different subsystems at the same privilege level with mutual isolation}.
This is the same concept as protected procedure calls in Plessey 250 and the enter instruction in the original capability system by \citeauthor{dennisProgrammingSemanticsMultiprogrammed1966}~\cite{winklerComputerCommunicationImpacts1972,dennisProgrammingSemanticsMultiprogrammed1966}: 
Subsystems (e.g., a network stack) can create execute-only capabilities for their public API.
The execute-only capabilities can then be provided to other subsystems like application software, allowing them to jump into the network stack.
The jump changes both control flow and \emph{sphere of protection}, i.e., the set of addressable segments.
Thereby, the subsystem and its caller (a user subsystem or possibly a different subsystem) are \emph{mutually isolated} by default.
A calling convention can be used to pass capability tokens between the user subsystem and the subsystem in registers, e.g., buffers for data to be sent or received by the network stack.
This primitive in conjunction with exclusive memory access facilitates the design of the Skadi operating system \emph{without run-time TCB.}

\paragraph{Indirect Capabilities Facilitate Exclusive Memory Access}
A second specialization of the \emph{indirect} capability is the \emph{lock-holder} capability, which is a capability that allows exclusive access to the capability it was created from.
It is created from a direct or indirect capability by the $lock$ operation described in~\cref{sec:operations}.
By design of our operations, each lock-holder capability can be recursively resolved to a direct capability, which we refer to as the \emph{base capability} of that lock-holder capability.
During the lifetime of the lock-holder capability, attempting to resolve any other (grand)child of the base capability or the base capability itself leads to an error.
Thereby, the subsystem(s) with knowledge of the lock-holder capability have \emph{exclusive access} to the base capability without being able to read or write outside of the bounds of the capability that they locked originally.
Lock-holder capabilities can also be tied to a specific subsystem using the restriction mechanism outlined below, which is orthogonal to exclusive access.

Crucially, Northcape uses \emph{locking} to solve the problem that other capability architectures like CHERI and CAPSTONE solve with \emph{revocation}: restoring exclusive access to a segment that was previously shared~\cite{xiaCHERIvokeCharacterisingPointer2019,yuCapstoneCapabilitybasedFoundation2023}.
Solving this problem with locking has the advantage that gaining exclusive access is reversible.
Also, the implementation of both locking and revocation is straightforward and requires no in-memory data structures outside of the CMT.
Indirect capabilities can be destroyed using the \emph{drop} operation, which as a side effect decreases the parent's reference count.
The same operation is also used for relinquishing exclusive access.

\ifarxiv
\else
\begin{figure}[t]
    \includegraphics[width=\linewidth]{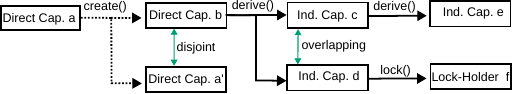}
    \caption{Types of capabilities in the Northcape system.}
    \label{fig:capability-types}
\end{figure}
\fi

\begin{figure*}[t]
    \centering
    \begin{minipage}[b]{0.45\textwidth}
        \includegraphics[width=\linewidth]{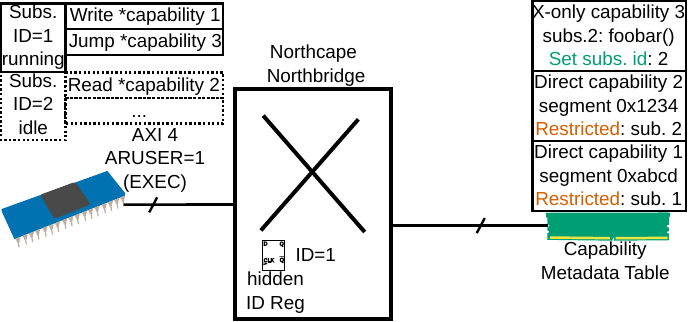}
        \caption{Exemplary scenario for two subsystems that hold subsystem-restricted capabilities: Subsystem 1 and 2 each hold a private data capability, and subsystem 2 has an execute-only capability for jumping into subsystem 2. Assume all capabilities also have read permission. Subsystem 1 can read capability 1 and perform a subsystem call into capability 3, but cannot read capability 3 or 2.}
        \label{fig:capability-subsystem-restrictions-before}
    \end{minipage}%
    \hfill
    \begin{minipage}[b]{0.45\textwidth}
        \includegraphics[width=\linewidth]{illustrations/overview_graphics/per_task_capabilities_1.pdf}
        \caption{The scenario depicted in \cref{fig:capability-subsystem-restrictions-before} after subsystem 1 has executed a subsystem call on the capability for subsystem 2. Subsystem 2 is now in control, it can read and execute capability 3 and read capability 2 but cannot read capability 1.}
        \label{fig:capability-subsystem-restrictions-after}
    \end{minipage}
\end{figure*}

\paragraph{Restriction Mechanism Prevents Guessing Attacks}
A shortcoming of relying on the nonce $\sigma$ alone for capability unforgeability is that an adversary can \emph{guess} the nonce for a capability with probabiliy $\frac{1}{{2^{16}}}$.
Note that the identifier $n$ of the capability might be predictable, as it is selected from a monotonically increasing sequence.
Thus, Northcape supports imposing \emph{restrictions} on the use of capabilities: Capabilities can be tied to both a \emph{device} (e.g., a DMA controller) and an individual \emph{subsystem} running on a device.
Devices are identified by their respective port, which means no changes to the device are necessary.
However, in order to restrict a capability to a specific \emph{subsystem}, the capability system needs a way of determining which subsystem is currently running.
This problem has already been approached by \citeauthor{bahmaniCURESecurityArchitecture2021} \cite{bahmaniCURESecurityArchitecture2021}: The authors suggest adding an additional privileged \emph{architectural} register to the CPU, which is programmed on context switch by trusted privileged software.
The register maintains the subsystem identifier of the currently running subsystem. The register value is then copied into the \emph{user} bits of the AXI system bus, allowing connected devices to identify the subsystem that is currently requesting access to the device.
We iterate on this concept in the following way. Northcape attempts to keep the amount of trusted software as small as possible. 
Hence, we did not want to rely on CPU privilege levels to protect the subsystem id register.
Instead, we make the subsystem id register \emph{non-architectural}, i.e., inaccessible to software, and also move it to the northbridge. The register is updated \emph{implicitly} during subsystem calls as explained in~\cref{sec:operations}. 
To this end, we introduce the concept of additional \emph{restrictions} for capabilities, consisting of a type and a 64-bit payload that depends on the type: 

For type \emph{subsystem\_id\_bound}, the restriction payload encodes a device and subsystem identifier that the capability is exclusively bound to.
If this restriction is set, the northbridge compares the device (identified by bus port) and the subsystem id set in the last subsystem call for a request with the restriction encoded in the capability, denying the request if it originates in a different subsystem.
Thereby, it is ensured that even if a different subsystem guesses the capability number and nonce, it cannot access the corresponding segment.

A type \emph{subsystem\_id\_set} can be indicated for execute-only capabilities, which instructs the CPU to set the subsystem id register correspondingly during subsystem calls.
This is how switching subsystems on the CPU is reflected in the capability system.
\Cref{fig:capability-subsystem-restrictions-before} and \cref{fig:capability-subsystem-restrictions-after} illustrate switching the subsystem id in an exemplary scenario.
Capabilities with \emph{subsystem\_id\_set} can also have both read and execute permissions. This is used for stateful subsystems; the capability can embed both code and data such as globals or stack.
In this case, the \emph{subsystem\_id\_set} restriction implies \emph{subsystem\_id\_bound} for read/write accesses.
Thereby, data contained on the segment can only be read \emph{after} switching the subsystem ID during the subsystem call operation, i.e., \emph{within} the called subsystem.
Hence, the caller of the subsystem can perform a subsystem call but is unable to read instructions and data of the subsystem.

Finally, there is a type \emph{device\_interpreted}, which transparently conveys the 64-bit restriction body into the AXI user bits to the target device, similar to the original concept in CURE~\cite{bahmaniCURESecurityArchitecture2021}.
This can be used to implement device-specific per-subsystem restrictions, such as enforcing a per-subsystem network policy in a NIC.

Note that the concept of capability \emph{restrictions} is orthogonal to the \emph{locking} operation: using \emph{indirect} capabilities and the \emph{derive} operation, the same segment can be accessible to different subsystems at the same time.
The $lock$ operation enables one of the subsystems to gain exclusive access.

Northcape restrictions can be used for the same purpose as \emph{sealed capabilities} in Capstone and CHERI~\cite{yuCapstoneCapabilitybasedFoundation2023,watsonCHERIHybridCapabilitySystem2015}: mutually protecting private data of subsystem caller and callee.
Both existing systems provide a \emph{seal} operation that allows sharing a capability with a different untrusted sphere of protection without the untrusted sphere of protection being able to use the capability.
Only the intended sphere of protection is able to use an \emph{unseal} operation, allowing it to access the associated memory.
This is used, e.g., for cookie arguments that are supposed to be passed back to a callback.
By using the subsystem ID restriction in Northcape, a capability token can be passed to an untrusted subsystem without the subsystem being able to use it.

The implementation in Northcape has the advantage that no explicit seal and unseal operations are needed, reducing overhead.
However, Northcape's solution has the disadvantage that an untrusted subsystem can attempt to manipulate the token.
This can be mitigated, e.g., by using our $inspect$ capability operation to verify the integrity of the token before using it and using a constant capability offset.

\subsection{Operations}
\label{sec:operations}
In order to guarantee security, subsystems are prevented from manipulating capabilities directly.
In fact, CMT entries are \emph{non-architectural} in Northcape.
Instead, subsystems can use the following operations to interact with the capability system:

    \noindent $create(c_a,l,r,p)\longrightarrow c_a',c_b$ 
    Splits a capability $c_a$ into two non-overlapping pieces. Creates one new capability $c_b$ and modifies the metadata of $c_a$ accordingly.
    Caller can specify a length $l$ as well as restrictions $r$ and permissions $p$ for the segment.
    $p$ and $r$ have to be as strict or stricter than the permissions and restrictions of $c_a$, and $l$ must be less than or equal the length of $c_a$.
    Invalidates $c_a$ if $l=l_a$.
    Can only be used for direct capabilities, requiring $c_a$ to not be locked and have a reference count of zero.

    \noindent $merge(c_a,c_b,r,p)\longrightarrow c_m$ 
    This operation is the opposite of $create$.
    It creates a new capability $c_m$ from physically adjacent capabilities $c_a$, $c_b$ with new restrictions $r$ and new permissions $p$.
    Permissions $p$ and new restrictions $r$ can be more permissive than those of the input capabilities.
    The input capabilities are destroyed.
    Requires the capabilities to have a reference count of zero and not to be locked.  
    
    A comprehensive study by \citeauthor{wilsonDynamicStorageAllocation1995}~\cite{wilsonDynamicStorageAllocation1995} has pointed out that splitting ($create$ in our model) and merging segments are the only operations that general-purpose slab memory allocation algorithms perform in practice.
    Merging is particularly useful for reducing fragmentation when an application allocates and returns many small segments at once.

    \noindent $derive(c_a,l,o,r,p)\longrightarrow c_i$ 
    The $derive$ operation is used to create indirect capabilities from an input capability $c_a$, which are exclusively used by all subsystems and devices except the memory allocator.
    An indirect capability has a (grand)parent that is a direct capability and possibly overlapping siblings.
    To this end, $c_a$ can be either a direct or another indirect capability.
    The output capability $c_i$ is always indirect and can have an additional offset $o$ and a reduced length $l$ in comparison to its parent.
    Thereby, $derive$ is suitable for sub-object protection akin to CHERI and RV-CURE\cite{watsonCHERIHybridCapabilitySystem2015,kimRVCURERISCVCapability2025}.

    \noindent $clone(c_a,r,p)\longrightarrow b$ A synonym for $derive$ that internally calls $derive$ with an offset $o$ of 0 and a length $l$ that equals the length of the segment.
    Useful for removing permissions from or adding restrictions to a capability a subsystem wants to share, especially in cases where the segment length of the capability is not known to the caller of $clone$. 
    
    \noindent $lock(c_a,r,p)\longrightarrow c_l$ 
    Attempts to gain \emph{exclusive access} to a capability.
    The input capability $c_a$ can be either direct or indirect.
    In case the \emph{lockable} permission is set for $c_a$'s grandparent direct capability, a lock-holder capability $c_l$ is created.
    The \emph{lockable} permission prevents abusing $lock$ for DoS.
    As long as it is alive, $c_l$ can be used to access the same (part of the) segment visible through $c_a$.
    At the same time, other capabilities that overlap with $c_a$ cannot be used.

    \noindent $drop(c_a)\longrightarrow b$ 
    Destroys capability $c_a$ if it does not have any references and returns whether this was successful.
    Can only be used on indirect and lock-holder capabilities.
    If $c_a$ is a lock-holder capability returned by $lock$, the parent direct capability is unlocked as a side effect.
    Thereby, the caller of $drop$ needs not know what type of capability a token resolves to.
    $clone$ and $drop$ are used for the cooperative reference counting mechanism on indirect capabilities, which is useful for shared segments.
    $drop$ can also free a capability after its parent was destroyed using $revoke$ to prevent leaking CMT entries.
    
    \noindent $revoke(c_a,r,p)\longrightarrow c_a'$ Destroys the CMT entry associated with $c_a$ and creates a new direct capability $c_a'$ for the segment identified by the capability with full permissions and removed restrictions.
    \emph{Overwrites} the segment with 0-bytes. Returns the new capability.
    Can only be used on direct capabilities.
    Permissions $p$ and new restrictions $r$ can be more permissive than those of the input capability.
    $revoke$ is intended for the allocator to reclaim memory on process crash or memory pressure, its use might lead to bus errors when subsystems continue using indirect capabilities for the segment.
    It only accepts \emph{direct} capabilities to prevent subsystems from \emph{stealing} memory.
    This is the main difference between Northcape and Capstone's linear capabilities: By overwriting the segment on revocation, we can ensure no secrets are leaked \emph{without} having to support uninitialized capabilities, making the implementation easier.
    Unfortunately, Capstone does not discuss these trade-offs or propose a hardware implementation of uninitialized capabilities~\cite{yuCapstoneCapabilitybasedFoundation2023}.

    Revocation in Northcape also differs significantly from the revocation extensions proposed for CHERI~\cite{xiaCHERIvokeCharacterisingPointer2019}.
    Most importantly, Northcape revocation is intended \emph{only for exceptional circumstances}, while CHERIvoke uses revocation for guaranteeing temporal memory safety, i.e., \emph{un-sharing} segments. 
    In both Northcape and CHERIvoke, revocation comes with the problem of initially leaving orphaned capabilities that clutter the CMT (Northcape) or physical address space (CHERIvoke).
    In CHERIvoke, this problem is solved with scanning \emph{the entire physical address space} for orphaned capabilities periodically, causing significant overhead.    
    Northcape provides more light-weight primitives for memory safety in the form of the $drop()$ and $lock()$ operations.
    The $drop()$ operation prevents use-after-reallocation and use-after-free vulnerabilities by destroying the indirect capability used by user applications.
    As it only needs to modify the released capability and its parent (in order to adjust its reference count), it comes with a low overhead.
    At the same time, the reference count mechanism ensures no indirect capabilities can become orphans (except when the direct capability that owns the memory they are pointing to gets revoked).
    The $lock()$ operation grants exclusive access to capabilities, akin to linear capabilities in Capstone~\cite{yuCapstoneCapabilitybasedFoundation2023}.
    It can be used in many scenarios in which CHERI needs to use revocation, e.g., to re-gain exclusive access to data in a segment after it was shared with a different subsystem or device.
    $lock()$ has a slightly higher overhead than $derive()$.
    In Northcape, $lock()$ always transparently modifies the direct capability that owns the storage - if this was not the case, anyone with a different indirect capability than the one locked originally would be able to circumvent the lock.
    Thus, in order to lock and unlock a capability, Northcape needs to walk its tree of parent capabilities to the direct capability each time.
    Still, due to the reference counting mechanism, $lock()$ never orphans capabilities as long as $revoke()$ is not also used.
    
    $revoke()$ in Northcape on the other hand destroys the direct capability immediately, orphaning all indirect capabilities and lock-holder capabilities derived from it.
    $revoke()$ is also expected to overwrite the data segment, giving it a large overhead depending on the segment size.
    It is only intended to re-gain memory in case of erroneous use of the capability operations, e.g., after subsystem crashes.
    Because all capabilities are resolved recursively until the direct capability is reached, the orphaned indirect capabilities allow no access to the associated segment after revocation, which is why they pose no security risk.
    Still, the orphaned capabilities clutter the CMT and might make it impossible to create new capabilities.
    There are two possible solutions for this problem.
    First, $drop()$ can be used to destroy all kinds of indirect capabilities which are still known to a running subsystem - the subsystems learn that the capabilities were orphaned when they get bus errors on attempting to use them.
    In case no running subsystem has access to the capability token that resolves to an orphaned capability any more, an exhaustive scan of the CMT is necessary.
    Note that the overhead of this is still lower than in CHERIvoke, as the CMT is expected to be far smaller than the entire physical memory.
    Such an exhaustive scan can be triggered by software when the CMT is full or nearly full and need not be executed after every execution.
    As $revoke()$ is expected to be executed very infrequently, executing this scan might almost never be necessary in real applications.

    \noindent $inspect(c_a)\longrightarrow b,l,r,p$ 
    Reads the metadata for any capability $c_a$.
    Resolves $c_a$ recursively to determine base and length if necessary (for lock-holder capabilities, which do not directly encode these information) and the lockable permission.
    This operation is required returning from a subsystem call or checking the admissability of a function pointer, see \cref{sec:subsystem-calls} for details.
    The operation acts slightly differently depending on any subsystem-id-bound restrictions on $c_a$: If there is no such restriction or it matches the caller of the operation, the operation returns all mentioned metadata.
    If $c_a$ has a set-subsystem-id restriction, the operation returns \emph{only the $R,W,X,I$ permissions} and the restrictions.
    This suffices to determine if it is safe to invoke $c_a$ with a subsystem call or use it as a return address, but does not leak any unnecessary information - see~\cref{sec:subsystem-calls}.
    In other cases, $c_a$ returns an error code.

    \noindent $restrict(c_a,r,p,o^+,l^-)\longrightarrow b$ 
    Makes a capability more restrictive in-place.
    This can be applied to all types of capabilities with slightly different semantics:
    For indirect capabilities, an additional offset $o^+$ and a length minuend $l^-$ can be specified.
    The bounds of the indirect capabilities are reduced accordingly.
    For other capabilities, these parameters are ignored.
    Also, all permissions that are not specified in \emph{both} $p$ and $c_a$'s metadata are dropped.
    The $L$ permission is only dropped if $c_a$ is a direct capability; otherwise, attempting to set or clear it has no effect.
    $r$ is only applied when $c_a$ does not currently have restrictions.
    
    Decreasing the accessible subset is especially useful for network buffers: Network stacks often remove network headers and footers one-by-one for each network layer until only application data remain.
    After processing in each layer is completed, there is no need to retain access to the metadata of the previous layer, so we can safely make it inaccessible.
    This also serves memory safety during parsing of network packets.
    Removing permissions and adding restrictions is useful for dynamic relocation; we discuss this when we introduce the Skadi operating system in \cref{sec:skadi-os}.

    \noindent $calls(c_a)$ This \emph{implicit} operation describes performing a subsystem call.
    On input a capability $c_a$ with set-subsystem-id restriction, set the \emph{non-architectural} subsystem identifier register to the indicated identity and transfer control to the code in the identified segment \emph{atomically}.
    $calls$ implements subsystem calls akin to Plessey System 250 or the original capability architecture proposed by \citeauthor{dennisProgrammingSemanticsMultiprogrammed1966}~\cite{dennisProgrammingSemanticsMultiprogrammed1966,winklerComputerCommunicationImpacts1972}.
    A key difference in Northcape is that there is no explicit instruction that implements $calls$; instead, the subsystem call is performed \emph{implicitly} as soon as a device (at this time, this is only supported for a CPU) executes the first instruction in a segment with set-subsystem-id restriction that points to a \emph{different} subsystem id then the one currently executing.
    To this end, normal jump instructions in conjunction with a modified calling convention can be used to trigger subsystem calls in software; our Skadi operating system uses jump-register (\emph{jr}) RISC-V instructions.
    We discuss Northcape subsystem calls in more detail in \cref{sec:subsystem-calls}.

The $restrict$ and $inspect$ operations are guaranteed to complete with a constant upper time bound.
The $calls$ operation is guaranteed to have a constant upper time bound that depends on the hierarchy height of the input capability.
The remaining operations do not have timing guarantees.

In Northcape, the northbridge implements the operations defined above (including $calls$, which we managed to implement \emph{without} adding a new instruction to the CPU, as it is an implicit operation).
It provides an MMIO interface which data users can utilize to invoke the aforementioned operations, conveying parameters and responses.
Thereby, Northcape provides higher flexibility than capability systems that encode operations in instructions.
Northcape-aware operating systems can use infrastructure such as device trees or ACPI tables to indicate the version of the Northcape MMIO interface that is present, and (possibly) further restrictions such as which operations are supported.
Hence, it is easy to make changes to the interface when necessary without having to provide backward compatibility.
The MMIO approach also makes Northcape's operation accessible to (firmware running on) non-CPU devices.

\subsection{Data Structures}
\label{sec:data-structures}

Northcape exclusively relies on the CMT for bookkeeping, with no additional data structures required.
This is an advantage over, e.g., RV-CURE~\cite{kimRVCURERISCVCapability2025}, which requires additional datastructures to maintain its equivalent of indirect capabilities.
The CMT is a hash table in main memory, which is maintained by the northbridge. It contains one entry for each active capability in the system, comprising bounds information and other metadata.
Two design goals were used when selecting a data structure for the CMT: First, the CMT should enable \emph{constant time lookups}.
This is critical for fulfilling the real-time guarantees promised by the Northcape system.
However, only the $calls$, $restrict$ and $inspect$ operations are guaranteed to complete in real time.
The remaining operations can take an arbitrary amount of time, which is why no timing guarantees for \emph{CMT entry insertion} are needed.
Second, the CMT should be \emph{physically contiguous} in memory.
Thereby, the table can co-exist with the Northcape memory management: it simply occupies a segment that is inaccessible outside the hardware TCB (specifically, the northbridge).

The CMT is implemented as a direct access table and uses linear probing for collision resolution.
In other words, the CMT is a hash table with a fixed bucket size of $1$.
Thereby, all token lookups require exactly one memory access for one capability.
The identifier $n$ contained in each capability token is used as key for the hash table.
When a newly allocated capability ID experiences a collision, the capability ID is incremented and another insertion is attempted (linear probing).
This causes a worst-case insertion time that is linear in the size of the capability ID.
However, for the operations that we wish to be executable in bounded time, Northcape needs not guarantee timing constraints for \emph{insertion}.

There are three reasons motivating this decision: First, all entries are padded to the same size, which allows us to match the size of each hash table directory with the entries.
Second, we can maintain the entire CMT in a single segment, which makes co-existence with the memory allocator in the OS significantly easier.
Finally, this allows us to perform a lookup in one memory access.
In order to allow a high number of active capabilities as well as memory efficiency, the CMT can be made \emph{resizable} using \emph{extendible hashing}~\cite{faginExtendibleHashingFast1979}.

The nonce $\sigma$ for each entry is stored in the table to facilitate fast comparison between the actual and the presented nonce.
Thereby, the northbridge can ensure that the token referring to the entry is valid \emph{at the time of the access}.
In conjunction with dropping and revoking capabilities, this is the mechanism used to provide \emph{temporal safety}.

In addition to the attributes given above, entries for indirect capabilities also contain a \emph{capability token} for their \emph{parent}, i.e., the capability they were derived from.
When validating the access to an indirect capability, the northbridge \emph{recurses} to its parent until it reaches the base \emph{direct capability}.
Thereby, when the base direct capability is revoked, all uses of the derived capabilities cause \emph{validation errors}.
The same recursive lookup is used to enforce exclusive access after $lock$.
Hence, no explicit sweeping of capabilities to revoked segments and no lifetime tokens are needed, which is an improvement compared to existing revocation schemes such as CHERIvoke~\cite{xiaCHERIvokeCharacterisingPointer2019}.

\subsection{Subsystem Calls}
\label{sec:subsystem-calls}
Subsystem calls generalize the concept of system calls commonly found in modern CPUs and operating systems.
A system call can be used by unprivileged software to transfer control to software running at a higher privilege, e.g., an operating system kernel.
The privileged software is then expected to perform an operation on behalf of the unprivileged software, e.g., to write data to a file on a hard disk.
To this end, following a calling convention, the unprivileged software usually configures the CPU registers with the arguments needed for the operation and triggers a synchronous interrupt.
The interrupt raises the privilege level that the CPU is currently running at and transfers control to an interrupt handler registered by the operating system.
The operating system can then perform the operation and transfer control back to the unprivileged software.
From a security standpoint, the hardware and operating system ensure that the unprivileged software has no access to code and data that belong to the privileged software.
However, code and data of the unprivileged software are not protected from the operating system, and unprivileged software generally has no access to devices.
A separate solution such as confidential computing is required for that purpose \cite{wangCAGEComplementingArm2024}.

Subsystem calls fundamentally serve the same purpose as system calls.
However, ringless capability systems such as Northcape (or, e.g., Plessey System 250 \cite{winklerComputerCommunicationImpacts1972}) do not have hardware privilege levels.
Instead, subsystem calls transfer control to a \emph{different subsystem that runs at the same privilege level}, ensuring \emph{mutual isolation} between the subsystems.
In particular, the caller is unable to access data structures of the callee and vice versa, with the exception of capabilities that are passed explicitly as arguments.
This allows unprivileged subsystems to, e.g., directly call into a driver for a network device without having to interface with the network stack.
This is particularly useful for operations that do not map well to the semantics of POSIX socket-based networking, e.g., time synchronization with Precision Time Protocol (PTP).
In addition to mutual isolation of the subsystems, the passing of capabilities in a subsystem call provides \emph{temporal and spatial safety} for the call handler.
In traditional operating systems such as Linux, arguments passed by the unprivileged software cannot be trusted and need to be sanitized, which usually involves expensive copying of buffers (\emph{copy\_from\_user()}, \emph{copy\_to\_user()}).

Subsystem calls in capability architectures are not a new concept.
In fact, they were already introduced in the seminal work on capability systems by \citeauthor{dennisProgrammingSemanticsMultiprogrammed1966}~\cite{dennisProgrammingSemanticsMultiprogrammed1966}.
Different capability architectures use different names for this concept, e.g., in Plessey System 250, subsystem calls are called protected procedure calls.
Within the context of Northcape, we believe the term subsystem call more accurately reflects the purpose of the operation: requesting some operation from a subsystem (e.g., file system, network stack) in application software.

For the purpose of implementing subsystem calls, the Northcape northbridge tracks the current subsystem that each connected device is running.
Initially, the trusted loader subsystem (0) is assumed to be running.
After that, each \emph{instruction fetch} from capability with a \emph{set\_subsystem\_id} restriction and a \emph{different} subsystem ID than the currently running one signifies a subsystem call.
An instruction fetch is triggered by a jump instruction on the CPU, using a capability token as address.
Thus, the northbridge can update the current subsystem.
The CPU registers can be used to pass scalar arguments or capability tokens between caller and callee; in order to prevent leaking information, caller and callee are expected to clear registers before call and return.

While Northcape is a ringless system, the trusted loader subsystem has a very limited amount of privilege:
As the only subsystem in the system, the loader is allowed to create \emph{set\_subsystem\_id} capabilities for \emph{subsystems other than itself}.
Any other subsystem with that power could trivially circumvent subsystem-id protections by creating a \emph{set\_subsystem\_id} token from its own code.
However, the power to create \emph{set\_subsystem\_id} capabilities for \emph{foreign subsystem ids} is required for bootstrapping - we require the loader to create at least one such capability to be able to start executing at that subsystem id.
Hence, we limit this power to the loader subsystem.
\Cref{sec:skadi-os} discusses how we prevent abuse of this privilege in our operating system design.
\ifarxiv
\else
\begin{figure*}
    \centering
    \includegraphics[width=.7\textwidth]{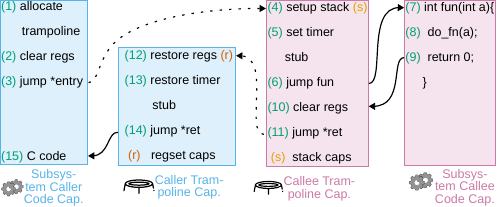}
    \caption{Subsystem call with return between two different subsystems.}
    \label{fig:subsys-call}
\end{figure*}
\fi
For the sake of this discussion, assume that there is a mechanism that creates \emph{set\_subsystem\_id} tokens for entry points explicitly exposed as subsystem calls by subsystems.
Further assume that there is a mechanism that provides subsystems with the \emph{set\_subsystem\_id} tokens for the subsystem calls that they are allowed to invoke.
Finally, assume that subsystems can generate \emph{set\_subsystem\_id} tokens for return addresses of all calls.
We will discuss how these mechanisms are implemented in~\cref{sec:skadi-os}.
Given a \emph{set\_subsystem\_id} for a subsystem call and a \emph{set\_subsystem\_id} return address token, subsystems can now call subsystems just as they would call normal functions.
To this end, they can use pseudo instructions such as \emph{jr} and \emph{ret} on RISC-V.
However, passing arguments and return values as well as being able to use a stack requires a special calling convention illustrated in \cref{fig:subsys-call}.

During a subsystem call, both the caller (after returning from the call) and the callee cannot rely on any registers except the program counter.
Thus, the Northcape operating system (Skadi) wraps subsystem call functions into trampoline code that saves and restores registers such as the stack pointer to and from a context capability.
The callee version of the trampoline code is run as soon as the subsystem callee gains control of the CPU.
The caller version of the trampoline code is run immediately after the callee returns.
The context capability is embedded within the trampoline code itself, such that it can be loaded using program counter-relative addressing.
Hence, no other register needs to be trusted to access it.
The trampoline codes also clear all registers before calling into and returning from the subsystem call to prevent information leakage.
Courtesy of the trampoline code, subsystems can be written in a high-level language like C and even maintain custom state between invocations.
Skadi also facilitates access to static and global variables from subsystem calls.

We would like to conclude the discussion of subsystems with three details of the subsystem call implementation which are relevant for providing security for subsystem calls.

First, \emph{before} changing the subsystem ID from caller to callee, the northbridge refuses any instruction fetch except from the beginning of the segment.
Thereby, the caller cannot simply jump over parts of the subsystems to force undesired behavior, such as the omission of security checks.
After changing the subsystem ID, all data and instructions in the segment can be read in the same way as from a segment with \emph{subsystem\_id\_bound} restriction.

Second, the callee trampoline uses the $inspect$ operation to determine whether the provided return address has a \emph{set\_subsystem\_id} restriction.
If this was not done, the caller would be able to execute arbitrary code in the subsystem context of the callee.
The same check is also done when dereferencing function pointers.

Finally, each subsystem can only create \emph{set\_subsystem\_id} restrictions with its own subsystem identifier.
Similar to the scenario in the previous paragraph, if this was not done, subsystems could spoof the identity of any other subsystem and circumvent \emph{subsystem\_id\_bound} restrictions.
We make an exception from this rule for the loader subsystem, identified by subsystem ID 0.
The loader needs to be able to create arbitrary \emph{set\_subsystem\_id} restrictions in order to start the other subsystems.
The loader is therefore theoretically able to circumvent \emph{subsystem\_id\_bound} restrictions and steal subsystem context pointers.
We therefore include the loader in the TCB. The loader can be small, reside completely in the hardware and it can terminate after starting the subsystems.
In case all subsystems are known at start time, the loader need not exist at run time.
If the loader does not provide the created subsystems with a \emph{set\_subsystem\_id} with subsystem ID 0, there is no way to start it again without resetting the system.

\ifarxiv
\fi

\section{Northcape Hardware Implementation}
\ifarxiv
\label{sec:northcape-hardware-appendix}
\else
\label{sec:northcape-hardware}
\fi
This section summarizes Bredi, our implementation of Northcape, a theoretical concept as of \cite{ackermannWorkinProgressNorthcapeEmbedded2024}.
We will start with an overview over the Bredi components and how they interface with the remaining system on chip.
We will then detail the individual Bredi hardware components we introduce.
Finally, we discuss Bredi's support for interrupts, which relies on both the new Bredi components and small modifications in the CPU we are using.

\subsection{Overview}

\ifarxiv
\else
\begin{figure}
    \includegraphics[width=\linewidth]{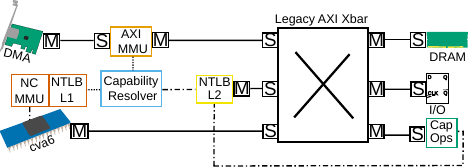}
    \caption{Bredi SoC architecture. Solid lines denote AXI, dotted lines AXI-Stream, dash-dotted lines custom buses.}
    \label{fig:northcape-soc}
\end{figure}
\fi

\Cref{fig:northcape-soc} contains the hardware components that comprise the Bredi SoC: the Northcape-aware CPU based on cva6~\cite{zarubaCostApplicationClassProcessing2019}, AXI Memory Management Unit (MMU), Capability Resolver Unit and Capability Operations Unit.
Bredi also uses two levels of Northcape Translation Lookaside Buffers (NTLBs) to speed up translation of capability tokens to bus addresses:
The CPU contains a small first-level NTLB, which is duplicated for instructions and data.
The AXI MMU omits the cache, reducing chip area.
The capability resolver and operations module share a larger second-level NTLB.

\subsubsection{AXI MMU}
The AXI MMU is responsible for the enforcement of memory access restrictions on DMA devices, akin to an IOMMU (Intel) or SMMU (arm).
It interfaces with the data user (DMA device) via its front-side bus, performs access control and translation from capability tokens to physical addresses and forwards bus transactions to the data store (DRAM or MMIO peripheral).
We currently support the standard AXI~4 bus including atomic transactions from AXI~5~\cite{armltd.AMBAAXIACE2011}.
We have verified our implementation with the proprietary AXI crossbar by Xilinx~\cite{advancedmicrodevicesinc.AXIInterconnectV212022} and the crossbar from the PULP platform~\cite{pulpprojectPulpplatformAxi2025} as well as the Xilinx AXI DMA~\cite{advancedmicrodevicesinc.AXIDMALogiCORE2022}.

\begin{figure}
    \includegraphics[width=\linewidth]{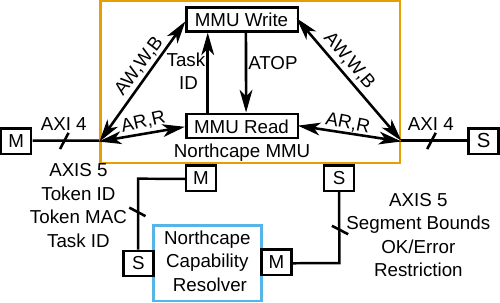}
    \caption{Architecture of the AXI MMU: two independent state machines for read and write channels with mutual communication for subsystem ID and atomic transaction support.}
    \label{fig:mmu-arch}
\end{figure}
The AXI MMU is implemented as two independent state machines, one for the read channels (AR and R) and one for the write channels (AW, W, B) of the AXI interface.
This is illustrated in \cref{fig:mmu-arch}.
For all requests that are not atomic, the state machines can process transactions fully independently.
AXI5 introduces atomic transactions that allow atomic modification of data in peripherals which support them, e.g., atomic-add or atomic-exchange operations \cite{armltd.AMBAAXIACE2011}.
The cva6 CPU uses AXI5 atomic operations to implement RISC-V atomic instructions \cite{zarubaCostApplicationClassProcessing2019}.
AXI5 atomic operations are always triggered on the write address (AW) channel, but depending on their type, they might also return data via the read channel.
Thus, there is a simple handshaking protocol between the read and write side state machine that provides the read state machine with the information it needs to forward an atomic transaction.

For all kinds of transactions, the MMU initially parses the capability token conveyed in the address lanes of the channel.
It forwards the capability ID and tag as well as the connected subsystem and device to the capability resolver, which performs a (possibly recursive) lookup in the CMT and returns the start and length of the segment as well as \emph{set-subsystem-id} or \emph{device-interpreted} restrictions associated with the segment.
The length of the segment is zero if tag or ID are invalid or the capability may not be accessed by the device and subsystem.
Otherwise, it adds the offset provided in the input token to the base address of the segment and determines whether the start and end addresses of the request are within the bounds of the segment.
Both start and end position of the request depend on the parameters of the AXI request.
For example, a fixed burst may read the single address many times, while an incrementing burst reads monotonically increasing addresses.
Bursts of type wrap can also wrap around and read addresses \emph{lower} than the address specified by capability offset and segment base address \cite{armltd.AMBAAXIACE2011}.
In case of an out-of-bounds request, the MMU generates a protocol-compliant error response.
Otherwise, it forwards the request to the data store on its master interface.
In case a segment's base and length are not exact multiples of the AXI bus size, the MMU needs to ensure that the connected device cannot read and write out of bounds.
To this end, the MMU implements the full AXI address decode logic in order to determine the physical address that each beat of a transfer accesses and computes masks to prevent out-of-bounds accesses.
The read state machine overwrites data outside of the segment with zeros to prevent information leakage, while the write state machine masks the AXI write strobes to prevent modification of data outside of the segment's bounds.

Finally, the read state machine also implements the subsystem call functionality for devices other than Bredi cva6.
To this end, it maintains a register that identifies the current subsystem.
This register is also readable by the write state machine.
Devices can indicate whether a transaction is an instruction or data fetch in the \emph{ARUSER} signal.
Hence, whenever an instruction fetch is executed on a capability with a \emph{set-subsystem-id} restriction, the current subsystem is updated.
By comparing the old and new values of the current subsystem, the MMU recognizes a subsystem call.
In this case, it refuses any request that does not start at the beginning of the segment. As discussed in \cref{sec:subsystem-calls}, this is crucial for the security of the subsystem call operation.

For each request, both state machines forward the current subsystem identifier, device identifier and the last device-specific restriction (if any) to the data store.
The capability operations module uses these information to enforce \emph{subsystem-id-bound} restrictions for capability operations.

\subsubsection{Bredi CPU}
We have integrated a simplified version of the AXI MMU into cva6.
Bredi's write-once global enable bit is used to multiplex between our Bredi MMU and the legacy sv39 MMU.
This allows us to retain backwards compatibility with legacy operating systems like Linux.
At the same time, changing the MMU control and status registers has no effect on Bredi.
The design of Northcape tokens alleviates the need for other modifications in cva6:
As tokens behave like 64-bit addresses, all assumptions cva6 makes about addresses before translation continue to hold and no extensive modifications are necessary.

Also, no special instruction for subsystem calls is necessary. Instead, subsystem calls are recognized by an instruction fetch on a capability with set-subsystem-id restriction.
Hence, Skadi can use standard RISC-V register jump instructions to jump to and return from subsystems.
If such an instruction fetch occurs, the Bredi instruction MMU updates the subsystem ID that it maintains internally and uses the updated subsystem ID for capability verification.
Thereby, \emph{control flow} and \emph{sphere of protection} are changed \emph{atomically}.
In the modified cva6, all instructions are annotated with the subsystem ID with which they were fetched.
This ensures that the data MMU always verifies the correct subsystem ID for an instruction, even if, e.g., a store issued before a subsystem call is executed only after the subsystem call was fetched.
At the same time, in case of a speculation mispredict, cva6 can roll back the subsystem ID to the one associated with the correct next instruction.

The current subsystem ID and device-specific restrictions are also conveyed to any MMIO device in the system.
For example, the capability operations module relies on the device and subsystem identifier to determine whether operations on provided capability tokens are permissible.
To this end, similar to \citeauthor{bahmaniCURESecurityArchitecture2021} \cite{bahmaniCURESecurityArchitecture2021}, we utilize the \emph{ARUSER} and \emph{AWUSER} user-defined signals contained in the AXI protocol~\cite{armltd.AMBAAXIACE2011}.

\subsubsection{NTLB L1}
\label{sec:ntlb_v1}
In order to speed up translation, both the data and instruction cva6 MMUs contain a small fully-associative first-level translation lookaside buffer (NTLB L1).
The NTLB L1 maps capability tokens to the response that the capability resolver would give, which includes segment bounds, permissions and restrictions for the capability.
This information is sufficient to decide whether a capability access is permissible.
The NTLB L1 is invalidated per-capability whenever a capability is updated, causing L1 cache misses for \emph{updated} capabilities only.
As per our operation design, operations with the exception of $revoke$, $lock$ and $drop$ on a lock-holder token can only modify the highest capability in the hierarchy.
Hence, after such an operation, it suffices to invalidate that specific entry in the NTLB L1---the cva6 MMU will then trigger a new lookup in the capability resolver and update its cache.
In case of $revoke$, $lock$ or $drop$, the NTLB L1 is invalidated entirely to maintain security:
If this was \emph{not} done, adversaries could \emph{circumvent exclusive access and revocation} using stale cache entries.
After clearing the cache, the NTLB L1 encounters a miss for all capability tokens, which is handled via the capability resolver.
As the NTLB L2 is fully invalidated for the same reason in that scenario, it performs a fresh CMT lookup, which causes an access fault.

\subsubsection{Capability Resolver Unit}
The AXI and cva6 MMUs interface with the Bredi Capability Resolver unit via an AXI-Stream bus~\cite{armltd.AMBAAXIStreamProtocol2021}.
The MMU provides the resolver with the capability ID and tag from the provided token as well as its device and subsystem.
The resolver then performs a lookup of the ID, determines whether the capability is valid and returns the bounds of the segment as well as set-subsystem-id or device-interpreted restrictions to the MMU.
The MMU makes the final decision whether the specific access is permissible---it needs to take the offset specified in the capability token into account as well as---for the AXI MMU---the burst type (fixed, incrementing, or wrapping), size and length~\cite{armltd.AMBAAXIACE2011}.
The MMU also recognizes whether an access overlaps with the CMT and refuses it, preventing privilege escalation by modification of capability metadata.

\begin{figure*}
    \includegraphics[width=\linewidth]{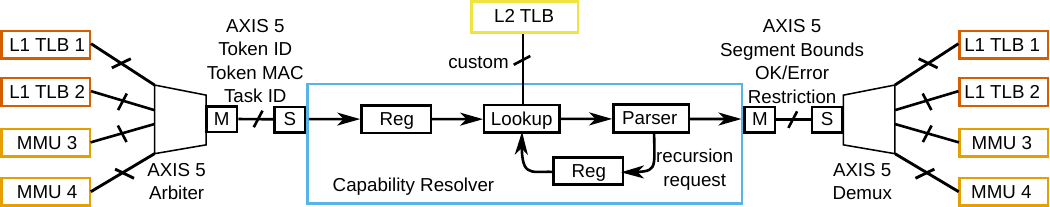}
    \caption{Architecture of the Bredi Capability Resolver: Three-stage pipeline with ID hashing, CMT lookup and CMT entry parsing stages.}
    \label{fig:resolver-arch}
\end{figure*}
The capability resolver is shared between all MMUs.
This facilitates an area-performance trade-offs:
As our L2 TLB is currently built around registers, all MMUs could theoretically access it at the same time and only be arbitrated at the missunit.
This would reduce access time in case of a high cache hit rate, at the expense of duplicating the capability parsing logic many times over.
In order to reduce this area overhead, we arbitrate the \emph{requests} from the MMUs and use a single capability resolver.
Under the assumption of a high hit rate in the L1 TLB and a moderate activity of the AXI MMUs compared to the CPU, which is the case throughout our envisioned usage scenarios, we believe this tradeoff is optimal.

The resolver is implemented as a three-stage pipeline, as depicted in \cref{fig:resolver-arch}.
The first pipeline stage captures the input in a register.
This is necessary to break the critical timing path, as our light-weight arbiter only captures the grant in a register, not the inputs.
The second pipeline stage is responsible for performing the actual memory lookups.
To this end, it uses a custom request/response-interface with the L2 TLB, providing the capability ID and tag and receiving the retrieved CMT entry and metadata as described in \cref{sec:northcape-hardware}.
The final pipeline stage accepts the MMU-provided metadata and CMT entry from the previous stage and verifies whether the access is permissible.
Optionally, an output pipeline register stage can also be enabled (omitted in the illustration).

In order to be able to do recursive capability resolutions, the parser stage loops back to the lookup stage, providing a recursion request.
The recursion request populates the same data structure as the MMU request.
For that reason, the data structure provides storage for metadata from the original request that are needed to formulate the final response, such as the bounds of the highest capability in the hierarchy.
MMUs just leave these fields blank, and the synthesis compiler will optimize them out where they are not needed accordingly.
The loopback request is buffered in a register to prevent a combinatorical loop and always takes precedence over MMU requests.
Thereby, we can issue one recursion request per cycle and thus resolve a recursion as fast as the L2 TLB will allow it without needing a complex state machine or pipeline blocker.

\ifarxiv
As we elaborate in \cref{sec:ntlb-l2-appendix}, recursive capability resolution can be skipped in case of an L2 TLB hit.
\else
As we elaborate in \cref{sec:ntlb-l2}, recursive capability resolution can be skipped in case of an L2 TLB hit.
\fi
An exception to this rule are lock-holder capabilities, which do not contain segment bounds due to size limitations.
If the first CMT entry in a recursion is a lock holder, we always have to start recursing and retrieve bounds from the CMT entry in the \emph{first} recursive call.

After the optional output register stage, a demultiplexer forwards a resolver response to the corresponding MMU.
In order to minimize the chip area of this component, we use the \emph{TDEST} AXI-Stream control signals to indicate which port a response belongs to.

\subsubsection{Capability Operations Module}
\begin{figure}
    \includegraphics[width=\linewidth]{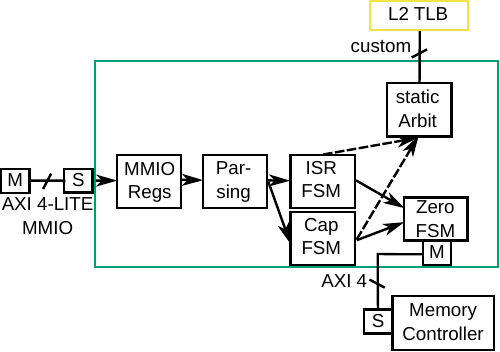}
    \caption{Architecture of the Bredi Capability Operations module: a finite state machine implements the operations while MMIO registers are responsible for providing it with inputs and returning the outputs back to the calling device.}
    \label{fig:ops-arch}
\end{figure}

The capability operations module implements the operations introduced in \cref{sec:operations} in hardware.
To this end, it provides a documented MMIO interface to consumers in software and hardware via an AXI4-Lite bus and implements the operations in a finite state machine (FSM).
\Cref{fig:ops-arch} gives an overview over the architecture of the operations module.
Consumers specify an input token in the MMIO input register, write the remaining operation parameters such as restrictions and segment lengths into the remaining input registers and start the operation by writing the operation code into the corresponding register.
All input registers are write-only to prevent, e.g., leaking input tokens to other subsystems and devices.
Devices can poll the operation register to check if the operation is complete or wait for an interrupt.
The operation output token (if any, depending on the specified operation) can then be read from the corresponding register.
The device and subsystem of the initiator of the operation (as provided by the MMU) are kept in registers in the operations module, and the operation module only responds with the output token if the read request originates in the same subsystem.
Otherwise, a zero response is returned.
As soon as the creator of the transaction has read the output register, the interrupt is cleared and the next operation may be issued.

The operations module is implemented as a hierarchical state machine with three sub state machines:
One state machine zeros the CMT and segments in main memory on revocation, one state machine is used to implement $inspect$ for ISRs and the final state machine implements the remaining operations.
The operations state machine also wraps the Qarma-64 block cipher which is used to compute the random number sequence that the capability tag $\sigma$ is chosen from.
The Qarma key and initial nonce are chosen from the output of a true random number generator which is also contained in the operations module.
The operations state machine also duplicates the logic for capability lookup that also exists in the capability resolver; thereby, it does not interfere with the time-critical lookups in the resolver.
This also allows the operations module to be synthesized at a lower clock frequency and while optimizing for chip area.

The operations state machine is partially duplicated: one copy implements all Bredi operations and is intended for use \emph{outside} interrupts.
It is controlled by a copy of the MMIO register set.
The second copy implements only the $inspect$ operation, it is intended for use during subsystem calls in interrupts.
This enables us to be able to accept an interrupt \emph{while a non-interrupt subsystem is interacting with the operations module}, which is crucial for the real-time guarantees of the system:
Bredi operations except $inspect$ are not required to complete in real-time.
If we had to disable interrupts while performing an operation using the operations module, this would mean a non-deterministic delay before the next interrupt can be accepted.
If interrupts were not enabled, the ISR would be unable to perform subsystem calls; if it did, it would change the internal state of the operations module, such as the input capability token, without being able to restore it to the state before the interrupt was accepted.
The operations module has its own interface with the L2 TLB, as documented in ~\cref{sec:northcape-hardware}, and uses a static arbiter to make sure the ISR FSM always takes precedence when it comes to accessing the L2 TLB.

We also facilitate non-maskable interrupts to perform Bredi operations while the interrupted subsystem might have been in the process of interacting with the operations module.
To this end, the ISR can use the ISR MMIO registers to commence operations just like in the non-ISR register set.
Operations like this start when the \emph{non-ISR} FSM is idle and are executed by the \emph{non-ISR FSM}.
Additional output value registers store the inputs specified by and outputs not yet read by any interrupted subsystem, preventing a breach of integrity or confidentiality by the ISR.

The capability operations module provides access to the TRNG via its register interface. To this end, after seeding the Qarma key, the operations module keeps requesting sequences of 64 bits of random bits from the TRNG.
We have wrapped the TRNG with a state machine that collects the output in a 64-bit shift register.
The operations module exposes each 64-bit sequence via a MMIO register.
Once the register is read, the sequence is \emph{immediately} discarded.
Thereby, a following read triggered by software or hardware will return a new random sequence, ensuring privacy of the sequence.
This feature is useful for both hardware devices and software that rely on secret random numbers:
They can retrieve a random number from the capability operations module without having to trust a separate device or driver.

Assume that two devices (e.g., two CPU cores, a CPU and an accelerator) want to perform a capability operation via the operations module at the same time.
As operations might consist of multiple reads and writes, this raises the question of how \emph{atomicity} of an operation is ensured:
Both devices could try to initiate different operations at the same time, potentially causing an operation to commence that neither device intended.

We implement \emph{locking} of the MMIO interface to enforce mutual exclusion between devices:
Due to arbitration on the access bus, at any given cycle, only one device can write the register interface at any time.
The first subsystem to write to the register file gains a \emph{lock} over the operations module.
As long as the lock is held, other devices or even subsystems on the same device receive an error when attempting to read the register file, and all writes are ignored.
Only the device and subsystem holding the lock can access the register interface.
The lock is released once the transaction result is read, which also clears the output in the same cycle.

Denial-of-service against the operations module with the help of this locking interface is currently accepted as a risk.
If this was a concern, one could add a \emph{timeout counter} to the register interface:
If the transaction is not completed within a timeout, the operations module destroys all inputs and outputs and lifts the lock.

The root capability is not treated specially in the Northcape system - the corresponding entry is created implicitly after each system reset, is resolved in the CMT by the resolver and can be used as input for operations.
The only difference is that ID and tag are hard-coded as $0$ instead of being computed.

The operations module tracks occupied and free CMT slots in a \emph{bitmap} that is adjusted when operations commit successfully.
The bitmap is maintained in an SRAM memory primitive of a configurable size, providing a technology-dependent performance-area tradeoff in the configurable \emph{row size}.
Therein, we utilize a design advantage of Northcape over, e.g., CHERI~\cite{woodruffCHERICapabilityModel2014}: capabilities are confined to the CMT, which occupies a small part of system DRAM ($2^{13}$ entries in our implementation).
This significantly limits the size of the bitmap, making it possible to store the \emph{entire bitmap} in SRAM rather than having to cache it.

Allocation of capability IDs uses the bitmap:
The capability operations FSM requests row by row from the SRAM until a row with non-occupied entries is found.
We use a parallel leading zero count implementation to select the first free slot in each row.
Thereby, assuming the SRAM has $r$ rows, we can either find a free slot in $\leq r$ cycles or reliably detect that the CMT is completely full in $r$ cycles.
Thus, Northcape operations on our platform \emph{complete in predictable time}, with a worst-case execution time that depends solely on the memory access latencies of loading the input CMT entries and writing the output CMT entries, the number of which is always the same for the same operation!

Note that due to the different \emph{offset sizes} in Northcape capability tokens, large capabilities like 32 and 24-bit offset capabilities can only occupy the lowest $2^6$ and $2^14$ capability IDs.
The same is not true for capabilities with smaller offsets.
Thus, we try to avoid allocating capability slots with low identifiers for small capabilities.
To this end, for each capability size, we start iteration not at zero but at the lowest capability ID of the next bigger capability type plus one.
We also \emph{remember the last row with an unallocated CMT slot for each offset type}, alleviating the need to iterate over fully occupied rows again during the next operation. 

\subsubsection{NTLB L2}
\ifarxiv
\label{sec:ntlb-l2-appendix}
\else
\label{sec:ntlb-l2}
\fi
Finally, the operations module and resolver share a fully associative write-through NTLB L2.
In contrast to the NTLB L1, the NTLB L2 maps capability tokens to the \emph{full CMT entry}.
The NTLB L2 also implements a full shared capability missunit and writeback unit.
An arbiter with static priority, always giving precedence to the resolver, multiplexes the missunit while the cache is read in parallel.

\begin{figure*}
    \includegraphics[width=\linewidth]{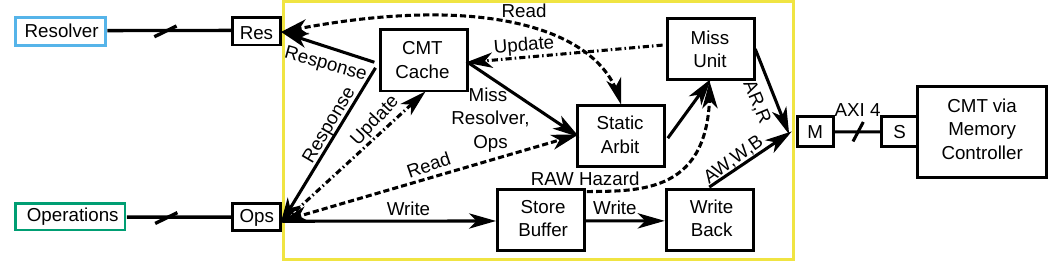}
    \caption{Architecture of the Bredi L2 TLB: a parametrized cache wrapped with a miss unit and writeback unit.}
    \label{fig:l2-arch}
\end{figure*}

The architecture of the L2 TLB is given in~\cref{fig:l2-arch}.
The L2 TLB provides a custom request-response interface to the capability resolver and operations module and has access to the CMT in main memory via an AXI 4 interface.
At its core, the L2 TLB uses a parametrizable cache: Using synthesis parameters, the algorithm and size can be set.
We implement a direct mapped cache that uses the capability tag to identify the cache line as well as n-times and fully associative caches, providing a user-configurable chip area-performance-tradeoff.
To this end, the memory implementation also differs: While the fully associative cache needs to use individual \emph{registers}, the n-times associative and direct mapped cache were built around more space-efficient \emph{dual-port SRAM primitives}.

The L2 TLB has two concurrent access ports, one used by the capability resolver and one used by the capability operations module.
Hence, a true dual-port SRAM or register-based implementation is used.
In case of a cache miss, a capability can be retrieved using the miss unit.
To this end, we use an arbiter with static arbitration, ensuring the capability resolver always takes precedence.
This is necessary, as stalling the resolver means possibly stalling multiple MMUs at once.
A resolved miss is written into the cache if the retrieved CMT entry has the corresponding cacheability permission (\emph{cacheable\_tlb}) and the request originates in the resolver.
This is needed for being able to skip hierarchical resolution in the capability resolver, as explained below.

While the capability resolver port never issues writes, the capability operations port does.
To this end, a memory write-back unit is connected to the operations module port.
If the operations module wishes to write a capability to memory, the writeback unit is started immediately (\emph{write-through}).
In the cycle during which the memory controller acknowledges the write, the written CMT entry is also provided to the cache.
If it is cacheable, it is written into the cache; otherwise, to ensure the cache has no stale data, the corresponding cache line is invalidated if it contains the old CMT entry for this capability.

There is a potential read-write hazard between the missunit and the write back unit: both could try to concurrently access the same capability.
In this case, we abort the missunit request if still possible. If not, we buffer the CMT entry to be written and use it instead of the response from the capability missunit for both the cache update and response generation.

Also, the cache is write-through---hence, the writeback unit is only used by the operations module.
In fact, all writes by the operations module are both committed to the cache and the memory before the current operation commences.
This design was chosen primarily to reduce chip area in contrast to a write-back cache.
As capabilities tend to be long-lived in our system, we believe a write-back architecture would only increase performance marginally.
We also optimize the configuration options of the Xilinx crossbar and memory controller---each CMT entry fits into 256 bits, which is also the maximum bus width of the memory controller.
Thereby, each CMT entry can be transferred and registered-in during a single cycle, reducing latency and complexity of the cache.

The design of our NTLB L2 allows the capability resolver to skip parsing the entire capability hierarchy in case the capability that was requested first encountered a cache hit.
This scheme is secure using the following invariant: \emph{In case a capability encounters a hit in the NTLB L2, its parents up until the final grandparent have been parsed by the resolver and not been updated since then}.
The invariant is maintained by the NTLB L2, resolver and operations module working together:
First, the NTLB L2 indicates to the resolver whether an access was a hit or not.
In case of a hit, the resolver can skip resolution of the (remainder of the) hierarchy and parse the capability immediately.
Otherwise, it needs to perform a complete recursive capability verification.
For each miss encountered during resolution for the \emph{resolver}, the missunit writes the retrieved CMT entry into the cache.
This causes a metadata bit \emph{speculative} to be set for the new cache entry.

Once the resolver has completed recursive resolution, in case of success, it can \emph{close the speculation window} causing the \emph{speculative} bit to be dropped across the cache.
The resolver pipeline ensures that recursive resolutions are always completed before the next resolution request from a different MMU is processed.
Thus, all CMT entries marked with \emph{speculative} must belong to the current hierarchical resolution, which was valid, and the speculative bit can be removed without violating the invariant.
In case of a recursive resolution error, the resolver indicates a \emph{flush} to the NTLB L2, causing all speculative entries to be invalidated.

The operations module uses the NTLB L2 for capability lookups.
However, in order to maintain the invariant, \emph{cache misses from the operations module are not written to the cache}.
As this is a write-through cache, all updated capabilities are written to both cache and memory.
Operations besides $revoke$, $lock$ and $drop$ can only modify the top capability in the hierarchy, which is always checked by the resolver.
Thus, for most operations, no additional work is needed to maintain the invariant.
Finally, like the NTLB L1, for $revoke$, $lock$ and $drop$, the NTLB L2 is invalidated in its entirety.

Committing capability writes to memory can take between dozens and hundreds of cycles on our platform.
Hence, waiting for capabilities to be commited to memory can slow down the operations module.
Thus, we employ a \emph{store buffer}: A synchronous first-in first-out (FIFO) memory buffers capabilities that are-to-be-written by the operations module.
The operations module can proceed as soon as space in the buffer is available, without having to wait for writes to commit fully.
The \emph{writeback unit} will drain the store buffer by actually committing writes to memory.

The store buffer introduces a \emph{read-after-write} hazard with the missunit:
In case a capability is first written into the cache and store buffer, then evicted from the cache, then re-loaded via the missunit and finally committed to memory via the writeback unit, the cache would contain a stale value for the capability.
We resolve this hazard in the following way:
We track the capability IDs for the capabilities in the store buffer using a \emph{ring buffer} implemented using registers.
We detect the read-after-write hazard by comparing the capability ID to be loaded from the missunit with all valid capability IDs in the store buffer.
In case of a match, we \emph{stall the missunit} until the write commits to memory.
Empirical analysis suggests that this hazard is very infrequent, causing no measurable slowdown to the system.
In case there is no hazard, both the missunit and the writeback unit can write one capability \emph{in the same clock cycle} by using the concurrent write ports.
Thereby, no arbitration between the units is needed, and a missunit write does not need to wait for the operations module in non-hazard scenarios.

\subsubsection{Performance Counters}
We have added performance counters for Northcape NTLB L1 and L2 misses.
They are integrated with RISC-V's run-time configurable performance event framework and can be read by all software components.
The performance counters exist for optimizing performance-critical software and can be disabled using the \emph{mcountinhibit} control register.

\subsection{Bredi Interrupt Architecture}
\label{sec:interrupt-architecture}
The subsystem call support we have described so far is capable of compartmentalizing an operating system in as many small, mutually isolated subsystems as desired.
However, it assumes \emph{voluntary} context switches between subsystems with the help of \emph{subsystem calls}.
Using this approach, it is not possible to support \emph{interrupts}, as they happen asynchronously.
In this section, we discuss how we solved the following challenges in supporting \emph{untrusted interrupt service routines}:
\begin{itemize}
    \item \emph{separation} of ISRs from subsystems and each other
    \item \emph{securely returning} to an interrupted subsystem from ISRs
    \item preventing \emph{sharing of registers} between subsystems and ISRs
    \item \emph{preemptive scheduling} with an \emph{untrusted scheduler}
    \item \emph{unconditional availability} of interrupt handlers
\end{itemize}

\ifarxiv
\else
\begin{figure}
    \centering
    \includegraphics[width=\linewidth]{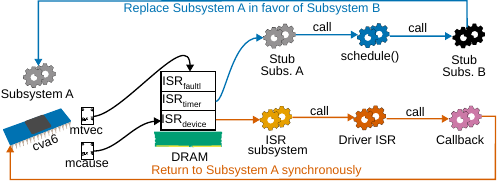}
    \caption{Vectored interrupts in Bredi cva6.}
    \label{fig:northcape-interrupts}
\end{figure}
\fi
\subsubsection{Separation of ISRs}
We add a non-standard interrupt vectoring scheme depicted in \cref{fig:northcape-interrupts} to cva6.
This vectoring methodology is similar to the ongoing work on RISC-V fast interrupts (FastIRQ)~\cite{risc-vinternationalCoreLocalInterruptController2025}.
However, as the fast interrupt specification has not been ratified or implemented in cva6, we implemented a simpler custom scheme.

Exactly as in the FastIRQ scheme, our vector table stores the \emph{address} of the ISRs for each interrupt.
When taking an interrupt, the CPU retrieves a 64-bit \emph{handler token} from the address $8\cdot \text{\emph{mcause}} + \text{\emph{mtvec}}$ and jumps to the handler.

We also change the vector table register \emph{mtvec} to be \emph{read-only} after our new vectored interrupt mode was first enabled.
Thereby, the trusted loader subsystem can configure the register at load time, with the remaining untrusted software subsystems being unable to change the value.

Vector table entries comprise capability tokens with \emph{set-subsystem-ID} restriction.
Thereby, accepting an interrupt causes a subsystem call into the ISR, ensuring isolation between ISR and subsystem as well as different ISRs.
Skadi provides an ISR subsystem that is permanently registered for device interrupts using our interrupt vectoring scheme.
Note that on cva6, all devices share one CPU interrupt line---the \emph{platform-level interrupt controller} (PLIC) tracks the actual interrupting device, implements IRQ prioritization and forwards interrupts from devices to the CPU.
Hence, in Skadi, the ISR subsystem always invokes the PLIC driver via subsystem call.
The PLIC driver determines which device triggered the interrupt and performs a subsystem call into the registered driver subsystem.

\subsubsection{Securely returning from ISRs}
In our IRQ mode, ISRs need to terminate execution using the return-from-interrupt instruction \emph{mret}, which causes the CPU to continue execution from the instruction where the interrupt was taken.
As long as this is not done, interrupt handlers cannot return to the interrupted subsystem and are only able to invoke \emph{subsystem calls} that were explicitly exposed.
In addition to re-setting the program counter, we also need to re-set the subsystem ID in the Bredi MMU when we return from the ISR subsystem.
As the \emph{interrupted instruction address mepc} generally is not at the beginning of the interrupted subsystem's code segment, we cannot rely on standard subsystem call semantics to return to the interrupted subsystem.
In order to solve this, the cva6 MMU distinguishes the current subsystem ID in the interrupt regime and the normal (i.e., non-interrupt) regime.
To this end, it contains two different subsystem ID registers.
The CPU enters the interrupt regime when taking an interrupt and exits it when a return-from-exception instruction \emph{mret} is executed.
Thereby, an \emph{mret} implicitly changes the active subsystem ID back to the one of the interrupted subsystem.

\subsubsection{Preventing sharing of registers}
In the standard RISC-V interrupt vectoring schemes, the ISR gains access to the CPU registers, facilitating \emph{rescheduling} in ISRs at the cost of leaking register contents.
Thus, we implemented a second register set in cva6 (\emph{register stacking}): whenever it processes a device interrupt, it changes to this second register set.
In order to prevent ISRs from modifying the CPU state, we also made changes to configuration and status registers:
\emph{mstatus}, \emph{mie} and \emph{mepc} are split into an ISR and a non-ISR copy, with the current execution regime determining which copy is used.
The respective second copy is not accessible to software until the regime of execution changes.
Thereby, the interrupted subsystem will always continue to execute with the same state, e.g., an enabled FPU, and at the correct instruction.

\subsubsection{Preemptive Scheduling}
The duplicated register file raises the question of how we can support \emph{preemption} of a subsystem that is currently executing:
As the ISR does not have access to the subsystem's registers, it cannot reschedule.
We utilize our \emph{interrupt vectoring} scheme to solve this problem:
Each subsystem is provided a writable, non-IRQ-accessible capability token that resolves to the capability token in the vector table that is associated with the timer interrupt.
Thereby, each subsystem can register a function \emph{within its code segment} that is called when the timer interrupt is accepted and performs a subsystem call into the \emph{scheduler}.
The scheduler can then determine whether the subsystem needs to be preempted and pick a different runnable subsystem. The scheduler can use a subsystem call on the capability token provided as return address in order to context switch to either subsystem.
Taking a timer interrupt with our non-standard vectoring mode \emph{is an exception to register stacking} and \emph{does not cause the CPU to transition into the IRQ regime}.
Thus, the scheduler has access to the non-IRQ register set as usual.
Obviously, this preemption scheme is \emph{voluntary}, as in it relies on the current subsystem for cooperating with the scheduler.

\subsubsection{Availability of interrupt services}
Our security model in \cref{sec:security-model} aims at guaranteeing the execution of interrupts handlers even in the presence of adversaries who can disable interrupts on the CPU.
In order to achieve this security goal without compromising confidentiality and integrity, we rely on \emph{non-maskable interrupts (NMIs)}, i.e., interrupts that cannot be disabled by software.

As cva6 lacks support for NMIs, we implement a custom scheme:
Using a custom CSR, software can \emph{mark} an ordinary interrupt as non-maskable.
Similar to our custom interrupt vectoring mode, there is no way of un-setting this bit, ensuring that the non-maskable attribute cannot be removed.
When the bit is enabled, our modified cva6 takes non-maskable interrupts irregardless of the interrupt enable flags, except if it is already executing an NMI.

\section{Skadi RTOS Implementation}
\ifarxiv
\label{sec:skadi-os-appendix}
\else
\label{sec:skadi-os}
\fi

This section discusses the architecture of the Skadi real-time operating system.
Skadi takes advantage of the Bredi architecture in order to maximize the compartmentalization of the operating system and eliminate the run-time TCB.
To this end, it does not rely on any trusted software to guarantee the confidentiality and integrity of subsystems' data.
It is also capable of meeting real-time deadlines as long as the subsystems involved in interrupt handling are not compromised.

The section starts with a discussion of the overall architecture of Skadi.
We will then specifically discuss the Skadi loader, which is the only trusted component in Skadi, in conjunction with the run-time relocation of subsystems.
After that, we will focus on how the security guarantees confidentiality, integrity and real-time compatibility are achieved.
Finally, we will briefly discuss how existing RTOS subsystems can be ported to Skadi.

\subsection{Overview}

\begin{figure}
    \includegraphics[width=\linewidth]{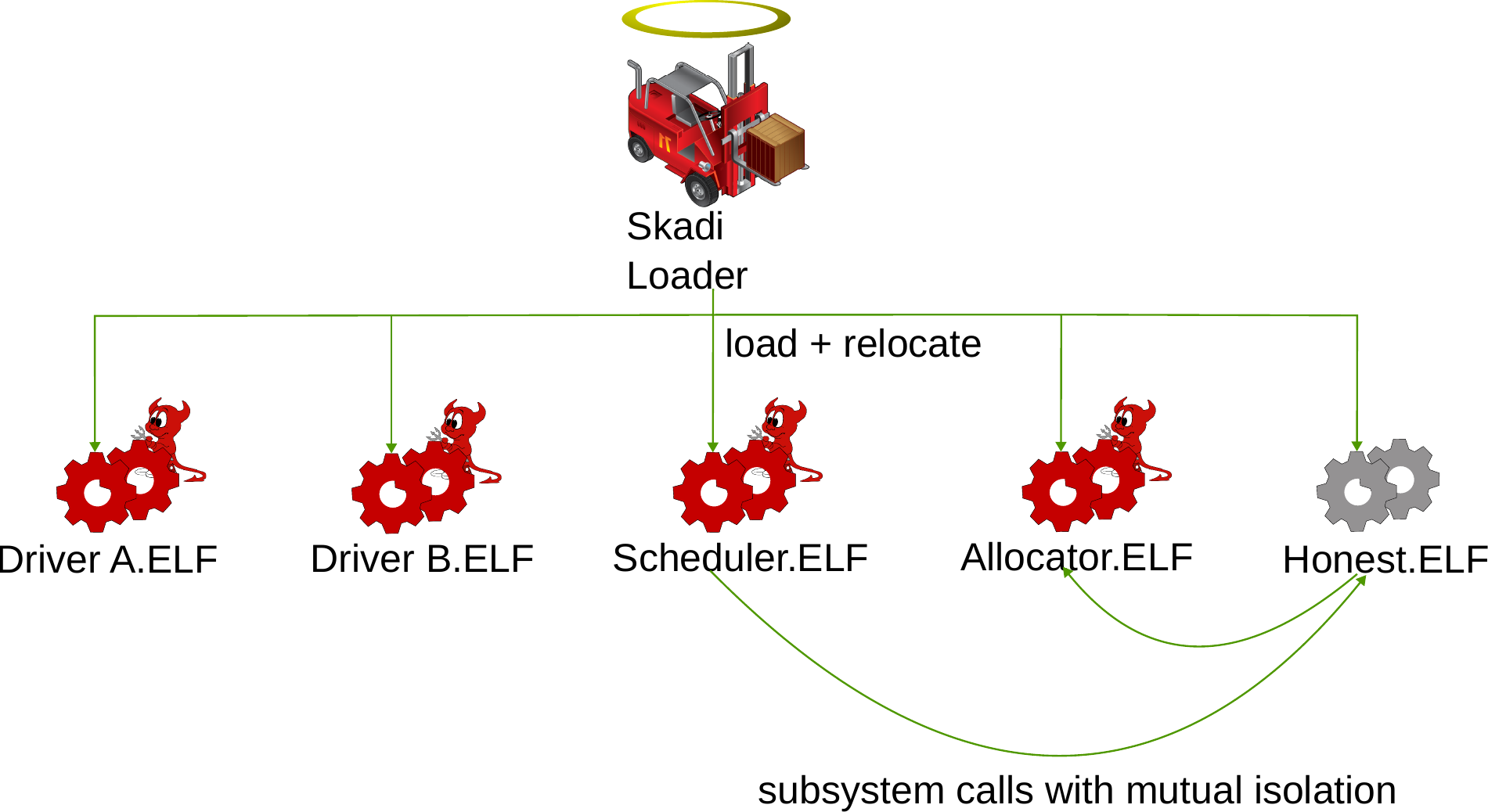}
    \caption{Decomposition of operating system components in Skadi.}
    \label{fig:os-structure}
\end{figure}

Skadi is based on Zephyr, a real-time operating system supported by the Linux foundation.
In comparison to other RTOS like RIOT and Contiki-NG, \citeauthor{silvaOperatingSystemsInternet2019} characterize Zephyr as a nanokernel~\cite{silvaOperatingSystemsInternet2019}.
Zephyr provides a minimal operating system kernel that implements crucial system facilities like scheduling, memory management and architecture-specific configuration.
In addition to that, a large selection of subsystems and device drivers can optionally be enabled at build time.
This includes a full network stack, filesystem support and device drivers for many SoC families and devices.
Zephyr also provides llext, a mechanism for relocating ELF binaries at run time, intended for loading extensions.
In the most basic configuration, all operating system components as well as user subsystems are executed with full access to all memory and privileged CPU instructions.
If the selected SoC supports this, user space subsystems can optionally be isolated from the operating system using a memory protection unit (MPU).
To this end, Zephyr includes system calls for user-space subsystems.

\Cref{fig:os-structure} shows how we structured Skadi.
Skadi takes advantage of Zephyr's highly modular design.
The RTOS uses the existing separation of Zephyr in kernel, drivers and subsystems and introduces run-time isolation of these components.
We would like to emphasize that this separation is enforced for \emph{all} components of the operating system that exist at run time with no exceptions.
For example, the interrupt service routines (ISRs) are encapsulated into \emph{three different} subsystems: one for handling interrupts from devices, one for handling exceptions and one for timer interrupts.
We have explained the utility of this approach in~\cref{sec:interrupt-architecture}.
Like all subsystems in Skadi, the ISR subsystems need not be trusted to ensure confidentiality and integrity of system subsystems, and only one of the two ISR subsystems (the device interrupt subsystem) needs to be trusted to achieve real-time execution.
All other operating system functionality, such as dynamic memory allocation, device management and networking, as well as all application software are likewise contained in mutually separated subsystems.

When booting Skadi, a trusted loader component allocates buffers for the operating system components using the Northcape $create$ and $derive$ operations.
It then \emph{relocates} each operating system component and application program into its individual buffers and uses the Northcape restriction system to \emph{restrict} usage of these buffers to the associated subsystems.
Conceptually, the loader initially has ownership over the root capability; to this end, it uses subsystem-id protection to prevent subsystems from reading or writing the root capability.
With each allocated buffer, the loader \emph{relinquishes} ownership of a piece of memory, until all subsystems are loaded.
After loading the last subsystem, the loader destroys the root capability using a final $create$, destroys the memory contents and hands the freed memory over to the allocator, making it part of the general-purpose heap.

At run time, Skadi uses Northcape subsystem calls to transfer control between components with mutual isolation.
This mimics message passing in a traditional microkernel design~\cite{kleinSeL4FormalVerification2009}, but relies on the Northcape hardware support rather than privileged software for secure transfer of control.
To this end, standard ELF symbols can be used to indicate imported and exported subsystem calls.
The Skadi loader is responsible for resolving imported subsystem calls from other components and relocating them.
Also, the loader derives capabilities for exported subsystem calls from the code segments based on the exported symbols.

\subsection{Skadi Structuring in Subsystems}

\paragraph{Skadi components are ELF binaries}
Skadi performs relocation of code and data conforming to the Executable and Linking Format (ELF) standard~\cite{tiscommitteeToolInterfaceStandard1995}.
To this end, each operating system component is compiled into a separate ELF binary.
Subsystem binaries are conveyed as part of the loader binary, using \emph{lz4 compression}~\cite{colletLz4Lz4Extremely2025} to reduce disk space.
The loader then performs standard ELF relocation of the binaries at run time.
For that purpose, the loader can mostly utilize the implementation of ELF loading and relocation in Zephyr's existing llext extension mechanism.
The primary innovation in the loader is that it includes a simple trusted allocator that uses the Northcape $create$ operation to allocate individual buffers for the ELF segments and provides them to llext as relocation targets.
The loader also uses the Northcape restriction mechanism to ensure code and data segments can only be accessed by the subsystem they are running.
Akin to the existing llext infrastructure, the loader also allocates a stack for the subsystem and performs subsystem calls into dedicated initialization functions of the subsystem, indicated using a special ELF region.

\ifarxiv
\else
\begin{figure*}
    \centering
    \includegraphics[width=.5\linewidth]{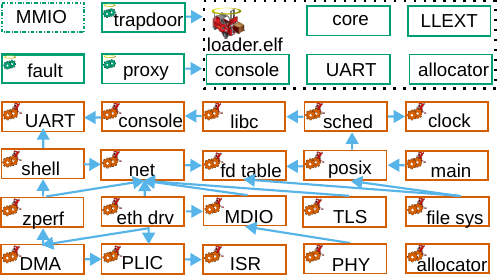}
        \caption{Compartmentalized subsystems in Skadi as well as loader architecture (dashed box). Arrows represent subsystem calls. First two rows of subsystems and loader are trusted: the fault subsystem is only trusted to halt the system after an exception, whereas \emph{trapdoor} and \emph{proxy} become inaccessible at the end of load time. MMIO capabilities are created by the loader and delegated to specific subsystems.}
        \label{fig:subsystems-skadi}
\end{figure*}
\fi

The key advantage of this approach is that it allows the integration of untrusted operating system components in a safe and secure way:
For example, device drivers may be (solely) provided as closed-source binaries by the device vendors.
Likewise, traditional user-space components can be purchased in binary form from a supplier.
System integrators can use Skadi's decomposition into subsystems to combine such untrusted closed-source binaries with their own proprietary drivers and devices and rely on subsystem isolation to ensure the integrity and confidentiality of their code.
In conjunction with Bredi, it is ensured that untrusted drivers cannot use devices as confused deputies to circumvent these protections - a key advantage over existing capability architectures like CHERI~\cite{esswoodCheriOSDesigningUntrusted2021}.
The subsystems that Skadi consists of as well as the loader architecture are illustrated in~\cref{fig:subsystems-skadi}.

Exporting and importing subsystem calls also utilizes the existing ELF symbol semantics to increase backwards compatibility:
Subsystems \emph{import} subsystem calls by defining a symbol with the name of the subsystem call and type \emph{undefined}.
Subsystems \emph{export} subsystem calls by defining a symbol with the name of the subsystem call and its location in a special, well-known region.
This can be automatically handled by any ELF-compliant linker and needs no toolchain modifications.
Also, the imported and exported symbols can be used to determine a load order of subsystems \emph{at build time}.
To this end, a simple script establishes imports-symbol-from dependencies between subsystems and computes a \emph{topological sort} of the subsystems, which is then used as initialization sequence.

In order to resolve dependency cycles between subsystems, the Skadi loader provides a subsystem call API for resolving symbols dynamically.
This API is proxied through the \emph{loader\_proxy} subsystem - this decreases the attack surface of the loader and can be used to stop execution if the API is used after the destruction of the loader.

\paragraph{Legacy compilers can generate code compatible with Northcape capability tokens}
For supporting arbitrary capability tokens during the relocation of ELF symbols, Skadi takes advantage of the \emph{large code model} specified in the RISC-V ELF specification~\cite{RiscvelfpsabidocRiscvelfadocMaster}.
The purpose of the large code model was to enable applications on 64-bit RISC-V processors to use the entire 64-bit (virtual) address space for code.
The medium-any (medany) and medium-low (medlow) code models rely on sequences of \emph{U-type} and \emph{I-type} RISC-V instructions with 20-bit and 12-bit immediates respectively\footnote{The U-type instructions like \emph{auipc} and \emph{lui} set the high 20 bits of a register as absolute or PC-relative value, while the I-type instructions like \emph{add} or \emph{ld} add a 12-bit offset to the low bits.}, enabling them to either indicate any 32-bit address (medlow) or any 32-bit \emph{offset} from the program counter (medany).
Standard ELF relocations are used to indicate to the ELF loader where to find these pairs of instructions and how to compute the immediates.
While medany and medlow are ideal for 32-bit programs, for 64-bit programs, they impose restrictions as to where code can be loaded.
Hence, the large code model was added.
In the large code model, the \emph{C compiler} allocates a table with one 64-bit entry for each external symbol used in a function directly before the function preamble.
It then uses U-type/I-type instruction pairs to access the \emph{per-function symbol table}, similar to the medany code model, and retrieves the actual value for each external symbol from the table.
While this mechanism was \emph{intended} for 64-bit virtual addresses, the code makes no assumption about the location of the external symbol in the table.
Because in the Northcape system, capability tokens have a width of 64 bits, the Skadi loader can directly specify Northcape tokens in the symbol table.
These tokens correspond with \emph{derived capabilities} that allow the importing subsystem to either execute a subsystem call onto an exported \emph{callee trampoline} or to access \emph{size-restricted global data}.
Hence, subsystems can import and use symbols for both code and data from other subsystems by importing the corresponding ELF symbols.
Thereby, Northcape needs no modifications to the compiler and linker, a notable advantage over RV-CURE and CHERI, which rely on compiler support~\cite{kimRVCURERISCVCapability2025,woodruffCHERICapabilityModel2014}.
Skadi uses this to provide both tokens for imported subsystem calls and tokens for functions and data from the same subsystem.
Crucially, this allows us to relocate code, data, bss and read-only data to different segments with only the required access permissions instead of having to include them in one buffer with read, write and execute permission.

\paragraph{Access to MMIO is requested via ELF undefined symbols}
The ELF symbol import mechanism is also used to give subsystems access to MMIO registers.
We have integrated this feature with Zephyr's existing build-time device tree parsing via macros - instead of resolving device registers to a scalar physical address, macros like \emph{DT\_INST\_REG\_ADDR} resolve to the \emph{address} of an external function.
The name of this function encodes the bounds of the requested MMIO region - for example, \emph{DT\_INST\_REG\_ADDR} on the device tree entry for the operations module on our test platform resolves to the function name \emph{\_\_skadi\_mmio\_0\_1107296256\_4096}.
This encodes an MMIO region that starts at $42000000_{16}$ and is $4096_{10}$ bytes long.
Using the \emph{address} of the function causes the compiler to declare an undefined (imported) symbol with the aforementioned name.
The compiler then embeds a 64-bit relocation that copies the resolved symbol address anywhere the \emph{DT\_INST\_REG\_ADDR} macro was used to resolve the device registers - by convention, this is usually contained in a struct \emph{device\_config} and referenced in the \emph{struct device} in the driver.
Finally, the referenced function is not actually defined anywhere. Instead, the Skadi loader checks whether unresolved imported symbols match the encoding scheme for MMIO entries and if this is the case creates a capability for the requested MMIO space, using the capability token as a symbol address.
This approach allows us to provide source-level compatibility with existing drivers that use the device tree macros, including embedding the capability token in a struct declared \emph{const}.

\paragraph{Generative macros automate subsystem call trampoline creation}
The calling convention for subsystem calls is provided using generative macros for caller and callee that contain a combination of preprocessor and compiler directives and inline assembly.
Therefore, modification of existing Zephyr components is limited to adding macros for invoking and exporting subsystem calls using the Skadi subsystem calling convention.

Skadi uses the following convention for subsystem calls.
Both the caller and callee provide trampolines for the subsystem calls.
The trampolines serve three purposes.
First, as discussed in \cref{sec:subsystem-calls}, Northcape only accepts subsystems calls to the \emph{first byte} of the target segment.
This is not generally true for function calls - the associated code can start anywhere in the code segment.
Second, before invoking the subsystem call in the caller and before returning from the callee, the trampolines save all registers if needed and overwrite them with zeros to prevent leaking information across subsystem boundaries.
Third, both the callee (immediately after the subsystem call) and the caller (immediately after \emph{returning} from the subsystem call) cannot trust any register except the program counter.
In particular, the stack, thread and global pointer registers might have been manipulated by the other subsystem in an attempt to steal or falsify data.
Thus, the trampolines restore a context token that was \emph{previously copied into the trampoline's code segment} and retrieve tokens for the stack, thread and global pointer registers.
The caller trampoline also restores register values from before calling the subsystem from the context.
The macros for caller and callee trampolines automatically register initialization subsystems that prepare the context pointers for caller and callee trampolines, such that this does not need to be done manually for all subsystem calls.
Caller and callee trampolines also configure the the CPU status in \emph{mstatus} depending on the subsystem's needs - e.g., for subsystems that want to use the floating point unit, the callee and caller trampolines ensure that the FPU is enabled before jumping into C code.

One crucial problem that we had to solve in the callee trampoline was preventing the stack from being overwritten while it is in use.
We refer to this problem as the stack-in-use problem.
We would like to explain this problem using an exemplary scenario: Consider subsystems A, B and C.
A invokes a subsystem call \emph{foo} that is implemented in B, causing B to use its subsystem call stack to save, e.g., the return capability token.
In order to compute \emph{foo}, B needs to invoke a subsystem call \emph{bar} in C.
While C uses its own stack to compute \emph{bar}, B needs to hold on to its own stack to return to A.
Now assume that during the execution of \emph{bar}, C calls a function pointer.
Also assume that the function pointer resolves to a subsystem call \emph{baz} in B.
In a na\"ive implementation where each subsystem has access to exactly one stack frame, during the execution of \emph{baz}, B overwrites the state it had when calling \emph{bar}.
Thus, after B returns from \emph{baz} and C returns from \emph{bar}, B attempts to continue execution of \emph{foo} with the stack frame at the time \emph{baz} returned, leading to incorrect results.

We use the following approach to solve the stack-in-use problem.
Each Skadi subsystem allocates a build-time configurable number of stack frames. At the time of writing, the default number is four.
The subsystem stores the allocated stacks in the subsystem context alongside a bitmap that indicates which stacks are currently in use.
Each subsystem callee trampoline can use the bitmap to find the first available stack for the subsystem.
Before returning from the subsystem call, we re-load the current bitmap to prevent stale data and mark the stack frame as available.
We implement both claiming and releasing stack frames using atomic instructions (amoor.d for claiming, amoxor.d for releasing a stack frame).
Thereby, our implementation is SMP-safe and IRQ-safe without having to disable IRQs.
A caveat of this approach is that we cannot be sure whether stack frames \emph{will} become available - our implementation currently spins through the bitmap indefinitely, waiting for a stack frame to become available.
If stack exhaustion is a concern, one can additionally count attempts and cause a controlled panic after a threshold.
 
Initialization uses a slightly simplified version of the trampolines; while the loader needs to isolate itself from the subsystems, the subsystems can leak their register contents to the loader.
Also, by convention, the loader calls the initialization function with a stack pointer that it allocated for the subsystem and the subsystem's identifier.

In Skadi, caller trampolines are mostly declared in \emph{system headers}.
This allows consumer subsystems to invoke subsystem calls transparently.
This has one disadvantage: a na\"ive implementation of this strategy that instantiates the entire trampoline in the header leads to duplicate caller trampolines as well as unused caller trampolines.
Thus, we provide an optimization: the caller trampolines declared in headers define an extern function that encodes the subsystem call name and signature in its name.
The extern function results in an undefined ELF symbol.
Before linking the final subsystem, we scan the object files using a python script and collect all undefined ELF symbols that correspond with a caller trampolines.
We then generate a source file that implements the according caller trampolines, compile it and link it into the subsystem, resolving the undefined ELF symbols.
Thereby, we deduplicate the caller trampolines.
Also, as the compiler only emits undefined ELF symbols for caller trampolines that are \emph{actually used}, we do not generate excess caller trampolines.

Caller trampolines save \emph{caller-saved registers} according to the RISC-V ABI in a \emph{register set struct} before commencing the call and restore them from the same structure later.
Before the call, the entire register set is also cleared.
Thereby, Skadi ensures no information can be leaked from caller to callee.
Access to the register sets uses the same allocation mechanism that the callee trampolines use to allocate stacks.

Finally, there is one additional issue that our trampoline implementation needs to work around:
In case a thread is canceled either synchronously or asynchronously, the thread can leak both register sets for caller trampolines and stacks for callee trampolines - the stack used by the thread function as well as all register sets and stacks involved in the thread's last call are never cleaned up, as there is \emph{no return} after the cancellation.
Note that the leaked trampolines are not necessarily restricted to the same subsystem as the thread's handler function - e.g., if the thread is waiting for a scheduler object like a semaphore when it is canceled, this blocks one stack in the scheduler forever.
In order to solve this issue, the scheduler exports a read-only capability for the identity of the currently scheduled thread via global data.
Caller and callee trampolines use this information to associate stacks and register sets with the current thread.
The scheduler invokes a callback in each subsystem when a thread is cancelled, allowing all subsystems to free register sets and stacks that are currently allocated to the thread.
\ifarxiv
\else
\begin{figure}[ht]
    \includegraphics[width=\linewidth]{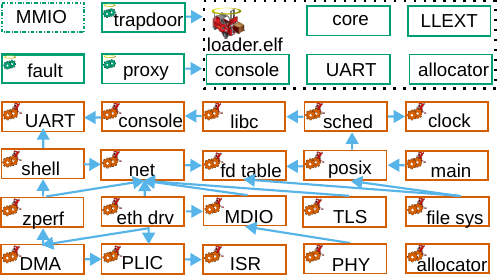}
        \caption{Implemented subsystems in the final Skadi operating systems as well as loader architecture (dashed box). Arrows represent subsystem calls.}
        \label{fig:subsystems-skadi}
\end{figure}
\fi
\paragraph{Trampolines support variadic functions}
C functions like \emph{printf}, \emph{open} and \emph{ioctl} are \emph{variadic}: they accept an \emph{arbitrary amount of function arguments}.
By convention, these functions have at least one \emph{mandatory argument} that informs the function how many arguments were passed:
for printf, the format string includes this information, while for open, this is indicated by the presence or absence of a bit in the flags argument.
Variadic functions are indicated by the presence of ellipses in the declaration.
C code can interact with variadic arguments with the help of the opaque type \emph{va\_list} and its associated accessors like \emph{va\_start} or \emph{va\_copy}.
\emph{va\_lists} can also be passed to other functions as arguments - by convention, functions like printf usually forward their argument to a function with a \emph{v} prefix such as \emph{vprintf}.
\emph{vprintf} takes a single \emph{va\_list} argument instead of the ellipses.
Irregardless, functions processing \emph{va\_lists} are expected to determine how many arguments of which types to extract from the list and use the stateful built-in \emph{va\_arg} to iterate the argument list.
How the variadic arguments are passed is architecture-dependent: on RISC-V, variadic functions expect to receive up to 8 arguments in the integer argument registers $[a0,a7]$ and floating point argument registers $[fa0,fa7]$, respectively.
Excess arguments are to be passed on the stack.
Likewise, \emph{va\_lists} are implemented as logically contiguous storage on the stack. Each storage element is as big as a machine register~\cite{RiscvelfpsabidocRiscvelfadocMaster}.

Variadic functions are a challenge for capability-based architectures~\cite{chisnallPDP11ArchitecturalSupport2015}: the compiler or run-time needs to ensure that the callee can access all \emph{intended} arguments.
At the same time, care needs to be taken to prevent leaking the private stack to the callee.
We provide a special macro to generate caller trampolines for variadic functions.
The macro generates a wrapper for the function call that uses gcc's \emph{\_\_gnu\_inline\_\_} attribute in conjunction with the builtin \emph{\_\_builtin\_va\_arg\_pack\_len} to determine the number of variadic arguments for each invocation of the function.
The wrapper function then uses the builtin \emph{\_\_builtin\_va\_arg\_pack} to convert the variadic function arguments into a \emph{va\_list}.
The \emph{va\_list} is then forwarded to a second, inner, wrapper function.
This inner wrapper then use the RISC-V ABI specification to our advantage: It knows the number of arguments as well as the base of the \emph{va\_list}.
Hence, it can use the $derive()$ capability operation to \emph{derive a capability that is exactly as large as the va\_list from the stack}.
To this end, remember that each argument in the \emph{va\_list} has as many bits as a machine register, ensuring that all intended arguments are forwarded while the remainder of the stack is inaccessible to the callee\footnote{This logic of course breaks when passing \emph{structures by-value} via variadic functions: in this case, each \emph{struct member} is padded to the machine register size and passed in the \emph{va\_list}. However, while this is technically permitted~\cite{RiscvelfpsabidocRiscvelfadocMaster}, we have not encountered it anywhere in Zephyr.}.
The \emph{va\_list} can then be forwarded like any other capability token and accessed using \emph{va\_arg} and the other typical macros.

Specifically for the \emph{printf} family of functions, there remains a final challenge:
The format specifiers \emph{\%s} and \emph{\%n} in the format string indicate the presence of a \emph{pointer argument}.
When \emph{\%s} is used, the corresponding argument is interpreted as a read-only pointer to a zero-terminated string, which is to be forwarded to the output.
When \emph{\%n} is used, the argument is interpreted as a write-only pointer to an integer, where the library will store the number of bytes that were output so far.
However, passing a capability token for a string in a subsystem's read-only data segment or an int on its stack as a printf argument will cause an \emph{access fault} in the libc, as these segments are subsystem-id protected.
Hence, our printf wrapper functions scan the format string and variadic argument list for such arguments and transparently derive suitable capabilities for these arguments.

\paragraph{Trampolines support the Floating Point Unit}
Zephyr optionally provides support for floating-point unit (FPU) operation.
We use this, e.g., for the computation of standard deviation in our benchmarks.
Hence, we implemented FPU support in the trampolines.
To this end, each subsystem can use a build-time flag to indicate whether it wants to use the FPU:
In case this is not the case, the FPU is ignored in all caller and callee trampolines to reduce the overhead of subsystem calls.
Otherwise, caller trampolines handle the FPU similar to how Zephyr handles it during context switching:
On the caller side, the trampoline uses RISC-V's \emph{FS.dirty} state in the machine status register to determine whether the FPU has been used since it was last saved.
If this is the case, the trampoline saves the FPU registers in the caller register set; otherwise, this is skipped.
The caller trampoline then zeros the FPU registers and proceeds with the call as usual.
After the return, it sets the FPU status to \emph{clean} (implying enabled) and restores the FPU registers from the caller set unconditionally.

The callee trampoline simply sets the FPU to \emph{clean} and zeros the registers with the exception of the floating point return registers on return.
Thereby, the callee trampoline prevents leaking private register information while supporting the floating point calling convention for floating point arguments and return values.

\paragraph{Dynamic allocator needs not be trusted}
The Skadi loader includes a basic allocator that uses $create$ and $derive$ operations.
Its primary purpose is allocating code and data segments for the relocated ELF binaries.
A more sophisticated allocator is implemented as a subsystem; it needs not be trusted.
Both implementations of the allocator use $create$ with word-aligned lengths to slice off direct capabilities from an unused capability.
The allocator subsystem also enforces a minimum length for the direct capabilities to limit fragmentation~\cite{yurchenkoAlgorithmDynamicSegmented1981}.
Both allocators use $derive$ to \emph{limit} the segment length to the number of bytes actually requested, providing memory safety.
Subsystems can use $derive$ or $restrict$ to decrease the length and/or permissions of an allocated capability if needed, and they can use $lock$ to gain exclusive access.
Finally, the Skadi loader creates non-overlapping capabilities for both allocators that serve as their initial heap.

Implementing an allocator that is aware of Northcape capabilities comes with additional challenges:
When a capability token is returned to the allocator using the \emph{free()} method, the allocator needs to first identify the \emph{direct} capability based on the \emph{indirect capability} that it handed out.
Second, it needs to determine whether any of the physically adjacent memory chunks are also freed and can be \emph{merged} to prevent increasing fragmentation of the memory.

We solve these additional challenges using the following algorithm.
The allocator maintains two lists: a \emph{free list} of chunks that can be allocated and an \emph{allocated list} of chunks that are currently in use by clients.
The chunks on the \emph{free list} are direct capabilities that start with a \emph{free list header}.
On the other hand, the allocated list consists of direct capabilities that comprise only the \emph{allocated list header}, which maintains a capability token pointing to the direct capability allocated to the caller.
Thereby, the allocator can \emph{traverse and manipulate the allocated list} while its callers are able to \emph{gain exclusive access to the allocated capability} via the \emph{lock} operation.
The \emph{allocated list header} also conveys the indirect capability token that the allocator handed out.
The \emph{free list header} also conveys bounds and sizes of the capability chunk, saving one \emph{inspect()} operation.
The \emph{allocated list} is implemented as a red-black tree that uses the \emph{indirect capability} that the allocator handed out as a key.
The \emph{free list} consists of \emph{two red-black trees}: one red-black-tree uses the \emph{chunk size}\footnote{We use the chunk base as a secondary key to enforce a strong ordering.} as a key, the second red-black tree uses the \emph{chunk base address} as the key.
We use Zephyr's intrusive implementation of red-black-trees that works without requiring dynamic memory for itself and guarantees $\mathcal{O}(log(n))$ complexity for key searches as well as real-time capability.
An \emph{allocation} operation can use a \emph{search operation} on the red-black-tree to implement a best-fit or first-fit allocation strategy in $\mathcal{O}(log(n))$.
As a side effect, the operation removes the allocated chunk from the free lists, converts the header to the allocated chunk header and inserts it into the allocated list.
A \emph{free} operation first uses a search operation on the allocated list to map the indirect capability to a direct capability and get the bounds information.
It then performs a lookup on the free list sorted by \emph{chunk base}, looking for the physical neighbors of the freed capability.
After that, the allocator linearizes the freed capability by merging the allocated list header chunk with the remainder of the capability.
If a neighbor is found in the free list, it is removed.
The allocator then performs a \emph{merge operation} and inserts one new entry, comprising the freed capability and its neighbor, into the free list.
Otherwise, the linearized freed capability is inserted into the free list.

\paragraph{Memory allocation does not always require a subsystem call}
The dynamic allocator subsystem has one crucial drawback: Allocating and freeing memory necessitates a \emph{subsystem call}, which is more costly than a simple function call.
In order to reduce this overhead, we provide a simple per-subsystem \emph{local allocator}.
The local allocator comprises a continuous memory arena in each subsystem's \emph{.bss} segment and an atomic bitmap.
The chunk is segmented into equal-size \emph{memory chunks}, whose allocation status is tracked by the bitmap.

For each allocation request, the per-subsystem allocator picks the first free chunk and derives a capability for it.
The derived capability has the exact length requested by the caller to provide memory safety.

On return, the per-subsystem allocator can use the $inspect$ operation to retrieve the physical base address of the corresponding chunk.
Using the chunk size and physical base address of the arena, the corresponding chunk can be determined efficiently.

The local allocator is used if three conditions are met:
\begin{itemize}
    \item the size of the memory request is smaller than the chunk size.
    \item the \emph{lockable} permission is \emph{not} requested.
    \item at least one chunk is free.
\end{itemize}
Note that as the memory arena is part of the .bss segment, attempting to lock an allocated chunk would lock the \emph{entire} .bss segment, possibly causing access faults in the subsystem.
All memory requests that are not eligible for the per-subsystem allocator are forwarded to the dynamic allocator subsystem.

The local memory allocator has an additional advantage:
Freed data \emph{need not be zerod out}.
By convention, the \emph{malloc} family of functions returns \emph{uninitialized} data.
Software needs to use a \emph{calloc}-style function to request \emph{memory to be overwritten with zeros}.
Thereby, costly \emph{memset}-operations during allocation can be provided.
However, if the remote allocator is used, this scenario creates a risk of \emph{leaking} information.
Thus, in order to improve security, freed memory is always overwritten before the free subsystem call.
On the other hand, for the local allocator, this need not be done.

\paragraph{The libc needs not be trusted}
Skadi subsystems have access to stateless C library functions including string functions and the \emph{printf} family of functions via the \emph{libc} subsystem.
To this end, the \emph{libc} subsystem comprises a relocated picolibc implementation, the same libc that Zephyr uses by default.
The \emph{libc} subsystem uses subsystem calls to interface with other components when necessary.
For example, it uses subsystem calls provided by the console subsystem for character input/output via \emph{printf}.
The \emph{libc} has its own data segment, but we have configured picolibc such that all processing occurs on the stack, such that data cannot be leaked between subsystems.
Hence, the data segment is only used to store constants.

Subsystems that are not security-critical, such as our shell subsystem, can choose to directly call libc functions.
To this end, they indicate undefined external symbols accordingly, relying on the Skadi loader to relocate the libc functions directly.
Thus, when calling libc functions, the subsystems \emph{leak} their subsystem identifier to the libc but gain a code size and latency advantage.

Security-critical subsystems can invoke C library functions via subsystem calls.
To this end, we statically link a libc stub called libSkadi into each subystem.
The stub declares caller trampolines for the standard C library functions.
The stub then provides implementations for the standard C library functions that invoke the caller trampolines.
Thus, when linking the subsystem, the linker uses the library functions defined in libSkadi and does not emit an undefined symbol.

Finally, if desired, subsystems can \emph{inline} selected C library functions such as string functions.
Thereby, we achive higher performance at the cost of higher code size.
This option can be selected at build time and is enabled by default.

\paragraph{Skadi scheduler}
In Zephyr, the nanokernel provides blocking data structures such as semaphores, queues and pipes.
The API for these data structures contains \emph{cancellation points} such as \emph{k\_sem\_take}:
When a thread calls such a function and no resources are available (e.g., the semaphore's count is 0), the thread \emph{blocks} until resources are available.

Under the hood, calling a cancellation point causes the calling thread to lose its \emph{runnable} status in the scheduler.
The scheduler picks a new runnable thread and performs a \emph{context switch} to continue execution.

Calling functions like \emph{k\_sem\_give} causes waiting thread(s) to re-gain their \emph{runnable} status.
Hence, in a subsequent context switch, they can continue execution.

Crucially, all Zephyr subsystems internally rely on the data structures provided by the scheduler for cancellation.
Hence, we chose to include blocking data structures in the scheduler subsystem and to make them accessible with \emph{subsystem calls}.
An alternative design can additionally move the \emph{core scheduling algorithm} into a different subsystem.
Thereby, integrity invariants for the data structures cannot be broken by a malicious scheduling strategy.

Blocking data structures' internal implementation contains data structures such as linked lists.
Hence, if, say, a \emph{struct k\_sem} was shared between a scheduler and a malicious component, the malicious component could manipulate the linked list to essentially gain a write primitive in the scheduler's sphere of protection, breaking compartmentalization.
Thus, the scheduler API \emph{maps} provided capability tokens to a dynamically allocated \emph{shadow copy} of the provided object.
This design resembles how \emph{file descriptors} in Linux or \emph{handles} in Windows grant access to kernel objects without revealing the object directly.
At the same time, we maintain full API compatibility with Zephyr subsystems.

\paragraph{Skadi subsystem initialization}
We provide two different means for calling initialization functions:
First, there is a new macro \emph{SKADI\_SUBSYSTEM\_INITIALIZATION\_FUNCTIONS}.
It designates a function to be run immediately after the subsystem was loaded, before permissions and subsystem-id protections are set.
This is used, e.g., by the Skadi trampolines to initialize private data that remains read-only for the remainder of the subsystem's life.

We have also ported Zephyr's \emph{SYS\_INIT} macro to Skadi.
This mechanism allows the user to designate a function to be run in a run level like \emph{EARLY} or \emph{POST\_KERNEL} and with a relative \emph{priority} in relation to other initialization functions in the run level.
To this end, for each run level, the Skadi loader enquires the next priority from each subsystem via a well-known callee trampoline.
It then determines which subsystem has the lowest priority and indicates via a second callee trampoline that the next initialization function should be run.
The subsystems use the same iterable section mechanism as Zephyr to find their own initialization functions.
The code responsible for returning the next priority and invoking the initialization function is contained in the statically-linked \emph{libSkadi}.

Initialization of the Skadi operating system itself is split into a part executed in the loader and a part executed in the scheduler.
The loader's initialization routines cover both the platform initialization, comprising subsystems such as zeroing the .bss segment and setting up an initial stack, as well as Zephyr's \emph{EARLY}, \emph{PRE\_KERNEL\_1} and \emph{PRE\_KERNEL\_2} initialization runlevels.
In other words, the loader is responsible for invoking initialization functions before the scheduler becomes active.
The early initialization functions comprise subsystems like setting up the initial allocator and early console as well as relocating the subsystems.
After completing \emph{PRE\_KERNEL\_2}, the loader invokes the scheduler via subsystem call.
The scheduler proceeds to create the main thread and perform its own initialization.
The scheduler can then invoke the initialization functions contained in higher runlevels like \emph{POST\_KERNEL} and \emph{APPLICATION}.
To this end, it invokes a subsystem call exposed by the loader via the \emph{loader\_proxy} subsystem.
After completing the last runlevel, the scheduler performs a subsystem call into the \emph{main()} function.

\paragraph{Skadi loader ceases to exist after initialization}
As we have laid out in~\cref{sec:security-model}, the Skadi loader is the only trusted software component.
It also is the only component that runs with subsystem id 0, giving only the loader the ability to create set-subsystem-id tokens with foreign subsystem identifiers.
This is crucial for relocating and bootstrapping the system, but there is no need for this facility after initialization is complete\footnote{A future version of Skadi might keep a part of the loader active to allow \emph{dynamic} loading of subsystems at run time.}.
In order to prevent the loader from becoming a security risk, it is destroyed after initialization.

The \emph{POST\_KERNEL} and higher initialization runlevels are triggered via subsystem call from the scheduler, which proceeds to call the \emph{main()} method after the last initialitzation runlevel was completed.
Thus, when the loader completes the last initialization runlevel, it can destroy itself and return to the scheduler.
This is facilitated with the help of the \emph{trapdoor} subsystem.

The trapdoor subsystem runs with subsystem id 0, allowing it to perform operations on the root capability.
Trapdoor is invoked via a subsystem call from the loader as soon as the last initialization runlevel is complete, receiving the original return address the scheduler subsystem provided as a parameter.
Trapdoor uses the $create$ Northcape operation on the root capability, slicing off a new capability that comprises the memory previously occupied by the loader.
Trapdoor then proceeds to use $revoke$ on the new capability to ensure the memory is overwritten with zeros and no secrets are leaked.
The new capability is then provided to the allocator subsystem, allowing it to be used as general-purpose heap.
For the same reason, the remaining heap from the initial allocator is forwarded to the allocator subsystem before trapdoor is invoked.
As a penultimate step, trapdoor uses $restrict$ to remove all permissions from the remainder of the root capability.
Thereby, the root capability cannot be used directly any more, and one cannot derive any new capabilities from it.
However, MMIO capabilities that were derived previously continue to provide device access.
As a final step, trapdoor clears all registers and returns to the scheduler subsystem via the provided subsystem call capability, changing the subsystem id in the process.

As the loader ensures that no subsystem can access trapdoor's subsystem call, and the loader ensures that no other subsystem executes at subsystem id 0, there is no way for executing code at subsystem id 0 going forward.
Thus, in addition to destroying the loader, the privileged subsystem id 0 is also inactive.
After trapdoor's return, Skadi exclusively executes the loaded subsystems.

\subsection{Custom ISA extensions for Skadi}
Skadi is functional and secure using just the standard RISC-V API and Bredi's MMIO register interface.
However, we introduce a significant loss of performance in three areas:
\begin{enumerate}
    \item Checking the validity of return addresses in subsytem calls via $inspect$
    \item zeroing out registers on subsystem call return
    \item accessing the MMIO register interface with loads and stores
\end{enumerate}

We introduce three ISA extensions to improve performance: a custom \emph{subsystem call instruction}, a custom \emph{register uninitialize instruction} and \emph{custom control-status registers (CSRs)} for accessing the operations module.

\subsubsection{Subsystem call instruction}
As discussed in \cref{sec:subsystem-calls}, both the caller and callee in a subsystem call need to verify that the capability they are about to jump into has a \emph{set\_subsystem\_id} restriction.
If this was not done, a jump could result in loss of control over one's subsystem ID, breaking compartmentalization.
This check can be done using the $inspect$ operation. While this provides adequate security, this costs a significant number of CPU cycles, especially for short subsystem calls.

Hence, our specialized subsystem call instruction performs this check \emph{as the jump commits}.
To this end, the subsystem call instruction causes an \emph{instruction access exception} for the first instruction to execute after the jump if it \emph{was not loaded from a capability that is a suitable subsystem call jump target}.
The implementation of this feature requires few changes to our hardware architecture:
First, our cva6 MMU tags each fetched instruction with a flag that indicates if this instruction is a suitable subsystem call target according to the security requirements detailed in \cref{sec:subsystem-calls}.
Second, we use a currently reserved function code in the encoding space of the jump-and-link-register (\emph{jalr}) instruction to indicate that this jump belongs to a subsystem call.
The jump instruction behaves like an ordinary \emph{jalr}, but when it commits, it sets a register in the CPU's commit stage, indicating that the next instruction needs to be a valid subsystem call target.
Finally, the commit stage triggers an exception when the instruction to which the subsystem call resolves to was \emph{not} a valid subsystem target.

This approach provides adequate security: As the commit stage triggers an exception in case of an invalid subsystem call, the invoked instruction never commits and the adversary cannot accomplish anything.
At the same time, the security check does not cost any cycles of its own. Also, this approach eliminates security checks for function points with $inspect$ that would otherwise be needed as well.

\subsubsection{Fast Register Zero Instruction}

In order to prevent information leakage between caller and callee, both caller and callee clear all registers besides function / return address and argument/return value registers.
Given RISC-V's large register set (32 integer and 32 floating point registers), this takes a considerable amount of cycles, especially for short subsystem calls.

In order to reduce this overhead, we introduce \emph{uninitialized registers} in cva6:
Using a custom instructions, an \emph{initialization mask} can be loaded, selectively marking registers as \emph{uninitialized}.
Unitinialized registers always read 0, irregardless of their value at the time of being uninitialized.
When an uninitialized register is written, it becomes initialized and reads back the original value.

Our implementation stores the initialization mask as a bitmap alongside the register file.
Every register read is zerod out when the according register is uninitialized.
Each register write clears the corresponding bit from the bitmask, making the register readable again.
Finally, the mask contains both integer and floating-point registers and is written at once from a source register.
Naturally, the mask is duplicated for the ISR and non-ISR register set.

Overall, this optimization reduces the worst-case time for clearing the register set from appx. 120 instructions to 4 instructions.

\subsubsection{Bredi Capability Operations Module Access via CSR}
Reading and writing the Bredi capability operations module via load and store instruction requires fence instructions to ensure access ordering, causing performance degradation.
Also, reads and writes need to traverse the cache subsystem and interconnect, costing multiple cycles.

In order to accelerate the access to the operations module, we implement custom control and status registers (CSRs).
Each CSR maps to a register in the MMIO interface.
However, in contrast to the MMIO register set, the CSRs are \emph{not duplicated}.
Instead, whether an instruction is executed in IRQ context decides which copy of the register interface is to be used.

We use a static arbiter to map the operations module between the AXI and CSR interfaces, prioritizing the CSR interface, as the CPU is the main user of this interface.

\subsection{Skadi Zero-Copy Networking}
By utilizing the Bredi capability system, Skadi manages to implement zero-copy networking, increasing performance and safety.
To this end, it eliminates two packet copy operations from Zephyr's network stack:
Zephyr copies each incoming packet from the DMA buffer into a \emph{struct net\_pkt} (or vice versa on TX).
For applications that use the sockets API, each packet is copied a second time into or out of the user-provided buffer.
Crucially, the \emph{struct net\_pkt} structure comprises one or more \emph{struct net\_buf} that hold (pointers to the) actual data of the packets as well as size information.
Thereby, network packets need not be physically contiguous. This is used to, e.g., add or remove network headers in separate \emph{struct net\_buf} in case there is no headroom in the existing buffer.

The first copy is eliminated by providing the DMA device with a \emph{indirect capability} derived from the buffer(s) in the \emph{struct net\_pkt}.
To this end, we use zehpyr's API for non-contiguous DMA to ensure network packets need not be linearized to be eligible for zero-copy DMA.
This is natively supported by DMA devices with scatter-gather capability, such as our Xilinx AXI DMA device~\cite{advancedmicrodevicesinc.AXIDMALogiCORE2022}.
After the DMA device indicates completion of a transfer, the indirect capability is \emph{dropped}, ensuring that the device loses access to the buffer.

The second copy is eliminated by modifying the sockets API for both datagram- and stream-sockets.
For transmission, we provide a custom setsockopt operation that clients can use to indicate that they would like to use zero-copy networking.
In this case, the client is expected to provide an indirect capability for the data it wishes to transmit.
The indirect capability is then wrapped into a \emph{struct net\_buf} and conveyed through the network stack.
To this end, we indicate a headroom of 0 such that additional network headers will be prepended in a new \emph{struct net\_buf}.
Thereby, the caller can give a \emph{read-only} capability to the network stack.
The caller can also register a \emph{deallocation callback} to be called after transmission of the packet, allowing it to drop the capability if desired.
This is also useful to wait for completed transmission of a packet, e.g., in order to know when TX timestamps can be queried.

On the RX side, we provide a custom \emph{ioctl}-based API to get access to network data.
Subsystems can invoke the \emph{ioctl} to receive a \emph{indirect capability} for the application data of the network packet currently on the head of the receive queue for the socket.
The packet is then moved to a special wait queue.
The caller can then process the application data and \emph{release} the packet via a different \emph{ioctl}, allowing the network stack to deallocate the packet and free the indirect capability it was given.

\subsection{Skadi Development}
We have added support for the development and testing of Skadi subsystems to the Zephyr toolchain.
Any set of source files can be declared as a Skadi subsystem using the \emph{create\_skadi\_subsystem} CMake function.
The function takes care of compiling the source files with the correct code model, architecture and include path flags and linking them into a relocatable ELF binary.
The function also adds the new extension to the set of loaded Skadi extensions.
As discussed earlier, during the build process, we compute an initialization order for the extensions and include them into the binary.

New Skadi subsystems can include header files which declare caller trampolines in order to import subsystem calls into other subsystems.
They can also export subsystem calls by providing subsystem callee trampolines as exported ELF symbols.

Finally, Skadi subsystems can be debugged using Zephyr's common wrapper commands (e.g., \emph{west debug}).
To this end, we utilize the debug module provided alongside cva6, which implements the RISC-V debug specification~\cite{newsomeRISCVExternalDebug2024}.
We will give a quick overview over how the debug module works before we explain how we incorporated it into Skadi and Bredi.

The debug module is a hardware component that is separate from the CPU. It is connected as an Axi Slave to the northbridge. 
It also has its own bus master interface, which it can use to access memory directly.
It can also invoke a debug interrupt on the CPU.
Finally, the debug module provides a JTAG interface that can be connected to a host system.

With the exception of the debug interrupt, all communication between CPU and debug module occurs in the form of memory accesses.
To this end, as soon as the debug interrupt is raised or a breakpoint is hit, the CPU sets its program counter to the program buffer MMIO register in the debug interface.
The default instructions in the program counter cause the CPU to spin.
Using JTAG, an external debugger can give the debug interface instructions such as inspecting register contents or reading memory~\cite{newsomeRISCVExternalDebug2024}.

We have added a custom CSR to the CPU that makes the MMIO base address of the debug module configurable.
Thereby, we can derive a capability token for the debug module before loading submodules.
By \emph{not} subsystem-restricting the capability, the debug module can be used \emph{irregardless of the active subsystem}.
Thus, we can protect the root capability and all subsystem code and data with subsystem-id restrictions without affecting the functionality of the debug module.
In case the debug module is not to be used, it can simply be deactivated in the device tree - this causes our loader to \emph{not} create a capability for it, rendering it inaccessible when the loader is not active.

Zephyr provides an integration of RISC-V debug modules with the \emph{west} build wrapper, using openocd as primary debugger.
Openocd can be used to load and invoke a Zephyr binary. It can also serve as a GDB server, allowing source-code level debugging with gdb.

Bredi can use the debug module without major modification.
We configure openocd to access the memory via CPU instructions and leave the debug module's AXI master interface unconnected.
Thereby, gdb can read and write capability tokens in the same way as it would access physical addresses.
This facilitates setting breakpoints, single-stepping and printing values within the context of Skadi subsystems.
As the debugger accesses registers and memory via the CPU, all memory requests triggered by it automatically are subject to Bredi token translation within the current subsystem context.
At the same time, the debug module cannot access memory without going through Bredi as its master interface is not connected.

We even provide symbol support for Skadi subsystems. To this end, the Skadi loader prints a GDB command for loading debug symbols from each subsystem's ELF binary at their final position after relocation.
Hence, programs can be debugged in the same way as they would on a legacy operating system without Bredi support, and no special support in hardware or software for debugging Skadi subsystems is needed.

\fi

\ifarxiv
\section{Detailed Evaluation Results}
\fi
\label{sec:extra-ops}

\ifarxiv
In order to facilitate transparency, we provide additional measurement values for our evaluation experiments.
\fi
\Cref{tab:compute_mem_benchmarks_full} and \Cref{tab:macrobenchmark_full} list detailed values for our compute and scheduler benchmark results.
\Cref{tab:real_time_performance_full} contains detailed results from our IRQ benchmarks.
In addition to Zephyr, Linux and Skadi on Bredi, we also include the upstream cva6 SoC as a reference.
\Cref{tab:system_network_performance_full} provides detailed results of our network benchmarks.
Due to the upstream cva6 SoC using a different networking subsystem, this reference is excluded here.
\balance

\ifarxiv
\else

\fi

\label{sec:extra-eval}

\begin{table*}[htbp]
    \centering
    \setlength{\aboverulesep}{0pt}
    \setlength{\belowrulesep}{0pt}
    \footnotesize
    \caption{Full results from compute benchmark (Coremark) and memory benchmark (Stream) for Skadi, \hlgray{Zephyr} and \hlgray{Linux}. Also includes \hlgray{Zephyr on the upstream cva6 FPGA SoC}~\cite{zarubaCostApplicationClassProcessing2019}. For Coremark, we provide the number of iterations, average duration and standard deviation from five runs as well as the standard metric, which is the median of the average numbers of iterations per second per run. For Stream, we provide results as reported.}
    \label{tab:compute_mem_benchmarks_full}
\ifarxiv
    \begin{tabular}{p{1.8cm}|>{\columncolor{gray!25}}r|>{\columncolor{gray!25}}r|>{\columncolor{gray!25}}r|r}
\else
    \begin{tabular}{p{3cm}|>{\columncolor{gray!25}}r|>{\columncolor{gray!25}}r|>{\columncolor{gray!25}}r|r}
\fi
    \textbf{Benchmark} & \textbf{Zephyr cva6 SoC} & \textbf{Linux Bredi} & \textbf{Zephyr Bredi} & \textbf{Skadi Bredi}\\
    \toprule
    Coremark Iterations & 2000 & 1100 & 2000 & 2000\\
    \hline
    Median Coremark Iterations/sec & \num{110.004950} & \num{92.1003055} & \num{110.017053} & \num{110.144289}\\
    \hline
    Coremark Duration avg / min / max / stddev/s & \num{18.181000} / \num{18.181000} / \num{18.181000} / \num{0} & \num{11.946} / \num{11.928000} / \num{11.960000} / \num{0.0099107125} & \num{18.179000} / \num{18.179000} / \num{18.179000} / \num{0} & \num{18.1576} / \num{18.157000} / \num{18.158000} / \num{0.00051639778}\\
    \hline
    Stream Copy Best Rate/ MB/s & \num{29.6} & \num{14.3} & \num{26.7} & \num{26.7} \\
    \hline
    Stream Copy avg / min / max / stddev/s & \num{5.411111} / \num{5.400000} / \num{5.500000} / \num{0.031427} & \num{11.226838} / \num{11.223349} / \num{11.234028}/- & \num{6.044444} / \num{6.000000} / \num{6.100000} / \num{0.049690} & \num{5.999885} / \num{5.999885} / \num{5.999945} / \num{0.000015} \\
    \hline
    Stream Scale Best Rate/ MB/s & \num{25.4} & \num{12.6} & \num{24.2} &  \num{24.1}\\
    \hline
    Stream Scale avg / min / max / stddev/s &    \num{6.344444} / \num{6.300000} / \num{6.400000} / \num{0.049690} & \num{12.753133} / \num{12.748885} / \num{12.759804}/- & \num{6.655556} / \num{6.600000} / \num{6.700000} / \num{0.049690} & \num{6.646107} / \num{6.646100} / \num{6.646116} / \num{0.000005}\\
    \hline
    Stream Add Best Rate/ MB/s & \num{27.6} & \num{13.7} &\num{26.7} & \num{26.6}\\
    \hline
    Stream Add avg / min / max / stddev/s & \num{8.733333} / \num{8.700000} / \num{8.800000} / \num{0.047140} & \num{17.534844} / \num{17.522104} / \num{17.595224}/- & \num{9.033333} / \num{9.000000} / \num{9.100000} / \num{0.047140}&  \num{9.015753} / \num{9.015753} / \num{9.015753} / \num{0.000000}\\
    \hline
    Stream Triad Best Rate/ MB/s &  \num{27.6} & \num{12.9} & \num{25.3} & \num{25.3}\\
    \hline
    Stream Triad avg / min / max / stddev/s & \num{8.766667} / \num{8.700000} / \num{8.800000} / \num{0.047140} & \num{18.619079} / \num{18.999640} / \num{18.630212}/-& \num{9.500000} / \num{9.500000} / \num{9.500000} / \num{0} &\num{9.500033} / \num{9.500031} / \num{9.500034} / \num{0.000001} \\
    \bottomrule
    \end{tabular}
\end{table*}

\begin{table*}[htbp]
    \centering
    \setlength{\aboverulesep}{0pt}
    \setlength{\belowrulesep}{0pt}
    \footnotesize
    \caption{Full results from networking benchmark for Skadi, \hlgray{Zephyr} and \hlgray{Linux}. }
    \label{tab:system_network_performance_full}
    \begin{tabular}{p{4cm}|>{\columncolor{gray!25}}p{4cm}|>{\columncolor{gray!25}}p{4cm}|p{4cm}}
    \textbf{Benchmark} & \textbf{Linux} & \textbf{Zephyr} & \textbf{Skadi}\\
    \toprule
    Ping latency min / avg / max / stddev ms & \num{1.767} / \num{1.916} / \num{2.623} / \num{0.152} & \num{0.834} / \num{0.876} / \num{1.080} / \num{0.021}& \num{2.464} / \num{2.555} / \num{3.964} / \num{0.075}\\
    \hline
    iperf min / max / avg / stddev throughput TCP in Kbits/s & \num{7580} / \num{14300} / \num{9031.1667} / \num{1733.1954} &\num{4380} / \num{4410} / \num{4396.6667} / \num{0006.8064443} & \num{1490} / \num{1500} / \num{1499.3333} / \num{002.5154887} \\
    \hline
    RX network stack processing durations ns min / max / avg / stddev & \num{1212720} / \num{1721960} / \num{1346524.210526} / \num{136558.719135} &\num{516193} / \num{564497} / \num{534142.810526} / \num{8120.032767} & \num{2111305} / \num{2180633} / \num{2138493.631579} / \num{13707.449448}\\
    \hline
    TX network stack processing durations ns min / max / avg / stddev & \num{765200} / \num{1827240} / \num{1050532.631579} / \num{203162.692163} &\num{514855} / \num{644431} / \num{530697.96} / \num{13202.113} & \num{1491231} / \num{1559519} / \num{1516145.021053} / \num{11670.623603}\\
    \hline
    MQTT-TLS Iteration duration ns min / max / avg / stddev & - & \num{1993429040} / \num{1993582320} / \num{1993484845.473684} / \num{30590.976095} & \num{2049993880} / \num{2065165720} / \num{2057174000.842105} / \num{5772516.868643}\\
    \bottomrule
    \end{tabular}

\end{table*}

\begin{table*}[htbp]
    \centering
    \setlength{\aboverulesep}{0pt}
    \setlength{\belowrulesep}{0pt}
    \footnotesize
    \caption{Results from real-time performance benchmarks for Skadi, \hlgray{Zephyr} and \hlgray{Linux}. Also includes \hlgray{Zephyr on the upstream cva6 FPGA SoC}~\cite{zarubaCostApplicationClassProcessing2019}.}
    \label{tab:real_time_performance_full}
    \begin{tabular}{p{3cm}|>{\columncolor{gray!25}}p{3cm}|>{\columncolor{gray!25}}p{3cm}|>{\columncolor{gray!25}}p{3cm}|p{3cm}}
    \textbf{Benchmark} & \textbf{Zephyr cva6 SoC} & \textbf{Linux Bredi} & \textbf{Zephyr Bredi} & \textbf{Skadi Bredi}\\
    \toprule
    IRQ Latency independent min / max / avg / stddev ns: & \num{4820} / \num{88140} / \num{12050.736842} / \num{21226.219697} & \num{1324020} / \num{9236700} / \num{3808793.684211} / \num{3194695.585604}  & \num{4940}  /  \num{89960}  /  \num{24127.157895}  /  \num{32149.031897} & \num{65720}  /  \num{72840}  /  \num{68423.157895}  /  \num{1529.023950} \\
    \hline
    IRQ Latency monotonic min period / min / max / avg / stddev ns: & \num{68000}  /  \num{4740}  /  \num{62820}  /  \num{6061.684211}  /  \num{7212.839233} & \num{1312640} / \num{16432080} / \num{6866233.684211} / \num{3989003.354999} / \num{5691090.280056}  & \num{70000}  /  \num{4860}  /  \num{69740}  /  \num{8392.421053}  /  \num{12120.038276} & \num{136000}  / \num{41300}  /  \num{114860}  /  \num{58826.315789}  /  \num{18714.180441} \\
    \bottomrule
    \end{tabular}
\end{table*}

\onecolumn
\topcaption{Detailed results from scheduler macrobenchmark for Skadi and \hlgray{Zephyr}. Also includes \hlgray{Zephyr on the upstream cva6 FPGA SoC}~\cite{zarubaCostApplicationClassProcessing2019}. Names are as used in original Zephyr benchmark.}
\label{tab:macrobenchmark_full}
\setlength{\aboverulesep}{0pt}
\setlength{\belowrulesep}{0pt}
\footnotesize
\begin{supertabular}{p{4cm}|>{\columncolor{gray!25}}p{4cm}|>{\columncolor{gray!25}}p{4cm}|p{4cm}}

\textbf{Macrobechmark} & \textbf{Zephyr cva6 min / max / avg / stddev ns} & \textbf{Zephyr Bredi min / max / avg / stddev ns} & \textbf{Skadi Bredi min / max / avg / stddev ns}\\
	\toprule
    	thread.\allowbreak yield.\allowbreak preemptive.\allowbreak ctx  & \num{4000 } / \num{ 4960 } / \num{ 4003.698492 } / \num{ 47.574217} & \num{4040 } / \num{ 4880 } / \num{ 4067.055276 } / \num{ 64.082798} & \num{13600 } / \num{ 15840 } / \num{ 13719.035176 } / \num{ 275.047139}\\
	\hline
	thread.\allowbreak yield.\allowbreak cooperative.\allowbreak ctx  & \num{4000 } / \num{ 4960 } / \num{ 4005.507538 } / \num{ 56.910568} & \num{4040 } / \num{ 4600 } / \num{ 4061.949749 } / \num{ 36.629580} & \num{13600 } / \num{ 16560 } / \num{ 13715.939698 } / \num{ 295.707985}\\
	\hline
	isr.\allowbreak resume.\allowbreak interrupted.\allowbreak thread.\allowbreak kernel     - Return from ISR to interrupted thread  & \num{5080 } / \num{ 7080 } / \num{ 5934.753769 } / \num{ 123.997952} & \num{5520 } / \num{ 5840 } / \num{ 5625.165829 } / \num{ 122.474683} & \num{176800 } / \num{ 198160 } / \num{ 187828.020101 } / \num{ 3154.815861}\\
	\hline
	isr.\allowbreak resume.\allowbreak different.\allowbreak thread.\allowbreak kernel       - Return from ISR to another thread  & \num{4800 } / \num{ 5080 } / \num{ 4839.919598 } / \num{ 60.177239} & \num{5080 } / \num{ 5400 } / \num{ 5102.753769 } / \num{ 51.759539} & \num{168520 } / \num{ 188720 } / \num{ 179111.035176 } / \num{ 3097.115668}\\
	\hline
	thread.\allowbreak create.\allowbreak kernel.\allowbreak from.\allowbreak kernel         - Create thread  & \num{1727320 } / \num{ 1805440 } / \num{ 1765850.251256 } / \num{ 11137.177842} & \num{2173200 } / \num{ 2266240 } / \num{ 2220408.442211 } / \num{ 13927.671061} & \num{2594040 } / \num{ 2692720 } / \num{ 2642536.442211 } / \num{ 15995.252104}\\
	\hline
	thread.\allowbreak start.\allowbreak kernel.\allowbreak from.\allowbreak kernel          - Start thread  & \num{8800 } / \num{ 15640 } / \num{ 12421.145729 } / \num{ 954.297918} & \num{9240 } / \num{ 15280 } / \num{ 12168.321608 } / \num{ 900.142198} & \num{87040 } / \num{ 103040 } / \num{ 95272.402010 } / \num{ 2539.697477}\\
	\hline
	thread.\allowbreak suspend.\allowbreak kernel.\allowbreak from.\allowbreak kernel        - Suspend thread  & \num{6520 } / \num{ 9520 } / \num{ 8426.090452 } / \num{ 488.194182} & \num{6000 } / \num{ 9120 } / \num{ 8254.190955 } / \num{ 420.369332} & \num{55480 } / \num{ 68040 } / \num{ 61603.738693 } / \num{ 1968.746481}\\
	\hline
	thread.\allowbreak resume.\allowbreak kernel.\allowbreak from.\allowbreak kernel         - Resume thread  & \num{4760 } / \num{ 6720 } / \num{ 5746.773869 } / \num{ 301.319617} & \num{4840 } / \num{ 6440 } / \num{ 5699.658291 } / \num{ 259.405358} & \num{52120 } / \num{ 64800 } / \num{ 57205.869347 } / \num{ 1813.721271}\\
	\hline
	thread.\allowbreak abort.\allowbreak kernel.\allowbreak from.\allowbreak kernel          - Abort thread  & \num{6480 } / \num{ 10480 } / \num{ 9197.829146 } / \num{ 585.672925} & \num{6520 } / \num{ 10440 } / \num{ 8873.567839 } / \num{ 576.065702} & \num{1745440 } / \num{ 1821880 } / \num{ 1784033.929648 } / \num{ 13360.655723}\\
	\hline
	fifo.\allowbreak put.\allowbreak immediate.\allowbreak kernel                - Add data to FIFO (no ctx switch)  & \num{3680 } / \num{ 4280 } / \num{ 3683.497487 } / \num{ 44.930194} & \num{4160 } / \num{ 4680 } / \num{ 4189.507538 } / \num{ 44.120196} & \num{32520 } / \num{ 35320 } / \num{ 32751.597990 } / \num{ 379.335784}\\
	\hline
	fifo.\allowbreak get.\allowbreak immediate.\allowbreak kernel                - Get data from FIFO (no ctx switch)  & \num{2200 } / \num{ 2240 } / \num{ 2220.180905 } / \num{ 19.999182} & \num{2280 } / \num{ 2800 } / \num{ 2280.683417 } / \num{ 17.234279} & \num{26400 } / \num{ 28000 } / \num{ 26517.025126 } / \num{ 222.975750}\\
	\hline
	fifo.\allowbreak put.\allowbreak alloc.\allowbreak immediate.\allowbreak kernel          - Allocate to add data to FIFO (no ctx switch)  & \num{11800 } / \num{ 12880 } / \num{ 11858.653266 } / \num{ 79.532525} & \num{11760 } / \num{ 12800 } / \num{ 11814.753769 } / \num{ 84.588360} & \num{32000 } / \num{ 34440 } / \num{ 32446.150754 } / \num{ 307.576615}\\
	\hline
	fifo.\allowbreak get.\allowbreak free.\allowbreak immediate.\allowbreak kernel           - Free when getting data from FIFO (no ctx switch)  & \num{8480 } / \num{ 9080 } / \num{ 8511.236181 } / \num{ 60.180461} & \num{8360 } / \num{ 8880 } / \num{ 8387.899497 } / \num{ 55.098906} & \num{26280 } / \num{ 29560 } / \num{ 26362.452261 } / \num{ 217.265825}\\
	\hline
	fifo.\allowbreak get.\allowbreak blocking.\allowbreak k\_to\_k                 - Get data from FIFO (w/ ctx switch)  & \num{5600 } / \num{ 6440 } / \num{ 5655.959799 } / \num{ 80.326960} & \num{5560 } / \num{ 6200 } / \num{ 5601.608040 } / \num{ 78.276839} & \num{18400 } / \num{ 21880 } / \num{ 19562.412060 } / \num{ 476.692764}\\
	\hline
	fifo.\allowbreak put.\allowbreak wake+ctx.\allowbreak k\_to\_k                 - Add data to FIFO (w/ ctx switch)  & \num{8720 } / \num{ 9320 } / \num{ 8751.718593 } / \num{ 58.598036} & \num{8720 } / \num{ 9760 } / \num{ 8747.899497 } / \num{ 62.044804} & \num{23600 } / \num{ 25640 } / \num{ 23807.115578 } / \num{ 266.447354}\\
	\hline
	fifo.\allowbreak get.\allowbreak free.\allowbreak blocking.\allowbreak k\_to\_k            - Free when getting data from FIFO (w/ ctx siwtch)  & \num{5720 } / \num{ 6680 } / \num{ 5831.195980 } / \num{ 129.821202} & \num{5360 } / \num{ 6000 } / \num{ 5368.361809 } / \num{ 62.285045} & \num{18800 } / \num{ 21640 } / \num{ 19073.688442 } / \num{ 457.809031}\\
	\hline
	fifo.\allowbreak put.\allowbreak alloc.\allowbreak wake+ctx.\allowbreak k\_to\_k           - Allocate to add data to FIFO (w/ ctx switch)  & \num{8880 } / \num{ 9480 } / \num{ 8928.884422 } / \num{ 53.016179} & \num{8840 } / \num{ 9360 } / \num{ 8858.371859 } / \num{ 59.924261} & \num{23400 } / \num{ 25720 } / \num{ 23683.899497 } / \num{ 280.993338}\\
	\hline
	lifo.\allowbreak put.\allowbreak immediate.\allowbreak kernel                - Add data to LIFO (no ctx switch)  & \num{3800 } / \num{ 4400 } / \num{ 3832.000000 } / \num{ 57.038469} & \num{3800 } / \num{ 4360 } / \num{ 3827.899497 } / \num{ 48.252997} & \num{31720 } / \num{ 33880 } / \num{ 31930.090452 } / \num{ 337.857708}\\
	\hline
	lifo.\allowbreak get.\allowbreak immediate.\allowbreak kernel                - Get data from LIFO (no ctx switch)  & \num{2040 } / \num{ 2160 } / \num{ 2059.939698 } / \num{ 20.239679} & \num{2120 } / \num{ 2640 } / \num{ 2121.045226 } / \num{ 23.290021} & \num{26520 } / \num{ 28240 } / \num{ 26596.542714 } / \num{ 189.741180}\\
	\hline
	lifo.\allowbreak put.\allowbreak alloc.\allowbreak immediate.\allowbreak kernel          - Allocate to add data to LIFO (no ctx switch)  & \num{11880 } / \num{ 12760 } / \num{ 11935.517588 } / \num{ 90.330540} & \num{11760 } / \num{ 12520 } / \num{ 11800.402010 } / \num{ 70.762105} & \num{31760 } / \num{ 34000 } / \num{ 31972.824121 } / \num{ 330.872403}\\
	\hline
	lifo.\allowbreak get.\allowbreak free.\allowbreak immediate.\allowbreak kernel           - Free when getting data from LIFO (no ctx switch)  & \num{8440 } / \num{ 9080 } / \num{ 8497.648241 } / \num{ 52.855134} & \num{7640 } / \num{ 8240 } / \num{ 7646.592965 } / \num{ 40.079696} & \num{26240 } / \num{ 28560 } / \num{ 26335.718593 } / \num{ 211.589251}\\
	\hline
	lifo.\allowbreak get.\allowbreak blocking.\allowbreak k\_to\_k                 - Get data from LIFO (w/ ctx switch)  & \num{5400 } / \num{ 6400 } / \num{ 5440.361809 } / \num{ 109.400696} & \num{5800 } / \num{ 6720 } / \num{ 5845.507538 } / \num{ 99.025496} & \num{18680 } / \num{ 21800 } / \num{ 19296.080402 } / \num{ 495.552393}\\
	\hline
	lifo.\allowbreak put.\allowbreak wake+ctx.\allowbreak k\_to\_k                 - Add data to LIFO (w/ ctx switch)  & \num{8240 } / \num{ 8840 } / \num{ 8260.261307 } / \num{ 46.742447} & \num{8240 } / \num{ 8760 } / \num{ 8267.698492 } / \num{ 48.318765} & \num{23360 } / \num{ 25120 } / \num{ 23523.618090 } / \num{ 262.686115}\\
	\hline
	lifo.\allowbreak get.\allowbreak free.\allowbreak blocking.\allowbreak k\_to\_k            - Free when getting data from LIFO (w/ ctx switch)  & \num{5080 } / \num{ 5720 } / \num{ 5084.984925 } / \num{ 48.462541} & \num{5120 } / \num{ 6280 } / \num{ 5371.015075 } / \num{ 250.415684} & \num{18280 } / \num{ 20920 } / \num{ 18530.251256 } / \num{ 421.763035}\\
	\hline
	lifo.\allowbreak put.\allowbreak alloc.\allowbreak wake+ctx.\allowbreak k\_to\_k           - Allocate to add data to LIFO (w/ ctx siwtch)  & \num{8040 } / \num{ 8640 } / \num{ 8043.618090 } / \num{ 44.071377} & \num{8480 } / \num{ 9040 } / \num{ 8485.467337 } / \num{ 36.255066} & \num{22960 } / \num{ 25360 } / \num{ 23072.884422 } / \num{ 247.258784}\\
	\hline
	events.\allowbreak post.\allowbreak immediate.\allowbreak kernel             - Post events (nothing wakes)  & \num{4320 } / \num{ 5280 } / \num{ 4373.226131 } / \num{ 62.513067} & \num{4280 } / \num{ 4600 } / \num{ 4317.587940 } / \num{ 38.720893} & \num{17720 } / \num{ 18880 } / \num{ 17798.150754 } / \num{ 181.437580}\\
	\hline
	events.\allowbreak set.\allowbreak immediate.\allowbreak kernel              - Set events (nothing wakes)  & \num{3280 } / \num{ 3880 } / \num{ 3300.703518 } / \num{ 44.697841} & \num{3360 } / \num{ 3920 } / \num{ 3363.979899 } / \num{ 44.349453} & \num{17720 } / \num{ 19440 } / \num{ 17806.311558 } / \num{ 196.353602}\\
	\hline
	events.\allowbreak wait.\allowbreak immediate.\allowbreak kernel             - Wait for any events (no ctx switch)  & \num{2480 } / \num{ 2680 } / \num{ 2497.447236 } / \num{ 20.631146} & \num{2560 } / \num{ 3040 } / \num{ 2560.482412 } / \num{ 15.209374} & \num{16560 } / \num{ 17960 } / \num{ 16655.115578 } / \num{ 188.940079}\\
	\hline
	events.\allowbreak wait\_all.\allowbreak immediate.\allowbreak kernel         - Wait for all events (no ctx switch)  & \num{2080 } / \num{ 2680 } / \num{ 2081.246231 } / \num{ 26.901172} & \num{2120 } / \num{ 2680 } / \num{ 2140.542714 } / \num{ 26.312901} & \num{16960 } / \num{ 18240 } / \num{ 17039.477387 } / \num{ 168.052780}\\
	\hline
	events.\allowbreak wait.\allowbreak blocking.\allowbreak k\_to\_k              - Wait for any events (w/ ctx switch)  & \num{5400 } / \num{ 6480 } / \num{ 5682.371859 } / \num{ 187.657344} & \num{5760 } / \num{ 6520 } / \num{ 5812.703518 } / \num{ 80.517344} & \num{19960 } / \num{ 23760 } / \num{ 20954.170854 } / \num{ 664.171522}\\
	\hline
	events.\allowbreak set.\allowbreak wake+ctx.\allowbreak k\_to\_k               - Set events (w/ ctx switch)  & \num{11480 } / \num{ 12120 } / \num{ 11548.783920 } / \num{ 77.558398} & \num{11000 } / \num{ 12120 } / \num{ 11302.391960 } / \num{ 123.527334} & \num{24120 } / \num{ 27400 } / \num{ 24982.954774 } / \num{ 500.724011}\\
	\hline
	events.\allowbreak wait\_all.\allowbreak blocking.\allowbreak k\_to\_k          - Wait for all events (w/ ctx switch)  & \num{5960 } / \num{ 6800 } / \num{ 5988.221106 } / \num{ 100.256712} & \num{5960 } / \num{ 7120 } / \num{ 6057.608040 } / \num{ 185.032680} & \num{19840 } / \num{ 24120 } / \num{ 20589.427136 } / \num{ 526.265284}\\
	\hline
	events.\allowbreak post.\allowbreak wake+ctx.\allowbreak k\_to\_k              - Post events (w/ ctx switch)  & \num{11120 } / \num{ 12160 } / \num{ 11462.753769 } / \num{ 165.085191} & \num{11080 } / \num{ 11800 } / \num{ 11110.874372 } / \num{ 87.481451} & \num{25360 } / \num{ 27480 } / \num{ 25949.427136 } / \num{ 407.869293}\\
	\hline
	semaphore.\allowbreak give.\allowbreak immediate.\allowbreak kernel          - Give a semaphore (no waiters)  & \num{2520 } / \num{ 3120 } / \num{ 2537.326633 } / \num{ 27.028532} & \num{2600 } / \num{ 2600 } / \num{ 2600.000000 } / \num{ 0.000000} & \num{17240 } / \num{ 19400 } / \num{ 17323.175879 } / \num{ 195.378679}\\
	\hline
	semaphore.\allowbreak take.\allowbreak immediate.\allowbreak kernel          - Take a semaphore (no blocking)  & \num{1600 } / \num{ 1640 } / \num{ 1615.718593 } / \num{ 19.536365} & \num{1680 } / \num{ 1680 } / \num{ 1680.000000 } / \num{ 0.000000} & \num{16440 } / \num{ 18400 } / \num{ 16545.366834 } / \num{ 202.809763}\\
	\hline
	semaphore.\allowbreak take.\allowbreak blocking.\allowbreak k\_to\_k           - Take a semaphore (context switch)  & \num{4560 } / \num{ 5480 } / \num{ 4594.251256 } / \num{ 98.458048} & \num{4640 } / \num{ 5280 } / \num{ 4650.613065 } / \num{ 74.179776} & \num{19480 } / \num{ 23600 } / \num{ 20730.894472 } / \num{ 655.335354}\\
	\hline
	semaphore.\allowbreak give.\allowbreak wake+ctx.\allowbreak k\_to\_k           - Give a semaphore (context switch)  & \num{8200 } / \num{ 8840 } / \num{ 8253.869347 } / \num{ 59.415943} & \num{7880 } / \num{ 8400 } / \num{ 7883.417085 } / \num{ 39.120763} & \num{23400 } / \num{ 27560 } / \num{ 24482.814070 } / \num{ 479.944847}\\
	\hline
	condvar.\allowbreak wait.\allowbreak blocking.\allowbreak k\_to\_k             - Wait for a condvar (context switch)  & \num{7960 } / \num{ 9800 } / \num{ 8300.422111 } / \num{ 248.805433} & \num{7920 } / \num{ 9720 } / \num{ 7997.105528 } / \num{ 139.089877} & \num{27800 } / \num{ 30960 } / \num{ 28254.874372 } / \num{ 557.202506}\\
	\hline
	condvar.\allowbreak signal.\allowbreak wake+ctx.\allowbreak k\_to\_k           - Signal a condvar (context switch)  & \num{10280 } / \num{ 11280 } / \num{ 10334.793970 } / \num{ 83.942028} & \num{10120 } / \num{ 10880 } / \num{ 10163.698492 } / \num{ 77.917554} & \num{23680 } / \num{ 26080 } / \num{ 23806.552764 } / \num{ 260.221158}\\
	\hline
	stack.\allowbreak push.\allowbreak immediate.\allowbreak kernel              - Add data to k\_stack (no ctx switch)  & \num{2680 } / \num{ 3240 } / \num{ 2701.748744 } / \num{ 32.251480} & \num{2760 } / \num{ 3600 } / \num{ 2761.608040 } / \num{ 31.175095} & \num{17960 } / \num{ 20960 } / \num{ 18153.809045 } / \num{ 334.966981}\\
	\hline
	stack.\allowbreak pop.\allowbreak immediate.\allowbreak kernel               - Get data from k\_stack (no ctx switch)  & \num{2800 } / \num{ 3160 } / \num{ 2852.582915 } / \num{ 48.891860} & \num{2840 } / \num{ 3120 } / \num{ 2861.025126 } / \num{ 54.691041} & \num{20880 } / \num{ 22640 } / \num{ 21023.959799 } / \num{ 200.786390}\\
	\hline
	stack.\allowbreak pop.\allowbreak blocking.\allowbreak k\_to\_k                - Get data from k\_stack (w/ ctx switch)  & \num{5120 } / \num{ 6200 } / \num{ 5178.090452 } / \num{ 173.919556} & \num{5240 } / \num{ 5760 } / \num{ 5242.894472 } / \num{ 35.524270} & \num{22440 } / \num{ 26760 } / \num{ 23344.402010 } / \num{ 822.046667}\\
	\hline
	stack.\allowbreak push.\allowbreak wake+ctx.\allowbreak k\_to\_k               - Add data to k\_stack (w/ ctx switch)  & \num{8440 } / \num{ 9040 } / \num{ 8451.095477 } / \num{ 54.208845} & \num{8200 } / \num{ 8960 } / \num{ 8452.301508 } / \num{ 53.828590} & \num{23280 } / \num{ 25480 } / \num{ 23410.613065 } / \num{ 273.153786}\\
	\hline
	mutex.\allowbreak lock.\allowbreak immediate.\allowbreak recursive.\allowbreak kernel    - Lock a mutex  & \num{1960 } / \num{ 2640 } / \num{ 1961.246231 } / \num{ 27.898860} & \num{2000 } / \num{ 2680 } / \num{ 2021.668342 } / \num{ 36.555603} & \num{16560 } / \num{ 18360 } / \num{ 16670.110553 } / \num{ 217.021148}\\
	\hline
	mutex.\allowbreak unlock.\allowbreak immediate.\allowbreak recursive.\allowbreak kernel  - Unlock a mutex  & \num{1440 } / \num{ 6680 } / \num{ 1462.512563 } / \num{ 168.138847} & \num{1520 } / \num{ 6320 } / \num{ 1525.869347 } / \num{ 153.833830} & \num{15640 } / \num{ 24840 } / \num{ 15763.296482 } / \num{ 369.326955}\\
	\hline
	heap.\allowbreak malloc.\allowbreak immediate                    - Average time for heap malloc  & \num{7120 } / \num{ 7120 } / \num{ 7120.000000 } / \num{ 0.000000} & \num{6800 } / \num{ 7360 } / \num{ 6805.894737 } / \num{ 57.151594} & \num{5360 } / \num{ 6120 } / \num{ 5420.210526 } / \num{ 127.930729}\\
	\hline
	heap.\allowbreak free.\allowbreak immediate                      - Average time for heap free  & \num{7120 } / \num{ 7120 } / \num{ 7120.000000 } / \num{ 0.000000} & \num{6800 } / \num{ 7360 } / \num{ 6805.894737 } / \num{ 57.151594} & \num{5360 } / \num{ 6120 } / \num{ 5420.210526 } / \num{ 127.930729}\\
	\bottomrule
  \end{supertabular}
\twocolumn

\end{document}